\documentclass[
  journal=pasa,
  manuscript=research-paper, 
  year=2024,
]{cup-journal}
\usepackage{microtype,siunitx,booktabs}
\usepackage[figuresright]{rotating}
\usepackage{amsmath} 
\usepackage{amssymb}

\def\degr{\hbox{$^\circ$}}

\def\farcm{\hbox{$.\mkern-4mu^\prime$}}
\def\farcs{\hbox{$.\!\!^{\prime\prime}$}}

\def\arcsec{\hbox{$^{\prime\prime}$}}

\usepackage{hyperref} 
\hypersetup{colorlinks,citecolor=blue,linkcolor=blue,urlcolor=blue}

\title{The GLEAMing of the first supermassive black holes: \newline III. Radio sources with ultra-faint host galaxies}

\author{J. W. Broderick}
\affiliation{International Centre for Radio Astronomy Research, Curtin University, GPO Box U1987, Perth, WA 6845, Australia}
\alsoaffiliation{SKA Observatory, Science Operations Centre, CSIRO ARRC, 26 Dick Perry Avenue, Kensington, WA 6151, Australia}
\alsoaffiliation{CSIRO Space and Astronomy, PO Box 1130, Bentley, WA 6102, Australia}
\email[J. W. Broderick]{jess.broderick@skao.int}

\author{N. Seymour}
\affiliation{International Centre for Radio Astronomy Research, Curtin University, GPO Box U1987, Perth, WA 6845, Australia}

\author{G. Drouart}
\affiliation{International Centre for Radio Astronomy Research, Curtin University, GPO Box U1987, Perth, WA 6845, Australia}

\author{D. Knight}
\affiliation{International Centre for Radio Astronomy Research, Curtin University, GPO Box U1987, Perth, WA 6845, Australia}

\author{J. M. Afonso}
\affiliation{Instituto de Astrof\'{i}sica e Ci\^{e}ncias do Espa\c co, Universidade de Lisboa, OAL, Tapada da Ajuda, PT1349-018 Lisboa, Portugal}
\alsoaffiliation{Departamento de F\'{i}sica, Faculdade de Ci\^{e}ncias, Universidade de Lisboa, Edif\'{i}cio C8, Campo Grande, PT1749-016 Lisbon, Portugal}

\author{C. De Breuck}
\affiliation{European Southern Observatory, Karl-Schwarzschild-Stra{\ss}e 2, D-85748 Garching bei M\"{u}nchen, Germany}

\author{T. J. Galvin}
\affiliation{CSIRO Space and Astronomy, PO Box 1130, Bentley, WA 6102, Australia}

\author{A. J. Hedge}
\affiliation{International Centre for Radio Astronomy Research, Curtin University, GPO Box U1987, Perth, WA 6845, Australia}

\author{M. D. Lehnert}
\affiliation{Universit\'e Lyon 1, ENS de Lyon, CNRS UMR5574, Centre de Recherche Astrophysique de Lyon, F-69230 Saint-Genis-Laval, France}

\author{G. Noirot}
\affiliation{Department of Astronomy \& Physics, Saint Mary’s University, 923 Robie Street, Halifax, NS B3H 3C3, Canada}
\alsoaffiliation{Space Telescope Science Institute, 3700 San Martin Drive, Baltimore, Maryland 21218, USA}

\author{S. S. Shabala}
\affiliation{School of Natural Sciences, University of Tasmania, Private Bag 37, Hobart, TAS 7001, Australia}

\author{R. J. Turner}
\affiliation{School of Natural Sciences, University of Tasmania, Private Bag 37, Hobart, TAS 7001, Australia}

\author{J. Vernet}
\affiliation{European Southern Observatory, Karl-Schwarzschild-Stra{\ss}e 2, D-85748 Garching bei M\"{u}nchen, Germany}

\received {dd Mmm YYYY}
\revised  {dd Mmm YYYY}
\accepted {dd Mmm YYYY}
\published{dd Mmm YYYY}


\begin{document}

\begin{abstract}
We present deep near-infrared $K_{\rm s}$-band imaging for 35 of the 53 sources from the high-redshift ($z > 2$) radio galaxy candidate sample defined in \citet[][]{broderick22}. These images were obtained using the High-Acuity Widefield $K$-band Imager (HAWK-I) on the Very Large Telescope. Host galaxies are detected for 27 of the sources, with $K_{\rm s} \approx 21.6$--23.0\,mag (2\arcsec\:diameter apertures; AB). The remaining eight targets are not detected to a median $3\sigma$ depth of $K_{\rm s} \approx 23.3$\,mag ( 2\arcsec\:diameter apertures). We examine the radio and near-infrared flux densities of the 35 sources, comparing them to the known $z > 3$ powerful radio galaxies with 500-MHz radio luminosities $L_{500\,{\rm MHz}} > 10^{27}$\,W\,Hz$^{-1}$. By plotting 150-MHz flux density versus $K_{\rm s}$-band flux density, we find that, similar to the sources from the literature, these new targets have large radio to near-infrared flux density ratios, but extending the distribution to fainter flux densities. Five of the eight HAWK-I deep non-detections have a median $3\sigma$ lower limit of $K_{\rm s} \gtrsim 23.8$\,mag (1\farcs5 diameter apertures); these five targets, along with a further source from \citet[][]{broderick22} with a deep non-detection ($K_{\rm s} \gtrsim 23.7$\,mag; $3\sigma$; 2\arcsec\:diameter aperture) in the Southern H-ATLAS Regions $K_{\rm s}$-band Survey, are considered candidates to be ultra-high-redshift ($z > 5$) radio galaxies. The extreme radio to near-infrared flux density ratios ($>10^5$) for these six sources are comparable to TN~J0924$-$2201, GLEAM~J0856$+$0223 and TGSS~J1530$+$1049, the three known powerful radio galaxies at $z > 5$. For a selection of galaxy templates with different stellar masses, we show that $z \gtrsim 4.2$ is a plausible scenario for our ultra-high-redshift candidates if the stellar mass $M_{\rm *} \gtrsim 10^{10.5}$\,M$_\odot$. In general, the 35 targets studied have properties consistent with the previously known class of infrared-faint radio sources. We also discuss the prospects for finding more UHzRG candidates from wide and deep near-infrared surveys. 

\end{abstract}

\section{INTRODUCTION }
\label{sec:int}

This paper is the third in a series describing an ongoing programme to find high-redshift radio galaxies with powerful radio emission \citep[HzRGs; redshift $z > 2$ and 500-MHz radio luminosity $L_{500\,{\rm MHz}} > 10^{27}$\,W\,Hz$^{-1}$; see review by][]{miley08}. In particular, to find new candidates, we have cross-matched the wideband 72--231\,MHz GaLactic and Extragalactic All-sky Murchison Widefield Array \citep[MWA;][]{tingay13} survey \citep[GLEAM;][]{wayth15} with the Visible and Infrared Survey Telescope for Astronomy (VISTA; \citealt[][]{dalton06}; \citealt*[][]{emerson06}) Kilo-degree Infrared Galaxy survey \citep[VIKING;][]{edge13}. In \citet[][]{drouart20}, we described a pilot study of four HzRG candidates, introducing a novel selection technique that considers both radio spectral steepness and curvature within the GLEAM band. This study led to the discovery of GLEAM~J0856$+$0223, a powerful radio galaxy at $z=5.55$, as well as another potentially ultra-high-redshift source that was further investigated in two companion studies \citep[GLEAM~J0917$+$0012;][]{drouart21,seymour22}. Then, in \citet[][]{broderick22}, henceforth B22, we defined a larger GLEAM--VIKING sample of 53 HzRG candidates (including GLEAM~J0856$+$0223 and GLEAM~J0917$+$0012). 

Our primary aim is that the B22 sample will provide high-quality candidate powerful radio galaxies at $z > 5$; we will henceforth refer to such sources as ultra-high-redshift radio galaxies (UHzRGs). Finding UHzRGs, particularly those within the Epoch of Reionisation (EoR; $z \gtrsim 6.5$), would provide further vital observational constraints for theoretical explanations of the formation and rapid growth of supermassive black holes (SMBHs) in the early Universe with mass $M_{\rm BH} \sim 10^{8-9}$\,M$_\odot$ (e.g. \citealt{volonteri12}; \citealt{johnson13}; \citealt{latif16}; \citealt*{smith17}; \citealt{wise19}; \citealt{smith19}; \citealt{inayoshi20}; \citealt{kroupa20}; \citealt*[][]{dimatteo23}). Identifying powerful radio galaxies within the EoR would also provide bright background radio sources to facilitate searches for redshifted 21-cm hydrogen absorption by the neutral intergalactic medium, including the 21-cm forest from intervening material along the line of sight \citep[e.g.][]{carilli04,mack12,ciardi15}, and also allow vital constraints to be placed on active galactic nucleus (AGN) feedback \citep[e.g. review by][]{hardcastle20} at early cosmic times.

While an increasing number of radio-loud quasars are now known at $z > 5$ \citep[e.g.][]{banados18,banados21,banados23,ighina21,ighina22,ighina23,ighina24a,ighina24b,momjian21,gloudemans22,belladitta23}, all but one of these sources are not radio powerful according to our aforementioned luminosity criterion. The exception is PSO~J352.4034$-$15.3373 at $z=5.84$ \citep[][]{banados18}.\footnote{Furthermore, depending on the extrapolation of the radio spectrum below 100\,MHz, PSO~J030947.49$+$271757.31 at $z=6.10$ \citep[][]{belladitta20} is potentially a radio-powerful blazar according to our radio luminosity selection criterion.} These radio-loud quasars are also not obscured (in the rest-frame ultraviolet and optical). The discovery of the obscured AGN COS-87259 at $z=6.85$ \citep[][]{endsley22,endsley23b} was therefore a particularly noteworthy development. This source has $M_{\rm BH} \gtrsim 1.5 \times 10^9$\,M$_\odot$ \citep[][]{endsley23b}, and is the only confirmed obscured and radio-loud (yet weakly radio emitting: $L_{500\,{\rm MHz}} \approx 6.9 \times 10^{25}$\,W\,Hz$^{-1}$) AGN currently known at $z > 6$. Additionally, the source COSW-106725 is a candidate $z \sim 7.7$ obscured AGN with $M_{\rm BH} \geq 6.4 \times 10^8$\,M$_\odot$ \citep[][]{lambrides24}. If the redshift is confirmed, and depending on the extrapolation of the radio spectrum below 100\,MHz, this source would have $L_{500\,{\rm MHz}} \approx 2.3 \times 10^{27}$\,W\,Hz$^{-1}$, i.e. it would be powerful in the radio. 

\citet[][]{gilli22} suggested that 80--90\,per cent of SMBHs at $z = 6$--$8$ are obscured, mainly by the dense interstellar medium (ISM) of the host galaxy. Moreover, \citet[][]{johnson22} presented a model where sustained super-Eddington growth remains hidden due to the extreme infall rate of material, preventing any radiation from escaping. These authors demonstrated that the obscured super-Eddington accretion rates can be maintained even for non-spherical accretion (i.e. more disc-like accretion), which would allow a pathway for radio jets to emerge. Therefore, a compelling picture is beginning to emerge in which radio selection, relatively unaffected by a dense ISM in the host galaxy, may play an especially vital role in finding AGN in the early Universe. Indeed, using wide-area radio surveys such as GLEAM to find obscured AGN provides a unique way to probe this population at high redshift. We note that if the radio lobes are sufficiently compact and within the host galaxy, there could be free--free absorption by the ISM, particularly at lower frequencies \citep*[e.g.][]{bicknell97}. By selection, the B22 sample will have low-frequency turnovers, which could be due to free--free absorption. Hence, it is of particular interest whether the B22 sample contains brighter radio analogues of COS-87259.

It is well known that there is a tight correlation between near-infrared $K$-band (2.2\,\textmu m) magnitude and redshift for powerful radio galaxies, which are among the most massive galaxies at a given redshift \citep[the $K$--$z$ relation; e.g.][]{lilly84,eales97,vanbreugel98,willott03,roccavolmerange04}. This $K$-band light from the host galaxy traces the old stellar population at low redshift, although it gradually probes the optical and then the ultraviolet with increasing redshift. Given the $K$--$z$ relation, an HzRG search can be made significantly more efficient by selecting only those sources in large radio catalogues with $K$-band magnitudes fainter than a given cutoff \citep[e.g.][]{ker12}. Dedicated $K$-band follow-up of HzRG candidate samples has been commonplace in the literature \citep[e.g.][]{chambers96,vanbreugel99,jarvis01,jarvis04,debreuck02,debreuck04,cruz06,bryant09,saxena18,saxena19}.    

The B22 HzRG candidate sample was partly selected on the basis of VIKING $K_{\rm s}$-band (2.15\,\textmu m) host galaxy non-detections: $K_{\rm s} > 21.2$\,mag ($5\sigma$ point-source limits presented in Table 5 in B22; median seeing full width at half maximum (FWHM) 0\farcs83; AB). As was described in B22, three of the 53 sources in our sample have deeper limits from the Southern H-ATLAS\footnote{The {\it Herschel} Astrophysical Terahertz Large Area Survey \citep[][] {eales10,valiante16,bourne16}.} Regions $K_{\rm s}$-band Survey \citep[SHARKS;][]{dannerbauer22}. A further four sources have $K_{\rm s}$-band host galaxy detections from Very Large Telescope \citep[VLT;][]{eso98} observations using the High-Acuity Widefield $K$-band Imager \citep[HAWK-I;][]{kissler08}, including both GLEAM~J0856$+$0223 and GLEAM~J0917$+$0012 from the \citet[][]{drouart20} pilot study. In this paper, we significantly increase the number of sources in our sample with deep $K_{\rm s}$-band imaging by presenting the results from a HAWK-I $K_{\rm s}$-band observing campaign for an additional 35 targets. Such deep HAWK-I imaging allows us to further narrow down the best UHzRG candidates compared to the VIKING selection alone; for this particular observing campaign, the improvement in sensitivity is $\sim$ 1--2\,mag at an equivalent detection threshold. Hence, we can significantly increase the efficiency of spectroscopic follow-up, which can be an expensive and often challenging process, particularly in terms of observing time given that (U)HzRGs have a very low surface density \citep[e.g.][]{miley08}.       

The layout of this paper is as follows. In Section~\ref{sec:obs}, we describe the HAWK-I observations as well as our method for detecting the host galaxies and measuring their magnitudes. We also outline how a compilation of radio and near-infrared flux densities was obtained for both the B22 sample and (U)HzRGs from the literature. We present our results in Section~\ref{sec:res}, including the host galaxy magnitudes as well as the $K_{\rm s}$-band--radio overlay plots for our best UHzRG targets. A discussion follows in Section~\ref{sec:dis}, in particular focusing on (i) how our data set compares to the literature in radio--near-infrared flux density space, and (ii) plausible scenarios for our best UHzRG candidates. We then present our conclusions in Section~\ref{sec:con}. Lastly, \ref{appendix} contains the remaining overlay plots as well as notes on individual sources.  

In this paper, we use the following conventions. Uncertainties are given as $\pm 1\sigma$. Magnitudes are given in the AB system \citep[][]{oke74}; to convert between Vega and AB magnitudes, we use the relation $K_{\rm s, AB} = K_{\rm s, Vega} + 1.85$ as given in e.g. \citet[][]{blanton07}. We also assume a flat, Lambda cold dark matter ($\Lambda$CDM) cosmology with Hubble constant $H_0=67.7$\,km s$^{-1}$ Mpc$^{-1}$, matter density parameter $\Omega_{\rm M}=0.31$ and vacuum density parameter $\Omega_{\Lambda}=0.69$ \citep{planck20}. Additionally, $\log$ refers to the decimal logarithm (base 10). 

\section{OBSERVATIONS, DATA REDUCTION AND ANALYSIS}
\label{sec:obs}

\subsection{HAWK-I observing campaign and data products}\label{sec:HAWKI overview}

Our HAWK-I observations of the B22 sample (49 targets without previous HAWK-I observations) took place from 2021 October 20 to 2022 March 30 in service mode (ESO period P108; programme ID 108.22HY.001). HAWK-I has a field of view (FoV) of $7\farcm5 \times 7\farcm5$ and an average pixel size of 0\farcs1066. There were some very small differences in pixel size from image to image, but only at the level of the fourth decimal place. For 33 sources, observations comprised 12 sets of $17 \times 10$\,s jittered exposures, where the jitter box width was set to 30\arcsec, i.e. 34\,min on-source per target. Shorter observations were possible for two sources: J0129$-$3109 (680\,s) and J0216-3301 (510\,s). To improve the seeing, we used the GRound layer Adaptive optics Assisted by Lasers (GRAAL); the angular resolution ranged from 0\farcs33--0\farcs78 FWHM, with a median value of 0\farcs46. While our programme was not fully completed by the end of P108, we obtained science-quality images for 71\,per cent of the targets (35/49 sources), with most observations having a quality control grade `A' and a handful `B'. Reduced data were mostly obtained from the automatic HAWK-I pipeline,\footnote{\url{https://doi.org/10.18727/archive/34}} although for J0042$-$3515, J0129$-$3109 and J0216-3301 some manual data reduction steps were necessary.  

\subsection{$K_{\rm s}$-band magnitude calibration}\label{sec:mag cal} 

An initial magnitude calibration was provided by the automatic HAWK-I pipeline. Then, as in B22, we used the Two-Micron All-Sky Survey \citep[2MASS;][]{skrutskie06} point-source catalogue \citep[][]{cutri03} to refine the magnitude calibration of the images. We used a sample of bright stars ($K_{\rm s} < 16.9$\,mag) within the HAWK-I FoV to calculate the inverse-variance-weighted mean magnitude offset to subtract from the host galaxy magnitude. These corrections were mostly small, with a median value of 0.13\,mag. The one large outlier was the correction required for J0042$-$3515: 3.25\,mag. However, given that the standard error of the weighted mean is only 0.01\,mag in this case, we have confidence that the calibrated magnitude for the host galaxy of this source is reliable. Galactic extinction was not corrected for given that the corrections are much smaller than the magnitude uncertainties for our targets.     

\subsection{Host galaxy magnitude measurements}\label{sec: host galaxy measurements}  

We adopted a conservative approach when searching for the $K_{\rm s}$-band counterparts of our targets. While some sources have multiple potential host galaxy candidates, we only assigned a host when a $K_{\rm s}$-band source is plausibly close to the radio centroid, as one might expect for a compact radio source. Otherwise, the source was classified as having a $K_{\rm s}$-band non-detection. We will further discuss this strategy in Section~\ref{sec:dis_alternate}.

Host galaxy magnitudes were determined as follows. Firstly, we used Source Extractor \citep[{\sc sextractor};][]{bertin96} to efficiently identify as many host galaxy detections as possible. To increase the reliability of the results, we used a signal-to-noise (S/N) detection threshold $\geq 3$ over at least three connected pixels. However, it became apparent that some sources were not detected by {\sc sextractor} using the above connected pixel criterion, yet there was visual evidence of a host galaxy (e.g. faint diffuse emission). Therefore, to ensure that the host galaxy magnitudes were determined in as consistent a manner as possible, we instead used {\sc photutils} \citep[][]{bradley21} to measure a magnitude in a 2\arcsec\:diameter circular aperture for each source. We chose an aperture of this size as it is one of the standard choices used in the literature and many of our host galaxy detections are compact (as one might expect for a source at high redshift). In some cases, a larger circular aperture would be more appropriate to enclose all of the flux density from a given source; further information is provided in \ref{appendix}.  

Each aperture was centred on, in order of preference, (i) the flux-weighted centre as determined by {\sc sextractor}, (ii) the $K_{\rm s}$-band centroid as estimated by eye if the S/N was insufficient for (i), or (iii) the radio centroid when (ii) was deemed to be inaccurate due to low S/N. For (iii), we used the high-resolution Australia Telescope Compact Array \citep[ATCA;][]{frater92} data from B22 where available, factoring in both the angular resolution and S/N to choose the most accurate position at either 5.5 or 9\,GHz. If high-resolution ATCA data were not available, we instead used the radio position from the 3-GHz VLA Sky Survey \citep[VLASS;][]{lacy20}.    

A host galaxy detection was regarded as significant if the integrated flux density determined with {\sc photutils} within the 2\arcsec\:diameter aperture had S/N $\geq 3$. The flux density statistical uncertainty was calculated by randomly placing $1000 \times 2$\arcsec\:diameter apertures within each image, avoiding sources. The $1\sigma$ uncertainty was then the standard deviation of these measurements. Each magnitude uncertainty was also calculated with the appropriate error propagation. In some cases (see the notes on individual sources in ~\ref{appendix}), we repeated this exercise with apertures of different sizes. For non-detections, we report a $3\sigma$ flux density (magnitude) upper (lower) limit. Moreover, for detections, we conservatively added a 2\,per cent flux density calibration error in quadrature with the statistical uncertainty (followed by error propagation to obtain the magnitude uncertainty). 

\subsection{Radio and near-infrared flux densities for the B22 sample and (U)HzRGs from the literature}\label{sec: flux flux analysis}

\begin{figure*}[t]
\begin{minipage}{1.0\textwidth}
\centering
\includegraphics[height=11cm]{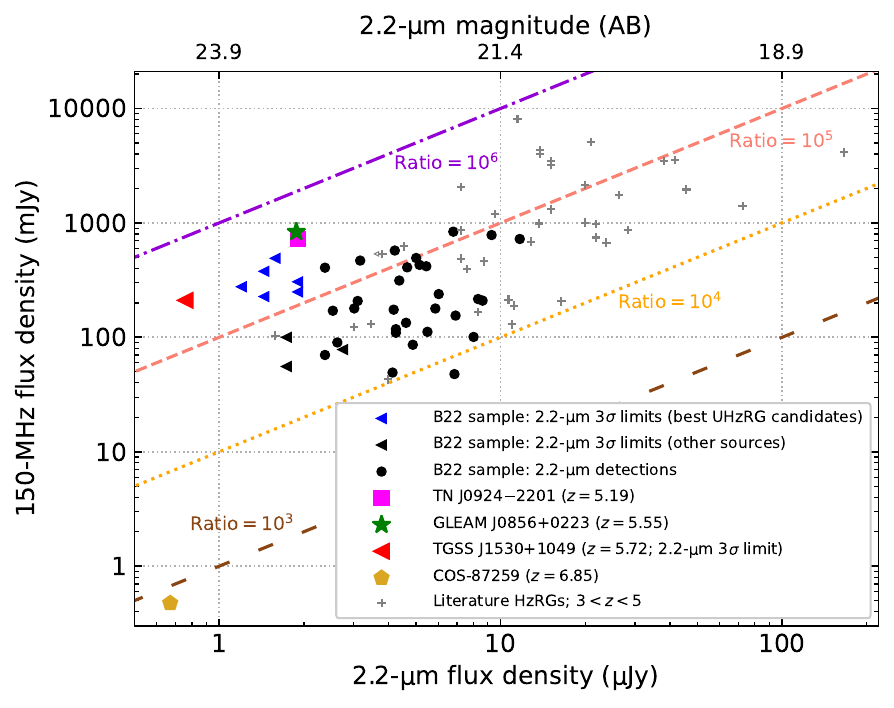}
\end{minipage}
\caption{150-MHz radio flux density (ordinate) versus 2.2-\textmu m near-infrared flux density (bottom abscissa) for 40/53 radio sources from the B22 sample as well as for other HzRGs and UHzRGs from the literature with $3.08 \leq z \leq 6.85$. Also plotted are four lines denoting various radio to near-infrared flux density ratios: $S_{150\,\rm{MHz}} / S_{2.2\,\mu\rm{m}} =10^{3}$--$10^{6}$. The top abscissa shows the AB magnitudes corresponding to a selection of the 2.2-\textmu m flux densities. Note that one of the grey `$+$' symbols is augmented with a triangle to denote a 2.2-\textmu m flux density upper limit. For the sake of clarity, error bars are not plotted. Further details on the figure, including a discussion of the data presented, can be found in Sections~\ref{sec:obs} and \ref{sec:dis}. The references used for the 150-MHz flux densities are as follows: \citet[][]{waldram96}, \citet[][]{hurleywalker17}, \citet[][]{intema17} and \citet[][]{endsley22}. Furthermore, the references used for the 2.2-\textmu m flux densities and magnitudes are as follows: \citet[][]{eales93}, \citet[][]{vanbreugel98,vanbreugel99}, \citet[][]{villani99}, \citet[][]{stern99}, \citet[][]{jarvis01}, \citet[][]{debreuck02}, \citet[][]{debreuck04}, \citet[][]{brookes06}, \citet[][]{cruz06}, \citet[][]{seymour07}, \citet[][]{jarvis09}, \citet[][]{parijskij14}, \citet[][]{saxena18}, \citet[][]{saxena19}, \citet[][]{drouart20}, \citet[][]{seymour22}, \citet[][]{endsley22} and B22.} 
\label{fig:flux_flux}
\end{figure*}

In Figure~\ref{fig:flux_flux}, we plot the 150-MHz radio flux density, $S_{150\,\rm{MHz}}$, versus 2.2-\textmu m near-infrared flux density, $S_{2.2\,\mu\rm{m}}$, for 40/53 sources from the B22 sample as well as (to the best of our knowledge) all of the known powerful radio galaxies at $z > 3$ with $L_{500\,{\rm MHz}} > 10^{27}$\,W\,Hz$^{-1}$ and $K_{\rm s}$-band data, plus the obscured, radio-loud AGN COS-87259 at $z=6.85$ (42 sources from the literature in total, one of which is also in the B22 sample: the UHzRG GLEAM~J0856$+$0223 at $z=5.55$ discovered by \citealt{drouart20}). 

From the B22 sample, we only considered the sources that have deep HAWK-I or SHARKS $K_{\rm s}$-band images presented either in this paper (35 sources) or by us previously (six sources including GLEAM~J0856$+$0223; \citealt{drouart20}; \citealt{seymour22}; B22). The literature sources were obtained from the compilation of $3 < z < 5$ radio galaxies in \citet[][with the exception of two sources which only had optical and not near-infrared identifications]{miley08}, with additional sources from \citet[][]{jarvis09}, \citet[][]{saxena19} and \citet[][]{yamashita20}. We also included the UHzRGs TN~J0924$-$2201 \citep[$z=5.19$;][]{vanbreugel99} and TGSS~J1530$+$1049 \citep[$z=5.72$;][]{saxena18}. However, we did not include the radio galaxy VLA J123642$+$621331 at $z=4.42$ \citep[][]{waddington99} as it is a weaker radio source that does not meet our radio luminosity criterion; the $3\sigma$ upper limit at 147.5\,MHz from the
Tata Institute of Fundamental Research (TIFR) Giant Metrewave Radio Telescope \citep[GMRT;][]{swarup91} Sky Survey (TGSS) Alternate Data Release 1 \citep[][]{intema17} is about 6\,mJy\,beam$^{-1}$.  

The values of $S_{150\,\rm{MHz}}$ and $S_{2.2\,\mu\rm{m}}$ per source were determined as follows.  

\subsubsection{Calculation of 150-MHz flux densities}\label{sec:dis_rad}

Apart from two cases, the 150-MHz flux densities were obtained from GLEAM or TGSS. Where GLEAM data were available from both the Extragalactic data release \citep[][]{hurleywalker17} as well as the deeper South Galactic Pole data release \citep[][]{franzen21}, we used the data from the latter. Furthermore, where GLEAM and TGSS data were available for a particular source, we used the GLEAM flux densities only. Some of the sources in the literature sample are too far north on the sky to have been catalogued in GLEAM, but TGSS data were available instead. The sources without GLEAM or TGSS data are as follows. Firstly, the $z=3.395$ radio galaxy B2~0902$+$34 is too far north to have been catalogued in GLEAM, but it was also not catalogued in TGSS despite being a bright radio source (possibly due to it being in one of the failed pointings in the survey; see \citealt[][]{intema17}\footnote{Further information can be found at \url{https://tgssadr.strw.leidenuniv.nl/doku.php?id=knownproblems}.}), and so we instead used the 151-MHz flux density from the 7C survey \citep[][]{waldram96}. Secondly, COS-87259 is too faint to have been catalogued in GLEAM and TGSS, and we instead used the 144-MHz flux density reported in \citet[][]{endsley22} from the Low-Frequency Array \citep[LOFAR;][]{vanhaarlem13} Two-metre Sky Survey \citep[LoTSS;][]{shimwell17,shimwell19,shimwell22}.      

To enable as consistent a comparison as possible between the B22 sample and the sample from the literature, we did not use the curved fits to the broadband GLEAM data for the former presented in B22, but instead took the 143-, 151-, 158- and 166-MHz GLEAM flux densities for each source and calculated the inverse-variance-weighted average. This step removes some of the scatter in the single-band GLEAM flux densities. At each frequency, a 2\,per cent internal calibration uncertainty (see \citealt{hurleywalker17}) was added in quadrature to the fitting error reported in the GLEAM catalogue before the weighted average was computed.

Note that for simplicity, we did not make any corrections for the slightly different central frequencies of the data discussed above, nor for the fact that these values have small offsets from a frequency of 150\,MHz. The scale offsets between the data sets used will affect the flux densities by $\lesssim 10$\,per cent (or $ \lesssim 0.04$\,dex).  

The inverse-variance-weighted average 150-MHz GLEAM flux densities for the 35 sources from the B22 sample with HAWK-I data presented in this paper are listed in Table~\ref{tab:magnitudes}.

\subsubsection{Calculation of 2.2-\textmu m flux densities}\label{section:dis_nir}

Regarding the 2.2-\textmu m flux densities, different strategies were used in the literature to obtain the most accurate measurement for a given source, particularly regarding the size of the aperture used. Where more than one measurement was available for the literature sample, we used the value from the largest-sized aperture considered in order to increase the likelihood that all of the flux density from the galaxy was enclosed in the aperture. In many cases, this was an 8\arcsec\:diameter aperture or a 64-kpc diameter aperture. Furthermore, to ensure consistency, we took the magnitude measurements from apertures larger than a 2\arcsec\:diameter for the sources J0007$-$3040, J0034$-$3112, J0048$-$3540, J0301$-$3132, J0326$-$3013, J0909$-$0154, J1032$+$0339, J1040$+$0150 and J1052$-$0318 from the B22 sample (\ref{appendix}; also see Section~\ref{sec: host galaxy measurements}).  

In \citet[][]{jarvis09}, the 2.2-\textmu m magnitude of the HzRG FIRST~J163912.11$+$405236.5 ($z=4.88$) was estimated to be $K = 24.2$\,mag (4\arcsec\:diameter aperture). This estimate was obtained by extrapolating from a 3.6-\textmu m detection with an assumed colour (2.2-\textmu m $-$ 3.6-\textmu m = 2.25\,mag). Alternatively, using the photometry reported in \citet[][]{jarvis09}, we followed a similar approach to the one that we used in \citet[][]{seymour22} and carried out broadband spectral energy distribution (SED) fitting with the {\sc p\'egase} code \citep[version 3;][]{fioc19} for different galaxy templates; the 2.2-\textmu m magnitude in a 4\arcsec\:diameter aperture is estimated to be $K \sim 23.4$\,mag (i.e. a flux density a factor of $\sim$ 2 brighter: $S_{2.2\,\mu\rm{m}}\sim 1.6$\,\textmu Jy). We have used our estimated magnitude in this study; in either case, this source is the faintest at 2.2\,\textmu m from our compilation of HzRGs at $3 < z < 5$. 

The SHARKS magnitude limit for J0008$-$3007 reported in B22 was converted from $5\sigma$ to $3\sigma$ (i.e. by adding 0.55\,mag; $K_{\rm s} \gtrsim 23.7$\,mag and $S_{2.2\,\mu\rm{m}} \lesssim 1.2$\,\textmu Jy; 2\arcsec\:diameter aperture). The SHARKS limit for J0007$-$3040 reported in B22 is superseded by the HAWK-I data presented in this study. Regarding the SHARKS data presented in B22 for J2340$-$3230, the interpretation of this source (and hence the validity of a $3\sigma$ limit) is uncertain (see Section 5.5 in B22), and deeper $K_{\rm s}$-band data are needed. We therefore did not include J2340$-$3230 in our analysis, i.e. the number of sources from the sample with previously presented deep $K_{\rm s}$-band data was reduced from six to five.    

For simplicity, we did not make any corrections for the slightly different central wavelengths of the image filters used in the various studies in the literature (e.g. $K_{\rm s}$ versus $K$), and considered all values as measurements at 2.2\,\textmu m. 
\newline
\newline
\section{RESULTS}
\label{sec:res}

\subsection{Host galaxy magnitudes}\label{sec:mag}

\begin{table*}
  \caption{$K_{\rm s}$-band properties of the 35 sources observed with HAWK-I. We present the host galaxy magnitudes and corresponding flux densities determined in 2\arcsec\:diameter circular apertures along with uncertainties. Magnitude (flux density) lower (upper) limits are $3\sigma$ values. Host galaxy positions are from {\sc sextractor} unless specified with a footnote. Where the $K_{\rm s}$-band position was sufficiently well determined, we also provide the $K_{\rm s}$-band--radio angular separation. Further details can be found in Sections~\ref{sec: host galaxy measurements} and \ref{sec:mag}. Furthermore, we list 150-MHz inverse-variance-weighted average flux densities from GLEAM; see Section~\ref{sec:dis_rad}. The uncertainties for these GLEAM flux densities include an extra 8\,per cent absolute flux density calibration uncertainty (see \citealt{hurleywalker17}) added in quadrature.}
  \begin{tabular}{crrrrrr}
  \hline\hline
  Source & \multicolumn{1}{c}{$K_{\rm s}$-band}  & \multicolumn{1}{c}{$S_{2.15\,\mu{\rm m}}$} & \multicolumn{1}{c}{$K_{\rm s}$-band position} & $K_{\rm s}$-band--radio & \multicolumn{1}{c}{$S_{150\,{\rm MHz}}$} \\ 
 \multicolumn{1}{c}{(GLEAM)} & \multicolumn{1}{c}{magnitude (AB)} & \multicolumn{1}{c}{(\textmu Jy)} & \multicolumn{1}{c}{(J2000)} & separation (\arcsec) & \multicolumn{1}{c}{(mJy)} \\ 
   \hline
J000216$-$351433 & $ 22.24 \pm 0.12 $ & $ 4.61 \pm 0.51 $ & 00:02:16.49\:$-$35:14:32.4 & $0.41$ & $134 \pm 12$ \\
J000614$-$294640 & $ 22.89 \pm 0.23 $ & $2.54 \pm 0.54 $ & 00:06:14.04\:$-$29:46:41.0\rlap{$^{\rm a}$} & $\cdots$\rlap{$^{\rm b}$}  & $171 \pm 14$  \\
J000737$-$304030 & $ 22.64 \pm 0.19 $ & $ 3.19 \pm 0.56 $ & 00:07:37.26\:$-$30:40:34.5 & $2.9$\rlap{$^{\rm c}$} &  $573 \pm 46$  \\
J003402$-$311210 & $ 22.78 \pm 0.24 $ & $ 2.81 \pm 0.62 $ & 00:34:02.00\:$-$31:12:12.1\rlap{$^{\rm d}$} & $3.5$ & $47.7 \pm 5.2$  \\
J004219$-$351516 & $ 22.96 \pm 0.21 $ & $2.38 \pm 0.46 $ & 00:42:19.55\:$-$35:15:23.4\rlap{$^{\rm a}$} & $\cdots$\rlap{$^{\rm b}$} & $70.1 \pm 7.5$  \\
J004828$-$354005 & $ 21.85 \pm 0.10$ & $6.61 \pm 0.61$ &  00:48:28.34\:$-$35:40:06.9 & $0.56$ &  $209 \pm 17$  \\
J005332$-$325630\rlap{$^{\dag}$} & $ >23.2 $ & $<1.9$ & $\cdots$ & $\cdots$ & $249 \pm 20$   \\
J010826$-$350157 & $ 22.30 \pm 0.12 $ & $4.37 \pm 0.48 $ & 01:08:25.97\:$-$35:01:57.2 & $0.32$ &  $314 \pm 26$  \\
J012929$-$310915 & $ 22.05 \pm 0.22 $ & $5.50 \pm 1.11 $ & 01:29:28.97\:$-$31:09:17.5 & $0.19$ &  $111.7 \pm 9.4$ \\
J020118$-$344100 & $ 22.85 \pm 0.20 $ & $2.63 \pm 0.48 $ & 02:01:18.39\:$-$34:41:03.4 & $0.43$ & $90.2 \pm 8.1$  \\
J021618$-$330148 & $ >22.8 $ & $<2.8$ & $\cdots$ & $\cdots$ & $78.2 \pm 7.6$ \\
J023937$-$304337 & $ 21.60 \pm 0.08 $ & $8.32 \pm 0.61 $ & 02:39:37.65\:$-$30:43:37.7 & $0.050$ & $216 \pm 18$   \\
J024019$-$320659 & $ >23.3 $ & $<1.7$ & 02:40:20.14\:$-$32:07:01.1\rlap{$^{\rm e}$} & $0.57$\rlap{$^{\rm e}$} & $55.6 \pm 5.7$  \\
J030108$-$313211 & $ 22.89 \pm 0.24 $ & $2.54 \pm 0.56 $ & 03:01:08.46\:$-$31:32:12.5\rlap{$^{\rm a}$} & $\cdots$\rlap{$^{\rm b}$} & $239 \pm 20$   \\
J030931$-$352623 & $ 22.18 \pm 0.14 $ & $4.88 \pm 0.63 $ & 03:09:31.98\:$-$35:26:28.5 & $0.81$\rlap{$^{\rm f}$} & $86.1 \pm 8.0$ \\
J032634$-$301359 & $ 21.91 \pm 0.09 $ & $6.25 \pm 0.52$ & 03:26:34.54\:$-$30:14:02.3 & $1.9$ & $723 \pm 58$   \\
J090942$-$015409 & $ 21.75 \pm 0.08 $ & $7.24 \pm 0.53$ & 09:09:42.34\:$-$01:54:05.9  & $0.71$ & $783 \pm 65$  \\
J103055$+$013519 & $ 22.67 \pm 0.20 $ & $3.10 \pm 0.57 $ & 10:30:55.26\:$+$01:35:24.1\rlap{$^{\rm d}$} & $0.32$ & $208 \pm 21$  \\
J103223$+$033933 & $ 22.24 \pm 0.13 $ & $4.61 \pm 0.55 $ & 10:32:23.63\:$+$03:39:41.6 & $0.51$ & $838 \pm 69$ \\
J103340$+$010725\rlap{$^{\dag}$} & $ >23.5 $ & $<1.4 $ & $\cdots$ & $\cdots$ & $377 \pm 33$  \\
J103747$-$032519 & $ >23.3$ & $<1.7 $ & $\cdots$ & $\cdots$ & $100 \pm 13$  \\
J104041$+$015003 & $ 22.64 \pm 0.20 $ & $3.19 \pm 0.59 $ & 10:40:41.51\:$+$01:50:08.3\rlap{$^{\rm d}$} & $0.73$ & $418 \pm 36$  \\
J105232$-$031808 & $ 22.12 \pm 0.12 $ & $5.15 \pm 0.57 $ & 10:52:31.95\:$-$03:18:04.0 & $1.1$ & $155 \pm 16$  \\
J111211$+$005607 & $ 21.98 \pm 0.11 $ & $5.86 \pm 0.59 $ & 11:12:10.23\:$+$00:56:25.1 & $0.72$ & $178 \pm 21$  \\
J112557$-$034203 & $ 22.15 \pm 0.11 $ & $5.01 \pm 0.51 $ & 11:25:57.38\:$-$03:42:04.2 & $0.89$ & $493 \pm 41$  \\
J112706$-$033210\rlap{$^{\dag}$} & $ >23.2$ & $ <1.9 $ & $\cdots$ & $\cdots$ & $306 \pm 27$  \\
J113601$-$035122 & $ 22.35 \pm 0.16 $ & $4.17 \pm 0.61 $ & 11:36:01.64\:$-$03:51:18.4 & $0.19$ & $174 \pm 18$   \\
J114103$-$015846 & $ 22.33 \pm 0.15 $ & $4.25 \pm 0.59 $ & 11:41:03.35\:$-$01:58:46.0 & $0.37$ & $110 \pm 14$  \\
J133531$+$011219 & $ 22.12 \pm 0.12 $ & $5.15 \pm 0.57 $ & 13:35:31.17\:$+$01:12:19.6 & $0.076$ & $430 \pm 38$  \\
J144305$+$022940\rlap{$^{\dag}$} & $ >23.5 $ & $ <1.4 $ &  $\cdots$ & $\cdots$ & $227 \pm 30$   \\
J221921$-$331206 & $ 22.70 \pm 0.19 $ & $3.02 \pm 0.53 $ & 22:19:21.92\:$-$33:12:10.7 & $0.33$ & $179 \pm 15$ \\
J231148$-$335918 & $ 22.36 \pm 0.13 $ & $4.13 \pm 0.49 $ & 23:11:48.80\:$-$33:59:24.0 & $1.3$ & $49.2 \pm 5.8$ \\
J231456$-$351721 & $ 22.33 \pm 0.14 $ & $4.25 \pm 0.55 $ & 23:14:56.72\:$-$35:17:29.2 & $0.54$ & $118 \pm 10$ \\
J232614$-$302839\rlap{$^{\dag}$}  & $ >23.4 $ & $ <1.6 $ & $\cdots$ & $\cdots$ & $490 \pm 40$ \\
J233020$-$323729 & $ 21.64 \pm 0.07 $ & $8.02 \pm 0.52 $ & 23:30:20.85\:$-$32:37:31.3 & $0.40$ & $100.8 \pm 8.6$ \\
\hline
\hline
\multicolumn{6}{p{135mm}}{Notes. $^{\dag}$UHzRG candidate. $^{\rm a}$Radio position where the aperture was centred; see \ref{appendix} for further details. $^{\rm b}$Separation uncertain due to an insufficiently well-constrained $K_{\rm s}$-band position. $^{\rm c}$This incipient double-lobed radio source was fitted with a single Gaussian in RACS-mid.  $^{\rm d}$Estimated by eye. $^{\rm e}$Host galaxy detected in a 1.5\arcsec\:diameter aperture; see the notes on this source in \ref{appendix} for further details. $^{\rm f}$Distance from the $K_{\rm s}$-band identification to the possible radio core visible at 5.5\,GHz.}  \\
\end{tabular}
\label{tab:magnitudes}
\end{table*}

The host galaxy magnitude measurements (2\arcsec\:diameter apertures) from our HAWK-I observing campaign are presented in Table~\ref{tab:magnitudes}, along with the corresponding flux density measurements, $S_{2.15\,\mu{\rm m}}$, as well as the positions on which the apertures were centred (see Section~\ref{sec: host galaxy measurements}). As can be seen in Table~\ref{tab:magnitudes}, 27 out of the 35 sources (77\,per cent) were detected in the HAWK-I images. For these detections, the host galaxy magnitudes range from $K_{\rm s} \approx 21.6$--$23.0$\,mag. The remaining eight non-detections have $3\sigma$ magnitude lower limits that are generally quite uniform (median depth $K_{\rm s} \approx 23.3$\,mag). We consider five of these non-detections (J0053$-$3256, J1033$+$0107, J1127$-$0332, J1443$+$0229 and J2326$-$3028; marked in Table~\ref{tab:magnitudes} with $^{\dag}$) to be our best UHzRG candidates, along with J0008$-$3007 from B22 (deep SHARKS limit). Notes on individual sources are given in \ref{appendix}.  

\subsection{Overlay plots}\label{sec:overlays} 

\begin{figure*}
\begin{minipage}[t]{0.48\textwidth}
\includegraphics[height=6.5cm]{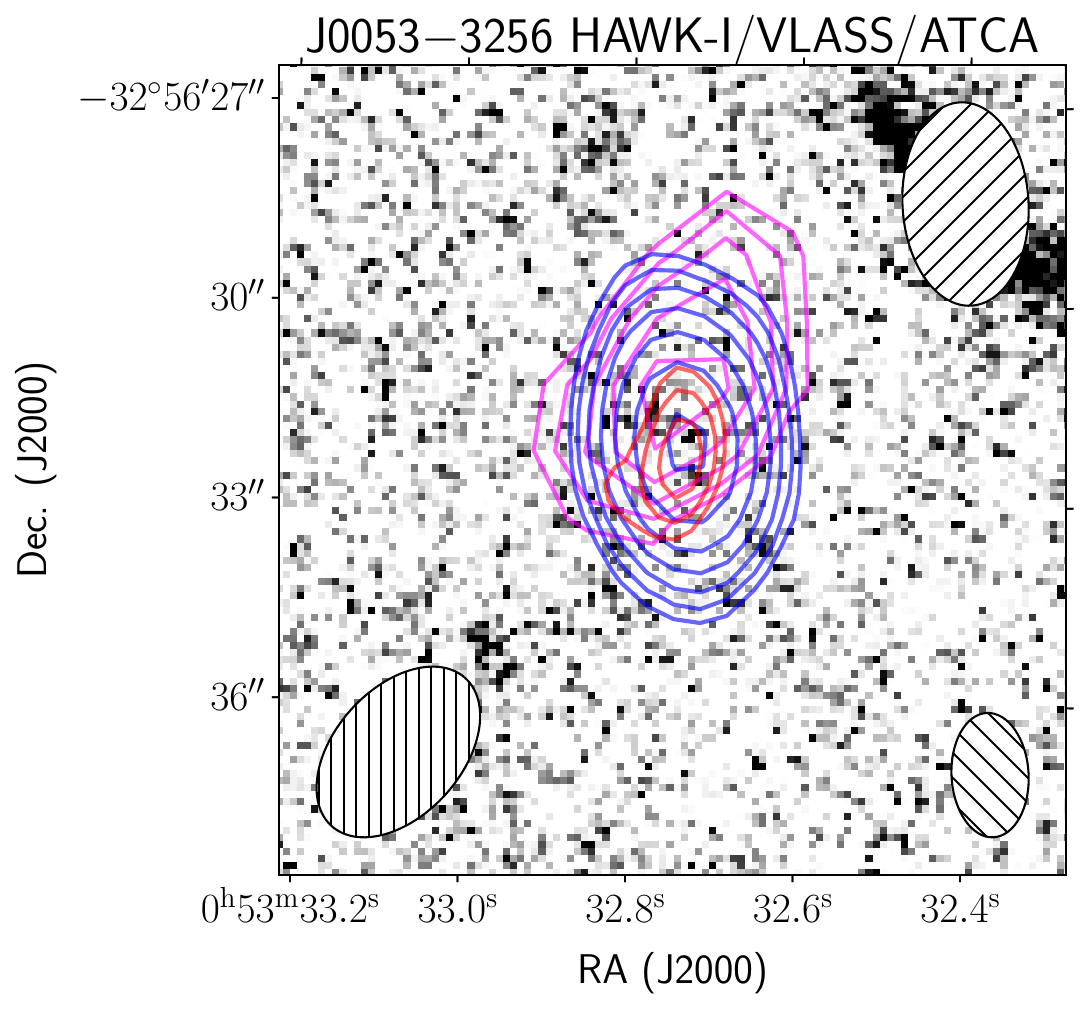}
\end{minipage}%
\hspace{0.02\linewidth}%
\begin{minipage}[t]{0.48\textwidth}
\includegraphics[height=6.5cm]{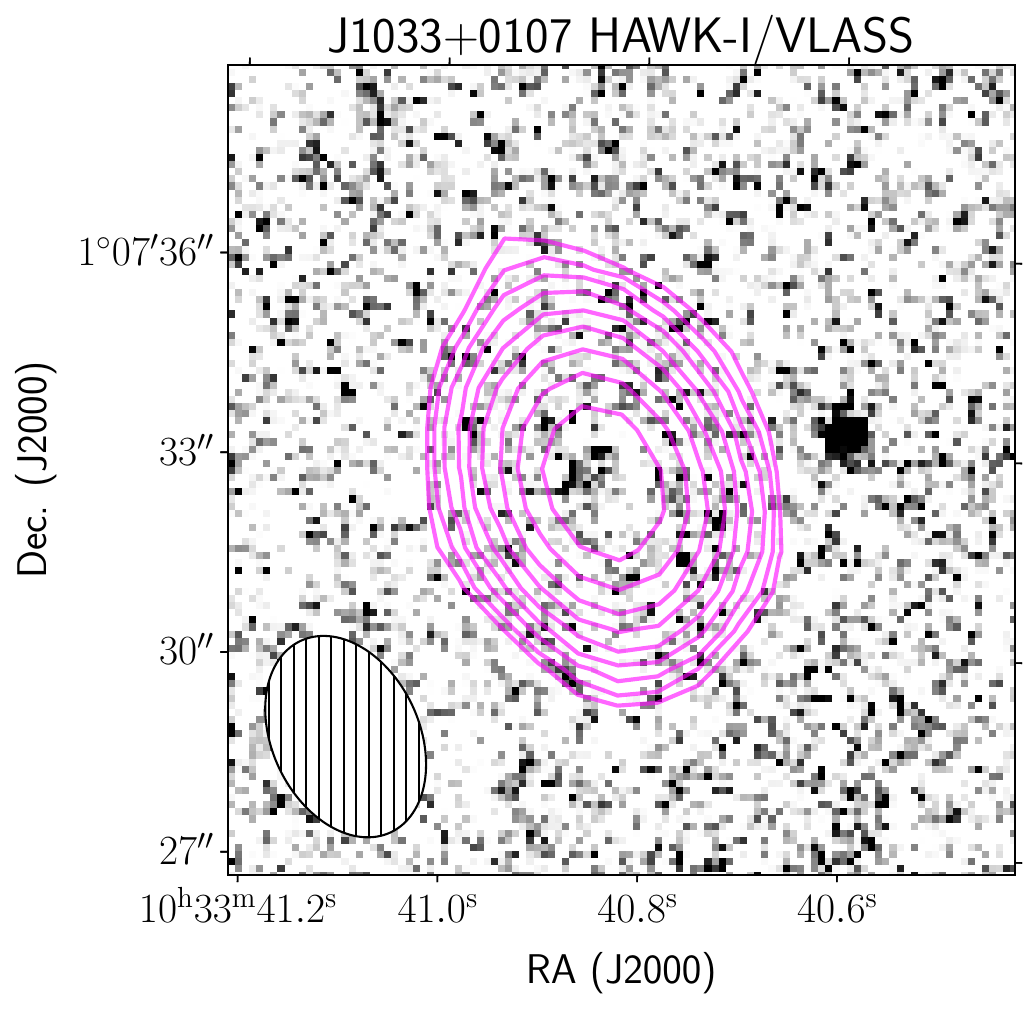}
\end{minipage}
\begin{minipage}[t]{0.48\textwidth}
\includegraphics[height=6.5cm]{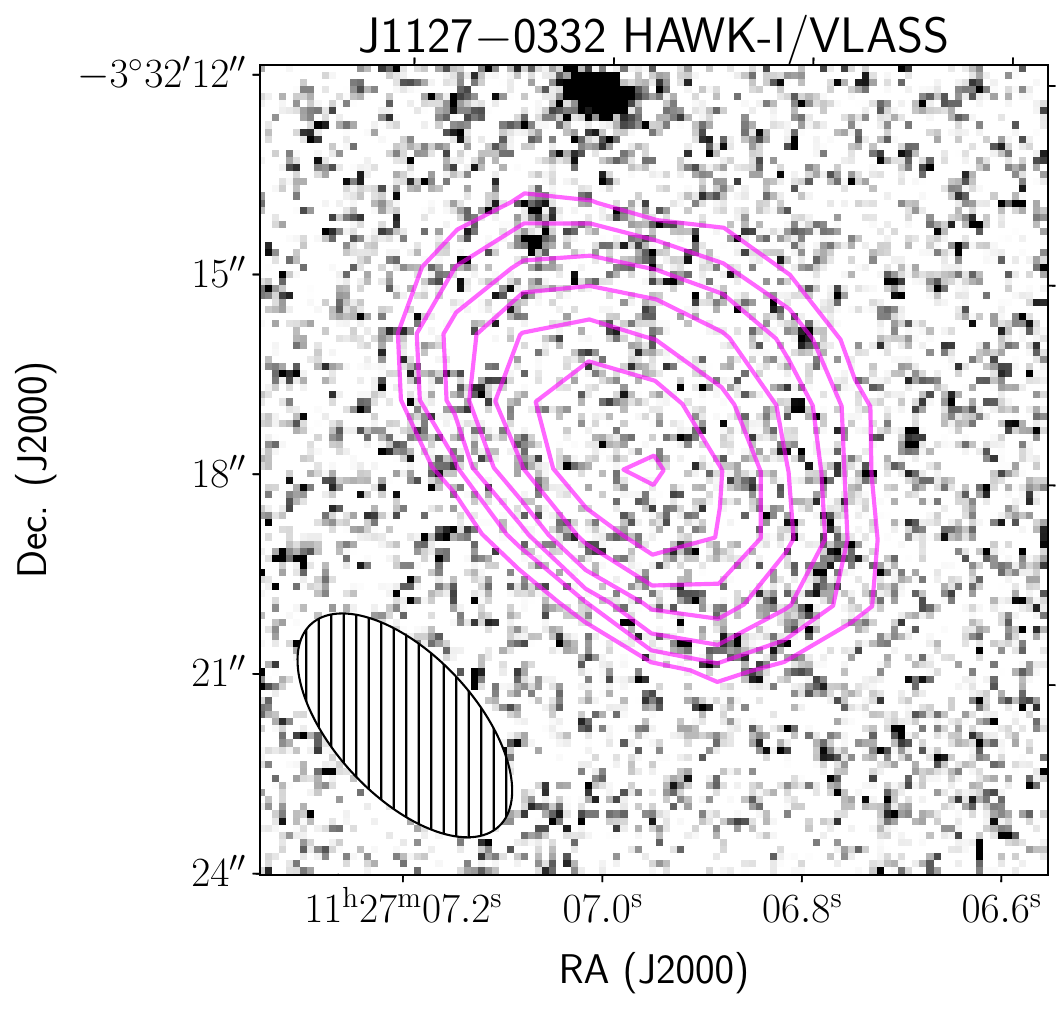}
\end{minipage}%
\hspace{0.02\linewidth}%
\begin{minipage}[t]{0.48\textwidth}
\includegraphics[height=6.5cm]{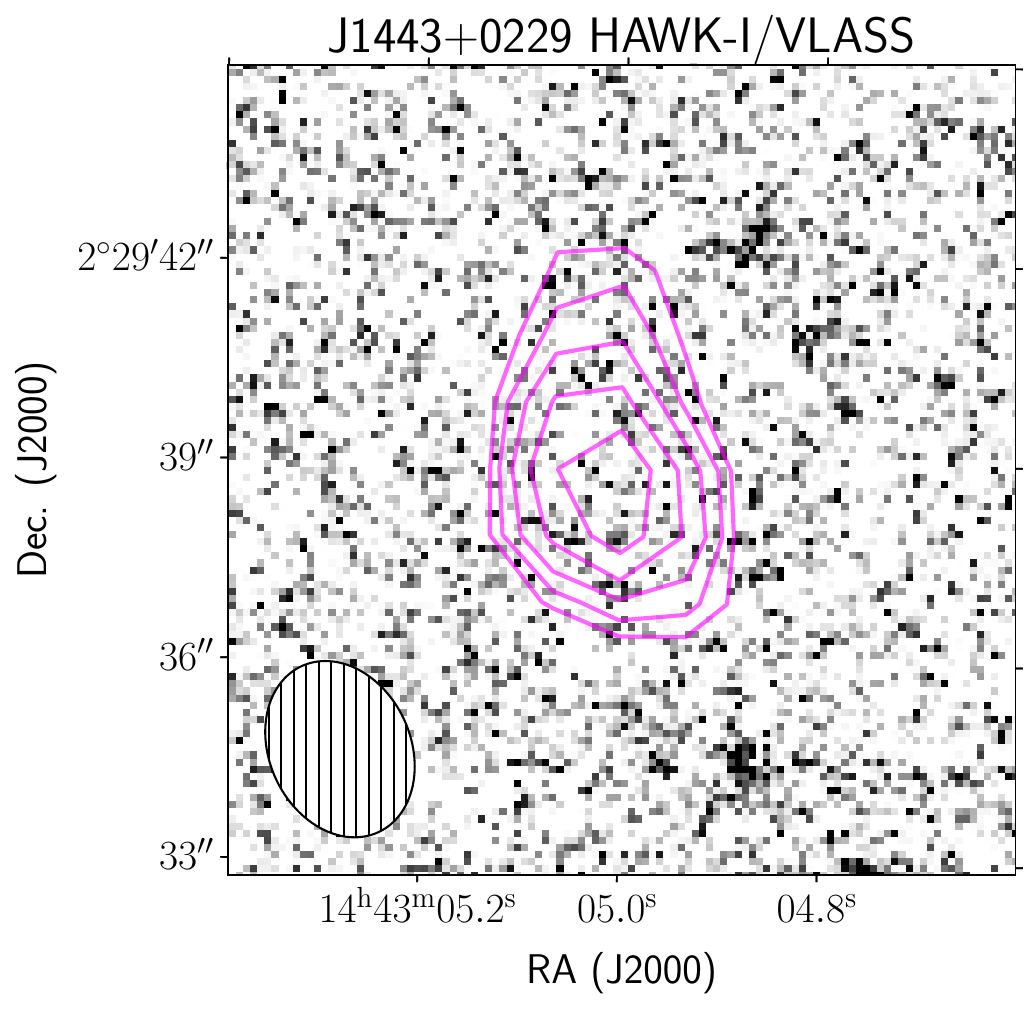}
\end{minipage}
\begin{minipage}[t]{0.98\textwidth}
\includegraphics[height=6.5cm]{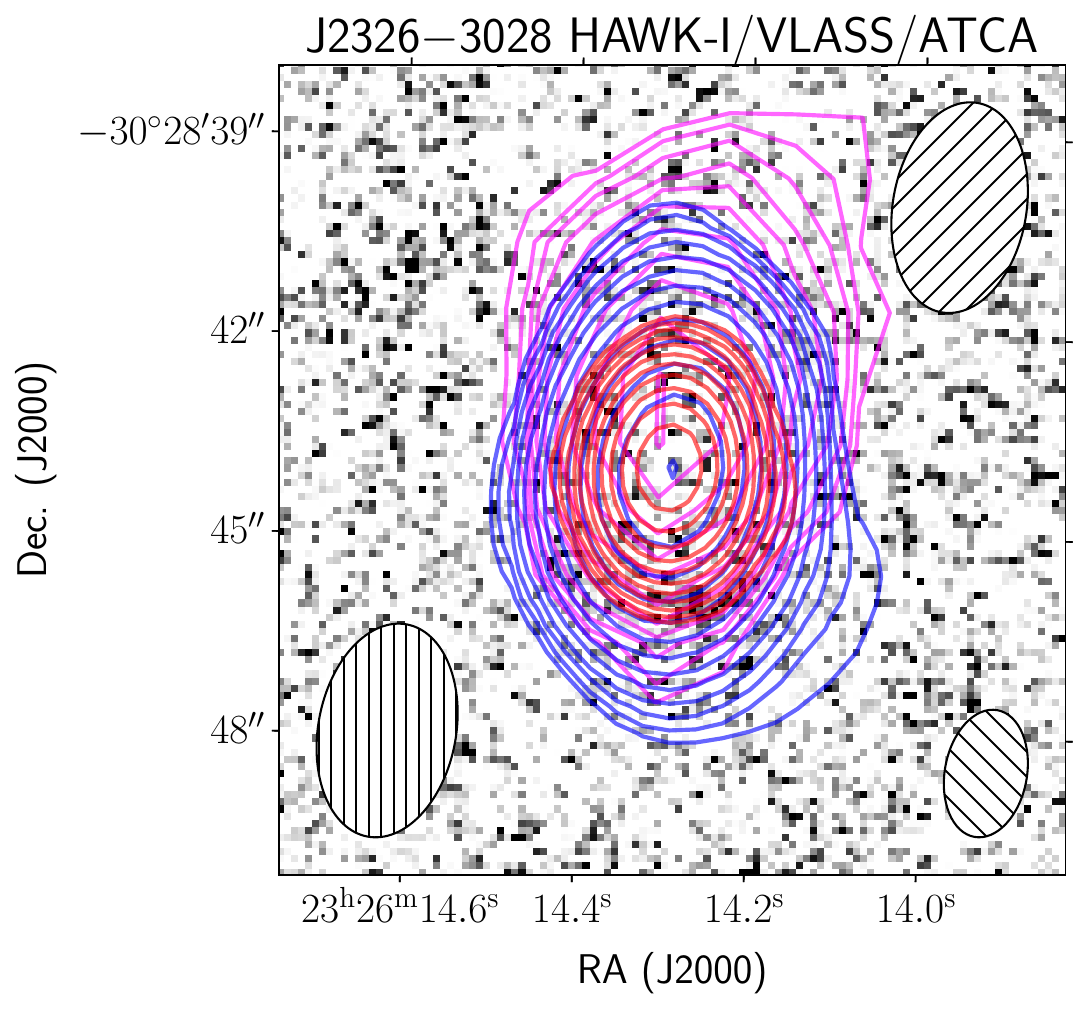}
\end{minipage}%
\caption{Radio contours overlaid on HAWK-I $K_{\rm s}$-band images for five of the six best UHzRG candidates from this study, ordered by right ascension. The VLASS (3\,GHz; magenta) and ATCA (5.5 and 9\,GHz; blue and red, respectively) contours were previously presented in B22, except for the new single-epoch contours from VLASS for J1033$+$0107. The contours are a geometric progression in $\sqrt{2}$, with the lowest contour at the $5\sigma$ level; a summary of the contour levels can be found in Table 5 in B22. For the single-epoch image from VLASS for J1033$+$0107, the updated lowest contour level ($5\sigma$) is 0.75\, mJy\,beam$^{-1}$. In each panel, we also show the radio synthesised beams with different hatching styles (VLASS: vertical; ATCA 5.5\,GHz: forward slash; ATCA 9\,GHz: backslash). The host galaxy magnitude lower limits are reported in Table~\ref{tab:magnitudes}. The overlay plots for the remaining 30 sources observed with HAWK-I can be found in Figure~\ref{fig:overlays_appendix} in \ref{appendix}, while the SHARKS/VLASS/ATCA overlay for J0008$-$3007, our sixth UHzRG candidate, can be found in Figure 2 in B22. Furthermore, notes on individual sources can be found in \ref{appendix}.}  
\label{fig:overlays}
\end{figure*}

In Figure~\ref{fig:overlays}, we present high-resolution radio contours overlaid on our HAWK-I images for five of the six best UHzRG candidates (see Figure 2 in B22 for the previously presented overlay plot for J0008$-$3007). The remaining 30 overlay plots can be found in Figure~\ref{fig:overlays_appendix} in \ref{appendix}.  

The radio data used in Figures~\ref{fig:overlays} and \ref{fig:overlays_appendix} were mostly presented in B22. Apart from four cases outlined below, we have only plotted contours that are at an angular resolution of a few arcsec or better: 3-GHz `quick-look' data from the first epoch of VLASS\footnote{Image cutouts were downloaded from \url{cutouts.cirada.ca}.} as well as 5.5- and 9-GHz ATCA data. For J1030$+$0135, J1032$+$0339, J1033$+$0107 and J1037$-$0325, we used the higher-quality, single-epoch (epoch 2.1) VLASS continuum images\footnote{Downloaded from \url{https://archive-new.nrao.edu/vlass/se_continuum_imaging/}.} instead, which had become available since the B22 study.

The overlays for J0007$-$3040, J0034$-$3112 and J1037$-$0325 presented in Figure~\ref{fig:overlays_appendix} also include data from the Rapid Australian Square Kilometre Array Pathfinder \citep[ASKAP;][]{johnston07,hotan21} Continuum Survey at 1367.5\,MHz \citep[RACS-mid; median angular resolution $11\farcs2 \times 9\farcs3$;][]{duchesne23,duchesne24}.\footnote{\url{https://doi.org/10.25919/6mr6-rd83} and \url{https://doi.org/10.25919/p524-xb81}} These RACS-mid data became available near the completion of our study and changed the interpretation of these three sources. Originally thought to be additional UHzRG candidates, these are larger radio sources than previously thought and therefore much more likely to lie at lower redshift. For both J0007$-$3040 and J0034$-3112$, the high-resolution VLASS and ATCA contours trace only one of the lobes. Further details can be found in the notes on individual sources in \ref{appendix}.  

For J2311$-$3359, also presented in Figure~\ref{fig:overlays_appendix}, given that this source is only faintly detected in the ATCA data presented in B22, and not detected in VLASS, we confirmed the host galaxy identification using the publicly available data (angular resolution 10\arcsec) from a recent study by \citet[][]{gurkan22}, who conducted deep 887.5-MHz ASKAP observations of the GAMA-23 field from the Galaxy and Mass Assembly survey \citep[][]{driver09,driver11,driver16}.\footnote{We downloaded the data from this study at \url{http://data.csiro.au}.} 

$K_{\rm s}$-band--radio angular separations are reported in Table~\ref{tab:magnitudes} for the 25 sources where the $K_{\rm s}$-band position is known to sufficient accuracy. We again used the high-resolution ATCA data where available (factoring in both the angular resolution and S/N to choose the most accurate position at either 5.5 or 9\,GHz), or otherwise the VLASS position. However, for J0007$-$3040, J0034$-$3112, J1037$-$0325 and J2311$-$3359, we instead used the ASKAP data from either RACS-mid or the \citet[][]{gurkan22} study. For multi-component radio sources, the geometric midpoint was calculated unless otherwise specified in Table~\ref{tab:magnitudes}. Given both the mostly compact radio morphologies of the targets and our conservative approach at assigning host galaxy identifications, it is not surprising that the offsets are small, with a median value of approximately 0\farcs5.    

\section{DISCUSSION}
\label{sec:dis} 

\subsection{Comparison to known HzRGs}\label{sec:dis_plot}

As can be seen in Figure~\ref{fig:flux_flux}, and as was initially discussed in Section~\ref{sec:mag}, our detection rate for the B22 sample is high in $K_{\rm s}$-band: 78\,per cent for the 40 sources plotted. We can see that the B22 sample overlaps with the $3 < z < 5$ powerful HzRGs, presenting an extension of the observed trend (i.e. flux density ratios $S_{150\,\rm{MHz}} / S_{2.2\,\mu\rm{m}}$ $\sim$ $10^{4}$--$10^{6}$) towards lower 150-MHz and 2.2-\textmu m flux densities. Extreme flux density ratios $\gtrsim 10^5$ are not surprising as they arise from a common approach in the various HzRG selection techniques applied in the literature: sufficiently bright radio flux densities to select luminous radio galaxies (most likely powered by SMBHs) combined with faint 2.2-\textmu m magnitudes to ensure that the host galaxy is distant. By and large, these extreme ratios are only seen at $z>1$ \citep[with a notable exception being Cygnus A; e.g. see Figure 5 in][]{drouart21}. Fainter 150-MHz and 2.2-\textmu m flux densities could, generally speaking, result from less massive galaxies with less massive BHs, or an increase in redshift. Which is the dominant of these two effects in the B22 sample is unclear given the lack of spectroscopic redshift information.    

It should be noted that the sources presented in Figure~\ref{fig:flux_flux} have been selected using a variety of multi-wavelength criteria. Additionally, as one moves from lower to higher redshift, $k$-corrections at both radio and near-infrared wavelengths affect the flux densities. As discussed in Section 3.3 of \citet[][]{drouart21}, these $k$-corrections account for separate segments from the SED that are both approximated by power laws with negative slopes; see that study for further details. 

\subsection{UHzRG candidates from the B22 sample}\label{sec:uhzrg_candidates}

\begin{figure*}[t]
\begin{minipage}{1.0\textwidth}
\centering
\includegraphics[height=8cm]{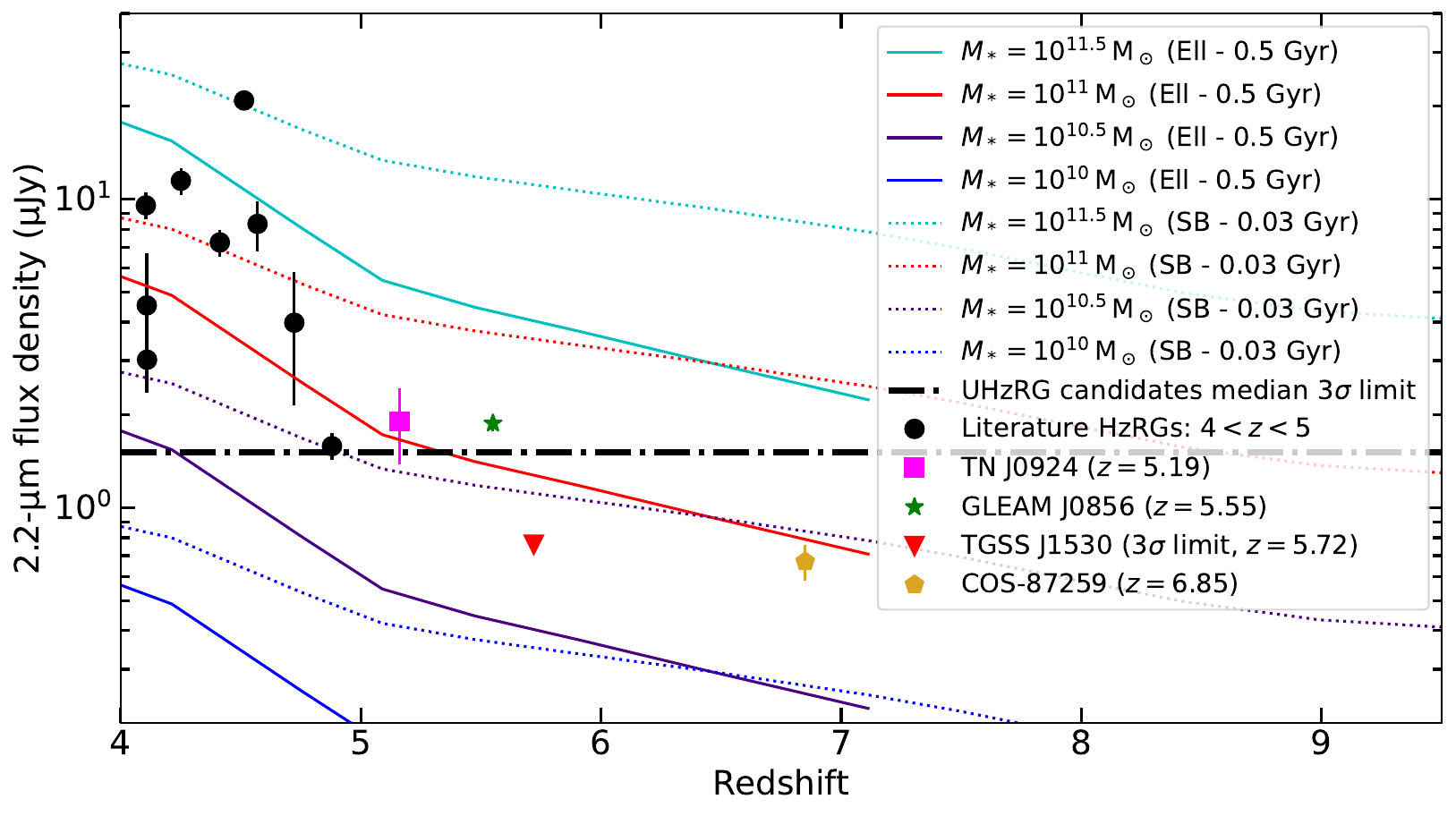}
\end{minipage}
\caption{Tracks of modelled 2.2-\textmu m (i.e. $K_{\rm s}$-band) flux density from {\sc p\'egase} for galaxies of different masses formed no earlier than $z=16$. `Ell' and `SB' refer to an old, 0.5-Gyr elliptical galaxy template and a young, 30-Myr starburst galaxy template, respectively. The elliptical tracks terminate when the age implies a formation redshift above $z=16$; see Section~\ref{sec:uhzrg_candidates} for more details. As in Figure~\ref{fig:flux_flux}, we overlay TN~J0924$-$2201, GLEAM~J0856$+$0223, TGSS~J1530$+$1049 and COS-87259, as well as the HzRGs with $4 < z < 5$ from our literature sample. We also indicate the median $3\sigma$ limit of our deep non-detections ($S_{2.2\,\mu\rm{m}} \lesssim 1.5$ \textmu Jy; 2\arcsec\:diameter apertures). Error bars ($\pm1\sigma$) are plotted (apart from the case of the $3\sigma$ limit for TGSS~J1530$+$1049), but sometimes are smaller than the symbols. When uncertainties were not available in the literature, we assumed a 10\,per cent error. Discussion of this figure can be found in Section~\ref{sec:uhzrg_candidates}.} 
\label{fig:flux_tracks}
\end{figure*}

Of particular interest in Figure~\ref{fig:flux_flux} are the six sources (marked in blue)  previously outlined in Section~\ref{sec:mag} as our best UHzRG candidates.  This subset of six sources has median $3\sigma$ limits of $K_{\rm s} \gtrsim 23.5$\,mag and $S_{2.2\,\mu\rm{m}} \lesssim 1.5$\,\textmu Jy (2\arcsec\:diameter apertures); moreover, $S_{150\,\rm{MHz}} > 200$\,mJy.  All of these sources have extreme radio to near-infrared flux density ratios, $S_{150\,\rm{MHz}} / S_{2.2\,\mu\rm{m}} >10^5$, similar to a number of the radio galaxies from the literature plotted in Figure~\ref{fig:flux_flux}, including the UHzRGs TN~J0924$-$2201 ($z=5.19$), GLEAM~J0856$+$0223 ($z=5.55$) and TGSS~J1530$+$1049 ($z=5.72$). While at least slightly fainter in $K_{\rm s}$-band than TN~J0924$-$2201 and GLEAM~J0856$+$0223, our UHzRG candidates have 2.2-\textmu m flux density upper limits that are about a factor of two brighter than the upper limit for TGSS~J1530$+$1049: $S_{2.2\,\mu\rm{m}} < 0.76$\,\textmu Jy and $K_{\rm s} > 24.2$\,mag \citep[$3\sigma$; 1.5\arcsec\:diameter aperture;][]{saxena18,saxena19}. For the five sources with HAWK-I data, repeating the limit calculations in 1.5\arcsec\:diameter apertures (see the notes on individual sources in \ref{appendix}) so as to better match the analysis for TGSS~J1530$+$1049 yielded deeper median $3\sigma$ limits of $K_{\rm s} \gtrsim 23.8$\,mag and $S_{2.2\,\mu\rm{m}} \lesssim 1.1$\,\textmu Jy. Henceforth, however, we will use the more conservative limits from the 2\arcsec\:diameter apertures. 

Only recently has an obscured, radio-loud AGN been detected at $z > 6$: COS-87259 with $z=6.85$, previously discussed in Section~\ref{sec:int}. This source has $S_{2.2\,\mu\rm{m}} = 0.67 \pm 0.09$\,\textmu Jy \citep[$K_{\rm s} = 24.33 \pm 0.15$\,mag; 1.2\arcsec\:diameter aperture;][]{endsley22}, i.e. too faint to have been detected in our deepest HAWK-I observations. It was found as part of the Cosmic Evolution Survey \citep[COSMOS;][]{scoville07}, which has unparalleled multi-wavelength data, albeit over 2\,deg$^{2}$ only. Assuming that the rest-frame ultraviolet and optical emission is dominated by star formation, the faint 2.2-\textmu m flux density is consistent with a massive galaxy with stellar mass $M_{\rm *} = (1.7 \pm 0.7) \times 10^{11}\,M_\odot$ \citep[][although those authors explain that the mass could lie anywhere between $10^{10}-10^{11}\,$M$_\odot$ depending on the star formation history]{endsley23b}. The host galaxy has a very high obscured AGN infrared luminosity, $L_{\rm IR} = (2.5 \pm 0.2) \times 10^{13}\,L_\odot$, which is not evident in the rest-frame ultraviolet/optical \citep[][]{endsley23b}. Its stellar mass and AGN luminosity are similar to powerful radio galaxies at $2 < z < 5$ \citep[e.g.][]{seymour07,drouart16}. While COS-87259 has a faint 144-MHz flux density ($S_{144\,{\rm MHz}} = 475 \pm 180$\,\textmu Jy; \citealt[][]{endsley22}) and is therefore much less luminous in the radio than the powerful radio galaxies referenced above, it nonetheless demonstrates the existence of detectable radio emission from an obscured AGN within the EoR. Potentially, some of the prime targets from the B22 sample are the radio-luminous analogues of COS-87259, but spectroscopic redshift determinations are needed. 

To obtain further evidence that our six deep non-detections are strong UHzRG candidates, in Figure~\ref{fig:flux_tracks} we have used {\sc p\'egase} to model 2.2-\textmu m flux density as a function of redshift for several galaxy templates (elliptical and starburst) with different stellar masses. We followed a similar method as that described in both \citet[][]{drouart21} and \citet[][]{seymour22}, but 
assumed that the galaxies formed no earlier than $z=16$ when the Universe was $\sim 250\,$Myr old. The elliptical template was assumed to be old (0.5\,Gyr: the oldest it could be at $z\sim 7$ given our formation redshift and therefore the tracks terminate at $z=7.1$) whereas the starburst template was assumed to be young \citep[0.03\,Gyr: chosen to match the youngest galaxies seen at $7<z<8$;][]{endsley23a}. We have overlaid the positions of the nine HzRGs with $4<z<5$ included in Figure~\ref{fig:flux_flux}, as well as TN~J0924$-$2201, GLEAM~J0856$+$0223, TGSS~J1530$+$1049 and COS-87259. In addition, we show the median $3\sigma$ limit of our deep non-detections. Several conclusions can be drawn from Figure~\ref{fig:flux_tracks} and are as follows. 
\begin{enumerate}
    \item If the host galaxies of our $K_{\rm s} \gtrsim 23.5$\,mag radio sources are massive like other known powerful HzRGs at $2<z<5$ \citep[$M_{\rm *} \gtrsim 10^{11}$\,M$_\odot$; e.g.][]{roccavolmerange04,seymour07,debreuck10,saxena19}, then they must lie at ultra-high redshift, $z\gtrsim 5.4$, to be so faint. 
    \item For a lower stellar mass, $M_{\rm *} \sim 10^{10.5}$\,M$_\odot$, the elliptical and starburst model tracks indicate that $z\gtrsim 4.2$ (elliptical) or $z\gtrsim 4.9$ (starburst) for $K_{\rm s} \gtrsim 23.5$\,mag. 
    \item A corollary of the two preceding conclusions is that for $M_{\rm *} \gtrsim 10^{10.5}$\,M$_\odot$ and $K_{\rm s} \gtrsim 23.5$\,mag, $z \gtrsim 4.2$.   
    \item Another possibility is that our deep non-detections are perhaps indicative of a class of powerful radio-loud AGN in undermassive ($M_{\rm *} \lesssim 10^{10}\,M_\odot$) host galaxies. Although not visible in the region of parameter space shown in Figure~\ref{fig:flux_tracks}, $z\gtrsim 2.4$ (elliptical) or $z\gtrsim 2.5$ (starburst) for $M_{\rm *} \sim 10^{10}$\,M$_\odot$ and $K_{\rm s} \gtrsim 23.5$\,mag. However, given the findings from previous studies of powerful HzRGs, undermassive host galaxies would be unexpected, at least at $z < 5$. Furthermore, radio-loud AGN in dwarf galaxies ($M_{\rm *} \lesssim 10^{9.5}\,$M$_\odot$) are at least a few orders of magnitude less radio luminous (e.g. \citealt*[][]{mezcua19}; \citealt[][]{davis22}). On the other hand, at $z>4$, there is emerging evidence from the {\em James Webb Space Telescope} \citep[{\em JWST};][]{gardner06,gardner23} of a population of AGN not detected in the radio with BH to stellar mass ratios up to several dex larger than is what usually observed \citep[e.g.][]{furtak23,furtak24,goulding23,pacucci23,kokorev23,maiolino24}. Assuming a fiducial value of $M_{\rm BH}$ $\gtrsim 10^9$\,M$_\odot$ for a powerful radio galaxy (e.g. \citealt[][]{nesvadba11}; \citealt[][]{drouart14}; also see \citealt[][]{willott99}; \citealt*[][]{kauffmann08}; \citealt[][]{martinezsansigre11}), then, in the case of an undermassive host galaxy scenario, our targets might have BH to stellar mass ratios exceeding the local relation \citep[e.g.][]{reines15} and potentially approaching the sources analysed in the above {\em JWST} studies.
    \item COS-87259 lies below both of the $10^{11}$\,M$_\odot$ tracks and the $10^{10.5}$\,M$_\odot$ starburst track; its estimated stellar mass from broadband SED fitting is in the range $M_{\rm *} \sim$ $10^{10}$--$10^{11}$\,M$_\odot$ \citep[assuming no AGN contribution to the ultraviolet/optical flux density;][]{endsley23b}. This large range is due to the uncertainty in the star formation history of this source, which demonstrates how uncertain stellar mass estimates can be even with extensive and accurate photometry. It is worth noting that from its BH mass estimate ($M_{\rm BH} \gtrsim 1.5 \times 10^9$\,M$_\odot$), COS-87259 is likely to have a relatively low stellar mass compared to local galaxies; see the discussion in Section 4 of \citet[][]{endsley23b}. 
    \item A suggested stellar mass of $M_{\rm *} < 10^{10.5}$ \,M$_\odot$ (starburst) or $M_{\rm *} < 10^{11}$ \,M$_\odot$ (elliptical) for TGSS~J1530$+$1049 is also broadly consistent with the modelling carried out by \citet[][]{saxena19}, who determined that $M_{\rm *} < 10^{10.5}$\,M$_\odot$. TN~J0924$-$2201 is best described by an elliptical template with $M_{\rm *} \sim 10^{11}$\,M$_\odot$ or a starburst with $M_{\rm *} \sim 10^{10.5}\,$M$_\odot$; from multi-wavelength SED fitting, \citet[][]{seymour07} determined the maximum stellar mass of this source to be $M_{\rm *} = 10^{11.1}$\,M$_\odot$. Lastly, GLEAM~J0856$+$0223 is potentially at least slightly more massive than TN~J0924$-$2201 (also see discussion in \citealt{drouart20}).  
    \item Though with some scatter, the inclusion of the $4 < z < 5$ HzRGs in Figure~\ref{fig:flux_tracks} demonstrates an increase in 2.2-\textmu m flux density with decreasing redshift, as expected from the $K$--$z$ relation, albeit with an upper envelope described by a $\sim$ $10^{11.5}\,$M$_\odot$ stellar mass compared to $\sim$ $10^{12}\,$M$_\odot$ at lower redshift \citep[e.g.][]{roccavolmerange04}. All but one of these HzRGs have 2.2-\textmu m flux densities at least a factor of two above the median $3\sigma$ limit for our deep non-detections.
\end{enumerate}

\subsection{Alternative scenarios for the UHzRG candidates}\label{sec:dis_alternate}

What is the nature of the $K_{\rm s} \gtrsim 23.5$ sources from B22 should they not be at $z \gtrsim 4$--$5$? With an undermassive host galaxy scenario unprecedented at $z<4$ given the powerful radio emission, these sources could instead be massive galaxies with heavy dust obscuration. As discussed in B22, none of our UHzRG candidates have detections in the mid-infrared {\it Widefield Infrared Survey Explorer} \citep[{\it WISE};][]{wright10} AllWISE data release \citep[][]{cutri14}. Additionally, no detections were found in the deeper unWISE \citep*[][]{schlafly19} data release. However, unWISE is much shallower than our HAWK-I data; the 50\,per cent completeness limits for the 3.37-\textmu m ($W1$) and 4.62-\textmu m ($W2$) bands are $W1 = 20.72$\,mag / $S_{3.37\,\mu\rm{m}} = 18.71$\,\textmu Jy and $W2 = 19.97$\,mag / $S_{4.62\,\mu\rm{m}} = 37.33$\,\textmu Jy, respectively \citep[][]{schlafly19}. {\em JWST} is opening up a new window on the study of dusty starburst galaxies at (ultra-)high redshift \citep[e.g.][]{zavala23,barrufet23}, suggesting that we would potentially need much deeper mid-infrared data to detect our targets. Moreover, COS-87259 has 3.6- and 4.5-\textmu m flux densities of $2.75 \pm 0.10$ and $2.77 \pm 0.12$\,\textmu Jy, respectively \citep[][]{endsley22}.     

Additionally, in B22, we investigated if all of the sources in the sample had been identified as pulsars in the literature, but none were. For the non-detections in the $K_{\rm s} \gtrsim 23.5$\,mag subset, a pulsar origin seems unlikely given that all of these sources are at least slightly extended in the radio (Table 2 in B22). None of the sources show evidence for variability at radio frequencies either. Furthermore, none of the sources are detected in the optical in the third data release of the SkyMapper Southern Sky Survey \citep[SMSS;][]{onken19},\footnote{DOI: 10.25914/5f14eded2d116} nor in the 18th data release of the Sloan Digital Sky Survey \citep[SDSS;][]{almeida23} for the equatorial targets. While our sources do not have X-ray data, the lack of radio variability, optical non-detections, and, additionally, the relatively steep spectral indices at GHz frequencies (B22 cf. e.g. \citealt{fender01} and \citealt*{trushkin03}) do not paint a compelling picture that we are seeing persistent radio emission from Galactic X-ray binaries \citep[e.g. see the radio--optical parameter space presented in][]{stewart18}.

Lastly, our non-detections could be radio sources with one-sided lobes, i.e. more extreme versions of J0007$-$3040 and J0034$-$3112 (as well as J2330$-$3237) in Figure~\ref{fig:overlays_appendix}. Therefore, the host galaxies may be located elsewhere in the HAWK-I images. For these sources, there is no evidence in RACS-mid of a second candidate lobe. While we cannot fully rule out that at least some of the $K_{\rm s} \gtrsim 23.5$\,mag sources have one-sided radio morphologies, the number density of $K$-band sources is $\sim$70\,arcmin$^{-2}$ for $23 < K < 24$ \citep[][]{fontana14}. Therefore, the probability of a host galaxy candidate within this magnitude range being within 2\arcsec\:of the radio centroid by chance is about 20\,per cent, i.e. for $\sim$1 of our 6 UHzRG candidates with non-detections. The corresponding values for a 5\arcsec\:search radius are 80\,per cent and $\sim$5 sources. Therefore, we can see that this quickly becomes a challenging exercise to identify the true host galaxy for increasingly large, one-sided radio sources. Very long baseline interferometry (VLBI) would allow a better assessment of whether the radio emission from the UHzRG candidates is lobe-like, possibly with backflow towards a host galaxy candidate, or whether the emission is resolved into multiple lobes and also potentially a core, which would strengthen the argument for a compact radio source with an undetected host galaxy in $K_{\rm s}$-band. Such observations are underway and will be presented in a future study. 

The above analysis of chance radio--near-infrared matches is also relevant for the host galaxy identifications presented in Section~\ref{sec:res}, although the offset from the radio centroid is typically much smaller (median 0\farcs5; Table~\ref{tab:magnitudes}). We would therefore expect at most one source of the 27 with $K_{\rm s}$-band counterparts in Table~\ref{tab:magnitudes} to be a chance radio--near-infrared match.     

\subsection{Comparison to IFRSs}

Our UHzRG candidates, and indeed the majority of the other sources from the B22 sample studied in this paper, have properties consistent with the class of infrared-faint radio sources (IFRSs; e.g \citealt{norris06,norris11}; \citealt{middelberg08,middelberg11}; \citealt{garn08}; \citealt*{zinn11}; \citealt[][]{collier14}; \citealt{maini16}). IFRSs have typically been characterised on the basis of their 1.4-GHz radio to 3.6-\textmu m mid-infrared flux density ratios; \citet[][]{zinn11} proposed standard IFRS selection criteria of (i) $S_{1.4\,\rm{GHz}} / S_{3.6\,\mu\rm{m}} > 500$ and (ii) $S_{3.6\,\mu\rm{m}} < 30$\,\textmu Jy. For the 40 sources from the B22 sample plotted in Figure~\ref{fig:flux_flux}, the 1.4-GHz flux densities from the National Radio Astronomy Observatory (NRAO) VLA Sky Survey \citep[NVSS;][]{condon98} span the range $<$1.6--98.9\,mJy, with a median of 18.7\,mJy (Table 2 in B22). Even after factoring in a $k$-correction between 2.2 and 3.6\,\textmu m, which will depend on the SED, we would expect the majority of our sources to (comfortably) meet the \citet[][]{zinn11} selection criteria. Indeed, this is suggested in Figure~\ref{fig:flux_flux}, albeit with plotted flux density ratios based on the radio value at 150\,MHz ($S_{150\,\rm{MHz}} / S_{1.4\,\rm{GHz}} \sim 10$ for a radio spectral index $\alpha = -1$; flux density $S_{\nu} \propto \nu^{\alpha}$).   

As mentioned above in Section~\ref{sec:dis_alternate}, none of our $K_{\rm s} \gtrsim 23.5$\,mag UHzRG candidates were detected in AllWISE or unWISE. However, despite the difference in sensitivity between unWISE and our HAWK-I data, seven sources (with $S_{2.2\,\mu\rm{m}} = 2.54$--$8.02$\,\textmu Jy in 2\arcsec\:diameter apertures)  from the B22 sample plotted in Figure~\ref{fig:flux_flux} have unWISE detections. In Table~\ref{tab:unwise}, we present the unWISE and NVSS flux densities for these sources, along with the values of $S_{1.4\,\rm{GHz}} / S_{3.37\,\mu\rm{m}}$. All seven sources meet the \citet[][]{zinn11} IFRS selection criteria, albeit considering a slightly different mid-infrared observing wavelength. An additional handful of sources from the B22 sample may also have unWISE counterparts (e.g. J1335$+$0112 and J2219$-$3312, which were discussed in Section 5.5 in B22 as having potential AllWISE detections), but it was not clear whether these detections are affected by confusion from nearby sources visible at 2.2\,\textmu m in the higher-resolution HAWK-I images. A detailed analysis was beyond the scope of this paper.      

\citet[][]{norris11} considered the scenarios listed in Section~\ref{sec:dis_alternate}, as well as others, to explain the nature of IFRSs. They concluded that most IFRSs are likely either radio-loud AGN at $z \gtrsim 3$ or radio-loud AGN at $1 < z < 3$ that are significantly obscured by dust. Subsequent studies have found that many IFRSs are at high redshift, up to $z=4.39$ (e.g. \citealt{collier14}; \citealt{herzog14}; \citealt{singh17}; \citealt*{orenstein19}). This bodes well for the B22 sample.

\begin{table}
  \caption{Sources in the B22 sample with unWISE detections. For each unWISE detection, we also report the 1.4\,GHz flux density from NVSS and the flux density ratio $S_{1.4\,\rm{GHz}} / S_{3.37\,\mu\rm{m}}$.}
  \begin{tabular}{crrr}
  \hline\hline
  Source & \multicolumn{1}{c}{$S_{3.37\,\mu\rm{m}}$}  & \multicolumn{1}{c}{$S_{1.4\,\rm{GHz}}$} & \multicolumn{1}{c}{$S_{1.4\,\rm{GHz}} / S_{3.37\,\mu\rm{m}}$} \\ 
 \multicolumn{1}{c}{(GLEAM)} & \multicolumn{1}{c}{(\textmu Jy)} & \multicolumn{1}{c}{(mJy)} & \\ 
   \hline
J012929$-$310915 & $10.5 \pm 1.9$ & $27.2 \pm 0.9$ & $2590 \pm 480 $ \\
J013340$-$305638 & $15.3 \pm 2.0$ & $43.3 \pm 1.4$ & $2830 \pm 380 $ \\
J030108$-$313211 & $12.0 \pm  1.9$ & $45.1 \pm 1.4$ & $3760 \pm 610 $ \\
J030931$-$352623 & $11.1 \pm 1.8$ & $6.0 \pm 0.5$ & $541 \pm 99 $ \\
J103223$+$033933 & $15.7 \pm 2.3$ & $76.4 \pm 2.3$ & $4870 \pm 730$ \\
J111211$+$005607 & $14.9 \pm 2.3$ & $17.1 \pm 0.7$ & $1150 \pm 180$ \\
J233020$-$323729 & $19.7 \pm 2.3$ & $15.0 \pm 0.7$ & $761 \pm 96$ \\
\hline
\hline
\end{tabular}
\label{tab:unwise}
\end{table}

\subsection{The role of wide and deep near-infrared surveys in the study of UHzRGs} 

From the above discussion of Figure~\ref{fig:flux_flux}, it is clear that the six sources from the B22 sample with the faintest $K_{\rm s}$-band magnitudes are the most promising UHzRG candidates and could potentially fall within the EoR. However, obtaining deep $K_{\rm s}$-band imaging is expensive in terms of telescope time, especially as these rare bright radio sources are well-spaced across the sky and therefore require individual observations. 

Various ongoing and future near-infrared surveys that are both wide and deep offer excellent prospects for improving the efficiency of UHzRG searches. For example, SHARKS, made use of in this project (B22), will cover 300\,deg$^2$ to a median $5\sigma$ limiting magnitude of $K_{\rm s} = 22.7$\,mag (2\arcsec\:diameter aperture; $S_{2.2\,\mu\rm{m}} = 3.0$\,\textmu Jy; \citealt{dannerbauer22}). The $K_{\rm s}$-band magnitudes of TN~J0924$-$2201 and GLEAM~J0856$+$0223 \citep[][]{vanbreugel99,drouart20} are at brightness levels approximately corresponding to the $3\sigma$ SHARKS depth. Referring to Figure~\ref{fig:flux_tracks}, SHARKS may be able to find UHzRGs with  $M_{\rm *} \gtrsim 10^{11}$\,M$_\odot$. 

The {\it Euclid} mission \citep[][]{racca16,euclid24} will also significantly advance searches for UHzRGs. The {\it Euclid} Wide Survey \citep[EWS;][]{laureijs11,euclid22} will cover approximately 15\,000\,deg$^2$ in a wide optical band (530--920\,nm) and three near-infrared filters ($Y$-, $J$- and $H$-bands; 0.95--2.0\,\textmu m). The near-infrared point-source depth is expected to be 24.5\,mag ($5\sigma$; radius of 50\,per cent encircled energy $< 0\farcs4$), i.e. approximately 1.5\,mag fainter (at an equivalent detection threshold) than the deepest HAWK-I imaging presented here. Such data will greatly speed up the discovery of UHzRG candidates going forward; as a lower limit, extrapolating from six UHzRG candidates over approximately 1200\,deg$^2$ from our study, we would expect $\gtrsim 75$ UHzRG candidates to be found in the EWS. Furthermore, while the {\it Euclid} data alone can be used to discover bright regular galaxies at $z>6$--$7$, having a strong radio detection as well will immediately rule out contamination by very low-mass stars or dwarf star-forming galaxies at $1<z<2$, which are not known to host powerful AGN. The required radio surveys are extant, with, for example, GLEAM complete and its successor GLEAM eXtended \citep[GLEAM-X;][]{hurleywalker22,ross24} observed. With {\it Euclid} having now launched and beginning to collect science data, further candidate UHzRGs can be identified quickly. 

Looking towards the second half of this decade, the Vera C. Rubin Observatory \citep[][]{ivezic19} will have an exceptionally large FoV (9.6\,deg$^{2}$) and survey approximately 18\,000 deg$^2$ in six optical and near-infrared bands across the wavelength range 320--1050\,nm. After a first pass of the survey area, the $5\sigma$ point-source sensitivity in $Y$-band will be approximately 23.0\,mag (expected median zenith seeing $\approx 0\farcs7$), with an expected depth of 24.8\,mag after 10 years of operations. Additionally, the {\it Nancy Grace Roman Space Telescope} \citep[][]{spergel15,akeson19} will have an unprecedented combination of FoV and sensitivity: the Wide-Field Instrument (WFI) will have a 0.281\,deg$^2$ FoV and a limiting $5\sigma$ point-source sensitivity in the $F213$ filter (2.13\,\textmu m) of 25.6\,mag (1\,h integration; point spread function FWHM $= 0\farcs169$). A combination of {\it Roman} and the upcoming Square Kilometre Array \citep[SKA; e.g.][]{dewdney09,braun19} would transform searches for UHzRGs, including those which are much less powerful at radio frequencies than the sources considered in this study.

\section{CONCLUSIONS}
\label{sec:con}

In this paper, we presented deep VLT/HAWK-I $K_{\rm s}$-band imaging for 35 of the 53 sources from the B22 HzRG candidate sample, bringing the total number of sources from this sample with deep $K_{\rm s}$-band imaging to 41. Our conclusions are as follows:
\begin{enumerate}
\item Host galaxies were detected for 27/35 sources (77\,per cent) using {\sc photutils}. The corresponding $K_{\rm s}$-band magnitudes in 2\arcsec\:diameter circular apertures range from approximately 21.6--23.0\,mag. The remaining eight sources were not detected in the HAWK-I images, with deep magnitude lower limits being obtained (median $3\sigma$ depth $K_{\rm s} \approx 23.3$\,mag; 2\arcsec\:diameter apertures). 
\item  In the $S_{150\,\rm{MHz}}$--$S_{2.2\,\mu\rm{m}}$ flux density parameter space, the distribution of known powerful radio galaxies across $3<z<6$ is extended to fainter flux densities by the inclusion of the B22 sample. Like these distant radio galaxies from the literature, the B22 sources have large to extreme radio to near-infrared flux density ratios ($S_{150\,\rm{MHz}} / S_{2.2\,\mu\rm{m}}$  $\sim$ $7.0 \times 10^3$ -- $3.1 \times 10^5$). 
\item Five of the eight targets with HAWK-I non-detections, along with J0008$-$3007 from B22 that was not detected in SHARKS, are of particular interest given that they are UHzRG candidates. These six sources have a median $3\sigma$ magnitude limit of $K_{\rm s} \gtrsim 23.5$\,mag in 2\arcsec\:diameter apertures; moreover, the five HAWK-I non-detections have a median $3\sigma$ limit of $K_{\rm s} \gtrsim 23.8$\,mag in 1.5\arcsec\:diameter apertures. All of these UHzRG candidates have extreme radio to near-infrared flux density ratios ($S_{150\,\rm{MHz}} / S_{2.2\,\mu\rm{m}} >10^5$), similar to the three known powerful radio galaxies at $z> 5$ (TN~J0924$-$2201, GLEAM~J0856$+$0223 and TGSS~J1530$+$1049).    
\item If the host galaxies of these $K_{\rm s} \gtrsim 23.5$\,mag sources have stellar masses $M_{\rm *} \gtrsim 10^{10.5}$\,M$_\odot$, then there is a distinct possibility that at least some of them lie at ultra-high redshift to be so faint, or at least at $z \gtrsim 4.2$. They could also perhaps be powerful radio sources at $z \gtrsim 4$ hosted by undermassive ($M_{\rm *} \lesssim 10^{10}$\,M$_\odot$) galaxies, i.e. the radio-loud analogues of the population of AGN with large BH to stellar mass ratios being uncovered with {\em JWST}. Other possibilities are lower-redshift sources that are extremely obscured by dust or have one-sided radio emission.  
\item The majority of the 35 sources studied in this paper have radio and near/mid-infrared properties consistent with IFRSs, many of which have been confirmed in the literature to be at high redshift. 
\item Wide and deep ($K_{\rm s} \gtrsim 23$--$24$\,mag) near-infrared surveys are key to efficiently selecting the most promising UHzRG candidates.
\end{enumerate}

For future work, deeper near-infrared observations are needed to identify the host galaxy for each of our UHzRG candidates. Given the tracks in Figure~\ref{fig:flux_tracks}, as well as the (very) faint near-infrared flux densities being measured with {\em JWST} for confirmed or candidate obscured AGN in massive galaxies at ultra-high redshift \citep[e.g.][]{labbe23a,labbe23b,akins23,lambrides24,barro24}, {\em JWST} or {\em Hubble Space Telescope} \citep[{\em HST}; e.g.][]{bahcall86}   observations would be the most efficient way of securing an identification for our targets. Additionally, efforts are underway to obtain spectroscopic redshifts for the B22 sample; these data will be presented in a future paper. 
 
\begin{acknowledgement}
We thank the referee for their constructive review, which helped improve the presentation and clarity of this paper.
We also acknowledge the Noongar people as the traditional owners and custodians of Whadjuk Boodjar, the land on which the majority of this work was completed. 

Based on observations collected at the European Southern Observatory under ESO programme  108.22HY.001. Based on data obtained from the ESO Science Archive Facility with DOI: \url{https://doi.org/10.18727/archive/34}. We thank Lowell Tacconi-Garman from the ESO User Support Department for helping us to finalise the observing strategy and for manually processing a number of datasets that could not be automatically reduced by the HAWK-I pipeline. We also thank Helmut Dannerbauer and Aurelio Carnero Rosell from the SHARKS team for providing us with available SHARKS data prior to the submission of the proposal for our HAWK-I observing campaign, allowing us to finalise the target list.  

We thank Anshu Gupta, Joe Grundy and Luca Ighina for useful discussions regarding the nature of our UHzRG candidates.

JMA acknowledges financial support from the Science and Technology Foundation (FCT, Portugal) through research grants PTDC/FIS-AST/29245/2017, UIDB/04434/2020 and UIDP/04434/2020. GN acknowledges funding support from the Natural Sciences and Engineering Research Council (NSERC) of Canada through a Discovery Grant and Discovery Accelerator Supplement, and from the Canadian Space Agency through grant 18JWST-GTO1.

This publication makes use of data products from the Two Micron All Sky Survey, which is a joint project of the University of Massachusetts and the Infrared Processing and Analysis Center/California Institute of Technology, funded by the National Aeronautics and Space Administration and the National Science Foundation. This publication makes use of data products from the {\it Wide-field Infrared Survey Explorer}, which is a joint project of the University of California, Los Angeles, and the Jet Propulsion Laboratory/California Institute of Technology, funded by the National Aeronautics and Space Administration.

This scientific work uses data obtained from Inyarrimanha Ilgari Bundara / the Murchison Radio-astronomy Observatory. We acknowledge the Wajarri Yamaji People as the Traditional Owners and native title holders of the Observatory site. CSIRO’s ASKAP radio telescope is part of the Australia Telescope National Facility (\url{https://ror.org/05qajvd42}). Operation of ASKAP is funded by the Australian Government with support from the National Collaborative Research Infrastructure Strategy. ASKAP uses the resources of the Pawsey Supercomputing Research Centre. Establishment of ASKAP, Inyarrimanha Ilgari Bundara, the CSIRO Murchison Radio-astronomy Observatory and the Pawsey Supercomputing Research Centre are initiatives of the Australian Government, with support from the Government of Western Australia and the Science and Industry Endowment Fund. This paper includes archived data obtained through the CSIRO ASKAP Science Data Archive, CASDA (\url{http://data.csiro.au}). 

The Australia Telescope Compact Array is part of the Australia Telescope National Facility (\url{https://ror.org/05qajvd42}) which is funded by the Australian Government for operation as a National Facility managed by CSIRO. We acknowledge the Gomeroi people as the Traditional Owners of the Observatory site. 

The National Radio Astronomy Observatory is a facility of the National Science Foundation operated under cooperative agreement by Associated Universities, Inc. 

The national facility capability for SkyMapper has been funded through ARC LIEF grant LE130100104 from the Australian Research Council, awarded to the University of Sydney, the Australian National University, Swinburne University of Technology, the University of Queensland, the University of Western Australia, the University of Melbourne, Curtin University of Technology, Monash University and the Australian Astronomical Observatory. SkyMapper is owned and operated by The Australian National University's Research School of Astronomy and Astrophysics. The survey data were processed and provided by the SkyMapper Team at ANU. The SkyMapper node of the All-Sky Virtual Observatory (ASVO) is hosted at the National Computational Infrastructure (NCI). Development and support of the SkyMapper node of the ASVO has been funded in part by Astronomy Australia Limited (AAL) and the Australian Government through the Commonwealth's Education Investment Fund (EIF) and National Collaborative Research Infrastructure Strategy (NCRIS), particularly the National eResearch Collaboration Tools and Resources (NeCTAR) and the Australian National Data Service Projects (ANDS).

Funding for the Sloan Digital Sky Survey V has been provided by the Alfred P. Sloan Foundation, the Heising-Simons Foundation, the National Science Foundation, and the Participating Institutions. SDSS acknowledges support and resources from the Center for High-Performance Computing at the University of Utah. The SDSS web site is \url{www.sdss.org}.

SDSS is managed by the Astrophysical Research Consortium for the Participating Institutions of the SDSS Collaboration, including the Carnegie Institution for Science, Chilean National Time Allocation Committee (CNTAC) ratified researchers, the Gotham Participation Group, Harvard University, Heidelberg University, The Johns Hopkins University, L'Ecole polytechnique f\'{e}d\'{e}rale de Lausanne (EPFL), Leibniz-Institut f{\"u}r Astrophysik Potsdam (AIP), Max-Planck-Institut f{\"u}r Astronomie (MPIA Heidelberg), Max-Planck-Institut f{\"u}r Extraterrestrische Physik (MPE), Nanjing University, National Astronomical Observatories of China (NAOC), New Mexico State University, The Ohio State University, Pennsylvania State University, Smithsonian Astrophysical Observatory, Space Telescope Science Institute (STScI), the Stellar Astrophysics Participation Group, Universidad Nacional Aut\'{o}noma de M\'{e}xico, University of Arizona, University of Colorado Boulder, University of Illinois at Urbana-Champaign, University of Toronto, University of Utah, University of Virginia, Yale University, and Yunnan University.

This research has made use of the CIRADA cutout service at URL \url{cutouts.cirada.ca}, operated by the Canadian Initiative for Radio Astronomy Data Analysis (CIRADA). CIRADA is funded by a grant from the Canada Foundation for Innovation 2017 Innovation Fund (Project 35999), as well as by the Provinces of Ontario, British Columbia, Alberta, Manitoba and Quebec, in collaboration with the National Research Council of Canada, the US National Radio Astronomy Observatory and Australia’s Commonwealth Scientific and Industrial Research Organisation.

This work made use of Astropy\footnote{\url{http://www.astropy.org}}: a community-developed core Python package and an ecosystem of tools and resources for astronomy \citep[][]{astropy13,astropy18,astropy22}. This research made use of Photutils, an Astropy package for detection and photometry of astronomical sources \citep[][]{bradley21}. 

This research has made use of the VizieR catalogue access tool, CDS, Strasbourg, France (DOI: 10.26093/cds/vizier). The original description of the VizieR service was published in A\&AS 143, 23 \citep*[][]{ochsenbein00}. This research has made use of the NASA/IPAC Extragalactic Database (NED), which is funded by the National Aeronautics and Space Administration and operated by the California Institute of Technology. This research has made use of NASA's Astrophysics Data System Bibliographic Services. This project also made use of {\sc kvis} \citep[][]{gooch95}, {\sc matplotlib} \citep[][]{hunter07}, {\sc numpy} \citep[][]{oliphant06}, {\sc scipy} \citep[][]{virtanen20} and {\sc topcat} \citep[][]{taylor05}.
\end{acknowledgement}


\bibliography{references}

\appendix

\section{Additional information from the HAWK-I observing campaign}\label{appendix}

In Figure~\ref{fig:overlays_appendix}, we present overlay plots for the 30 sources not included in Figure~\ref{fig:overlays}. For four of the host galaxy detections in Figure~\ref{fig:overlays_appendix}, additional greyscale plots without radio contours overlaid can be found in Figure~\ref{fig:overlays_appendix_2}, allowing the reader a clearer view of the $K_{\rm s}$-band emission. 

\begin{figure*}
\begin{minipage}[t]{0.48\textwidth}
\includegraphics[height=6.5cm]{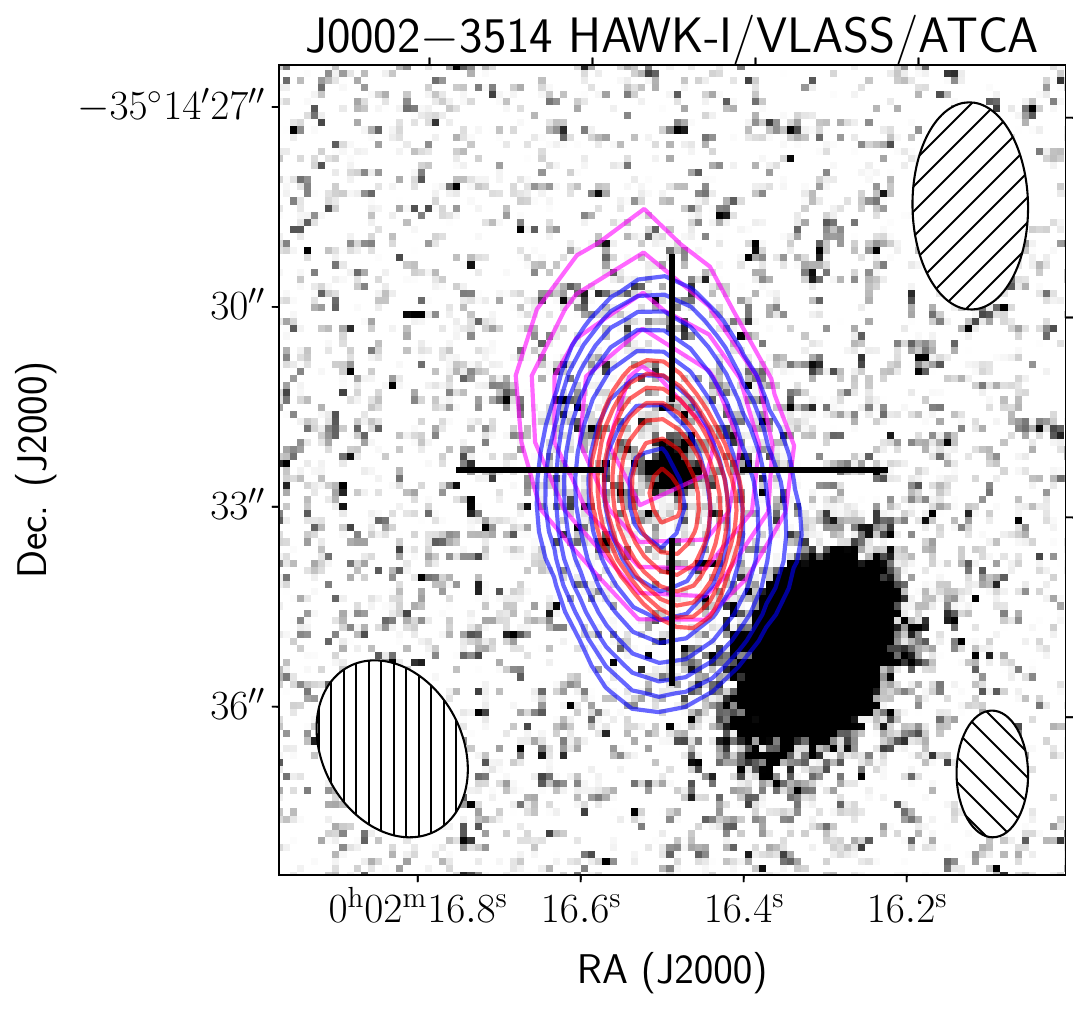}
\end{minipage}%
\hspace{0.02\linewidth}%
\begin{minipage}[t]{0.48\textwidth}
\includegraphics[height=6.5cm]{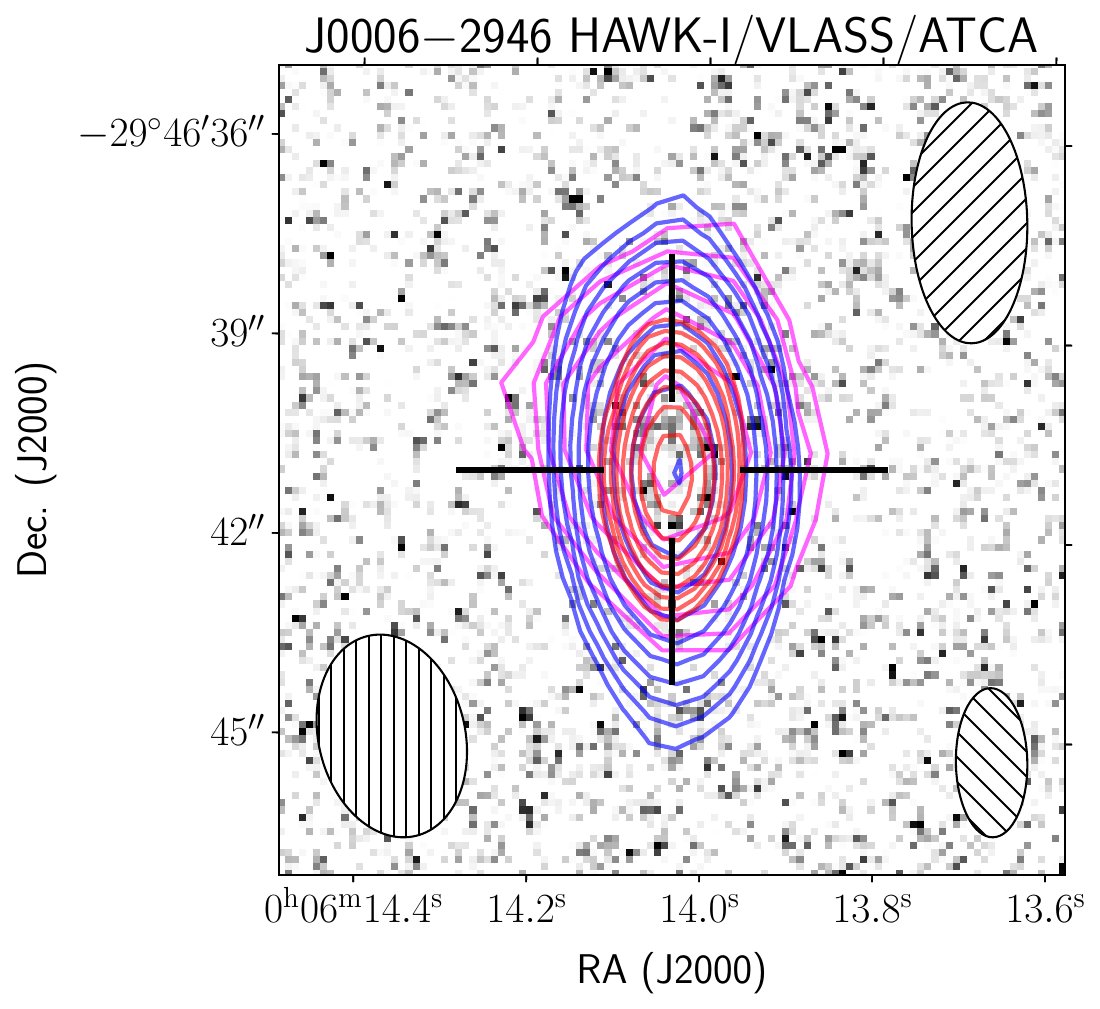}
\end{minipage}
\begin{minipage}[t]{0.48\textwidth}
\includegraphics[height=6.5cm]{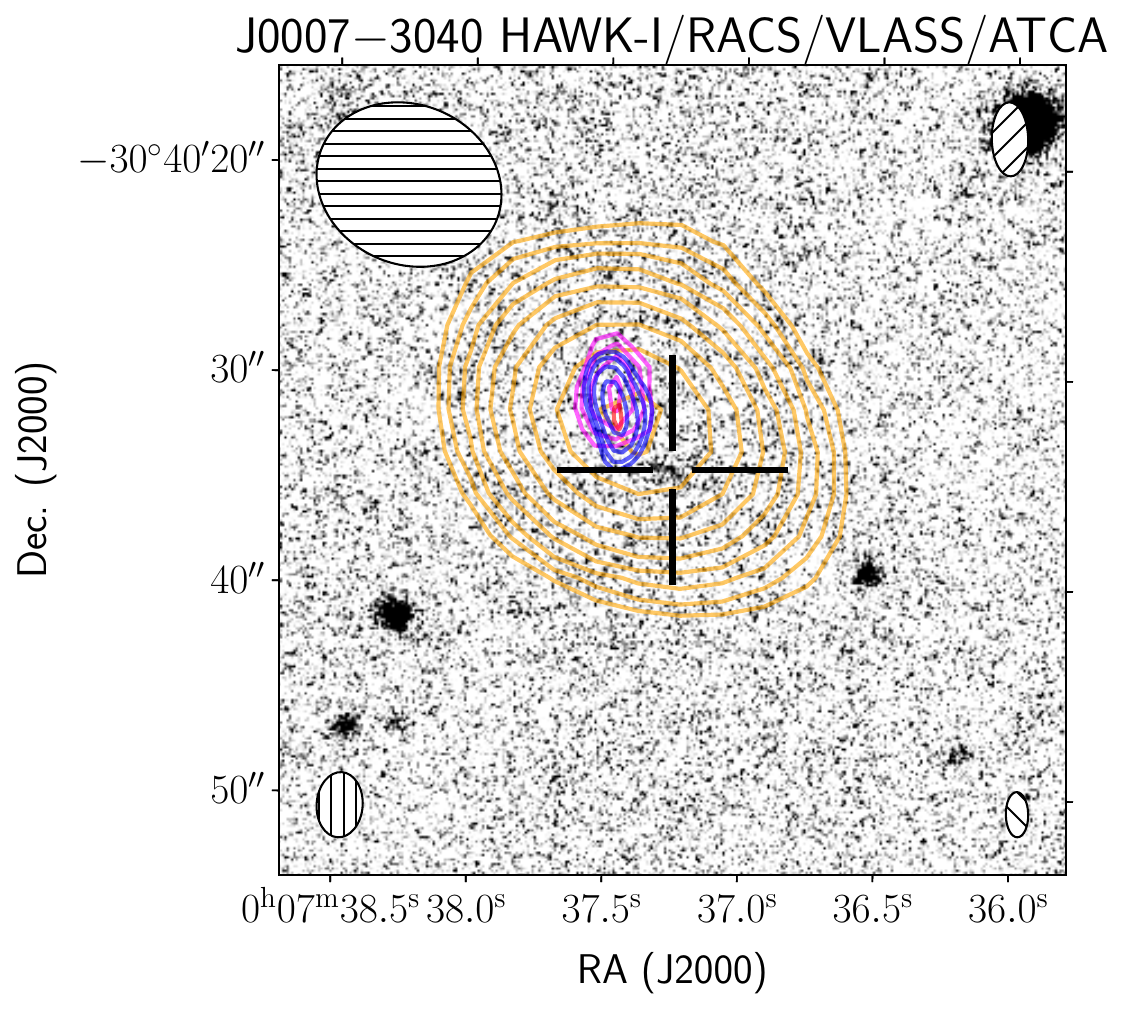}
\end{minipage}%
\hspace{0.02\linewidth}%
\begin{minipage}[t]{0.48\textwidth}
\includegraphics[height=6.5cm]{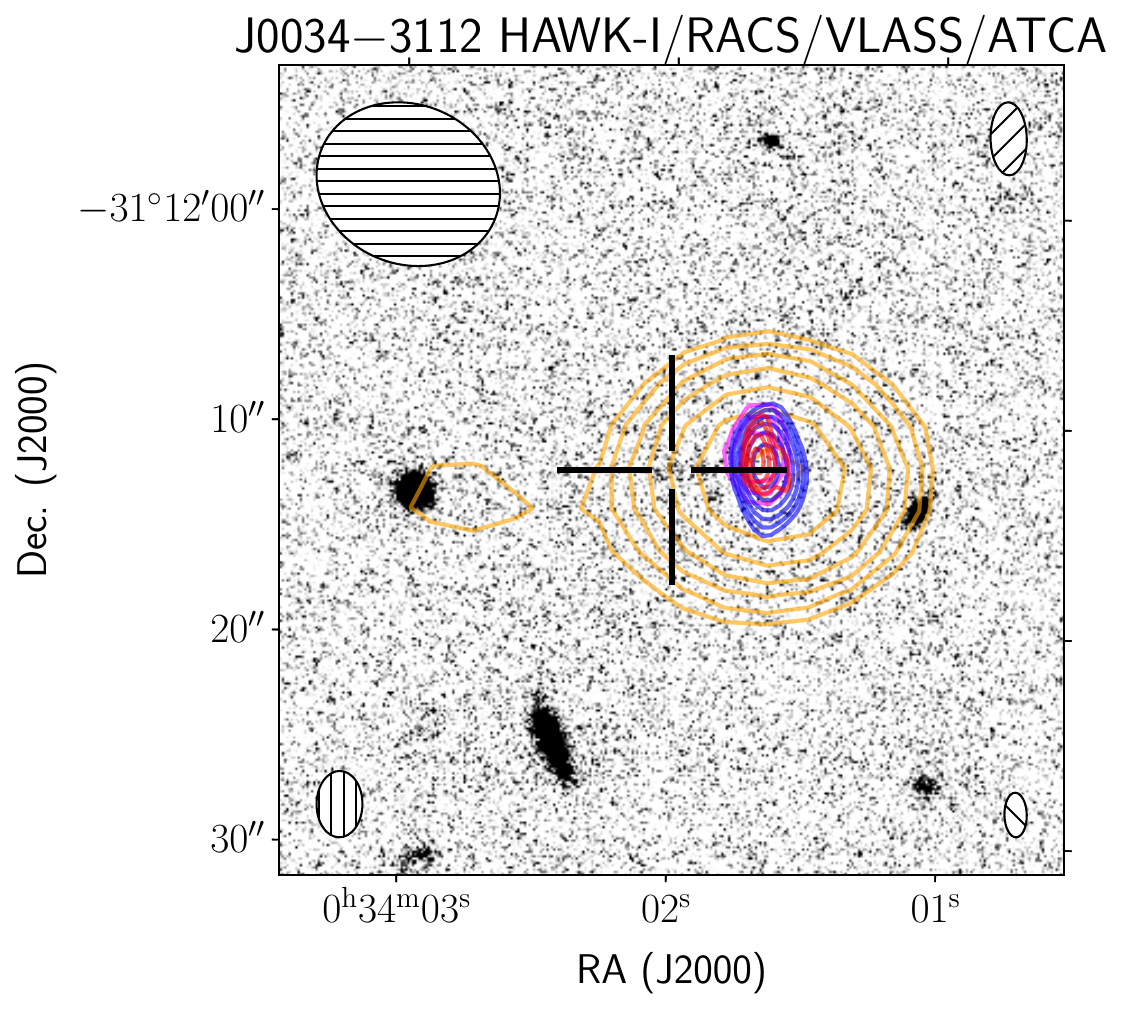}
\end{minipage}
\begin{minipage}[t]{0.48\textwidth}
\includegraphics[height=6.5cm]{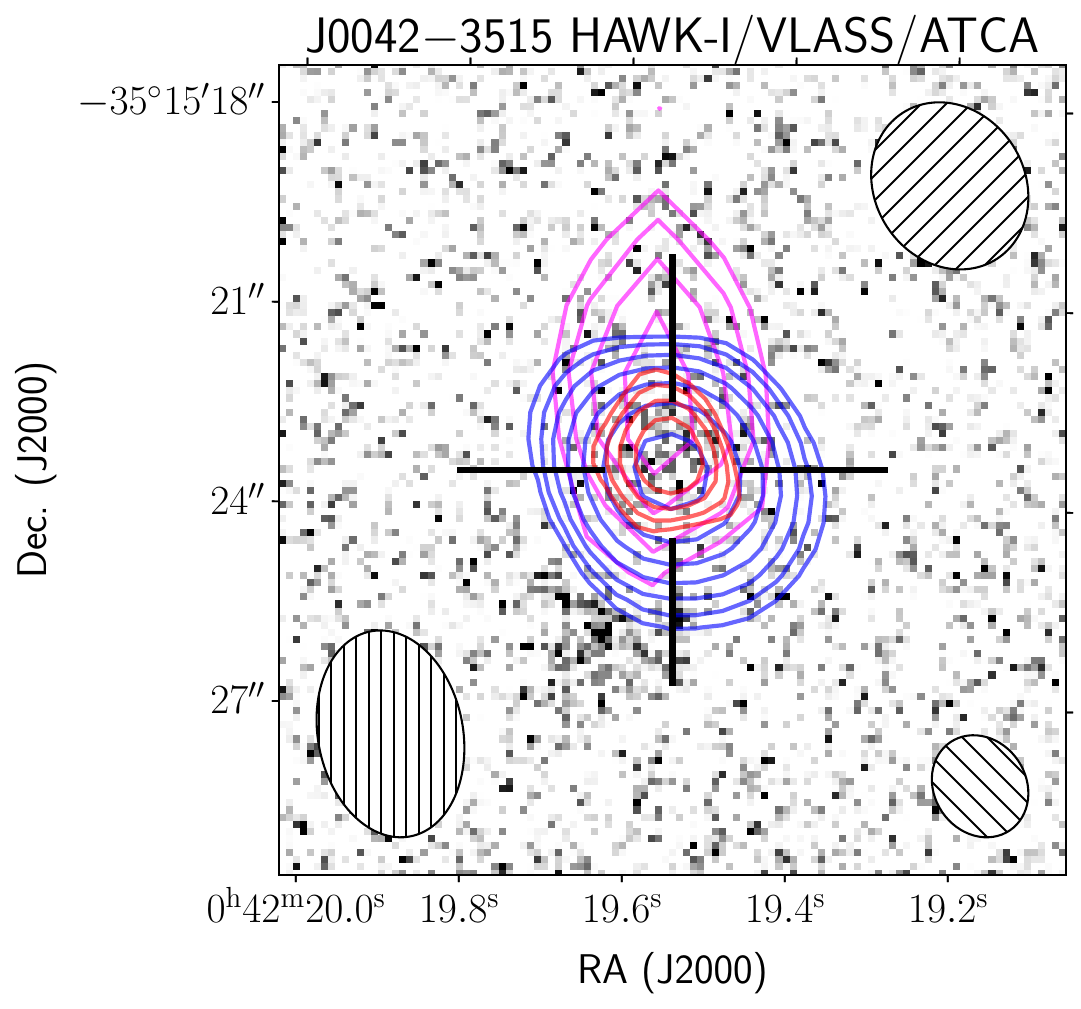}
\end{minipage}%
\hspace{0.02\linewidth}%
\begin{minipage}[t]{0.48\textwidth}
\includegraphics[height=6.5cm]{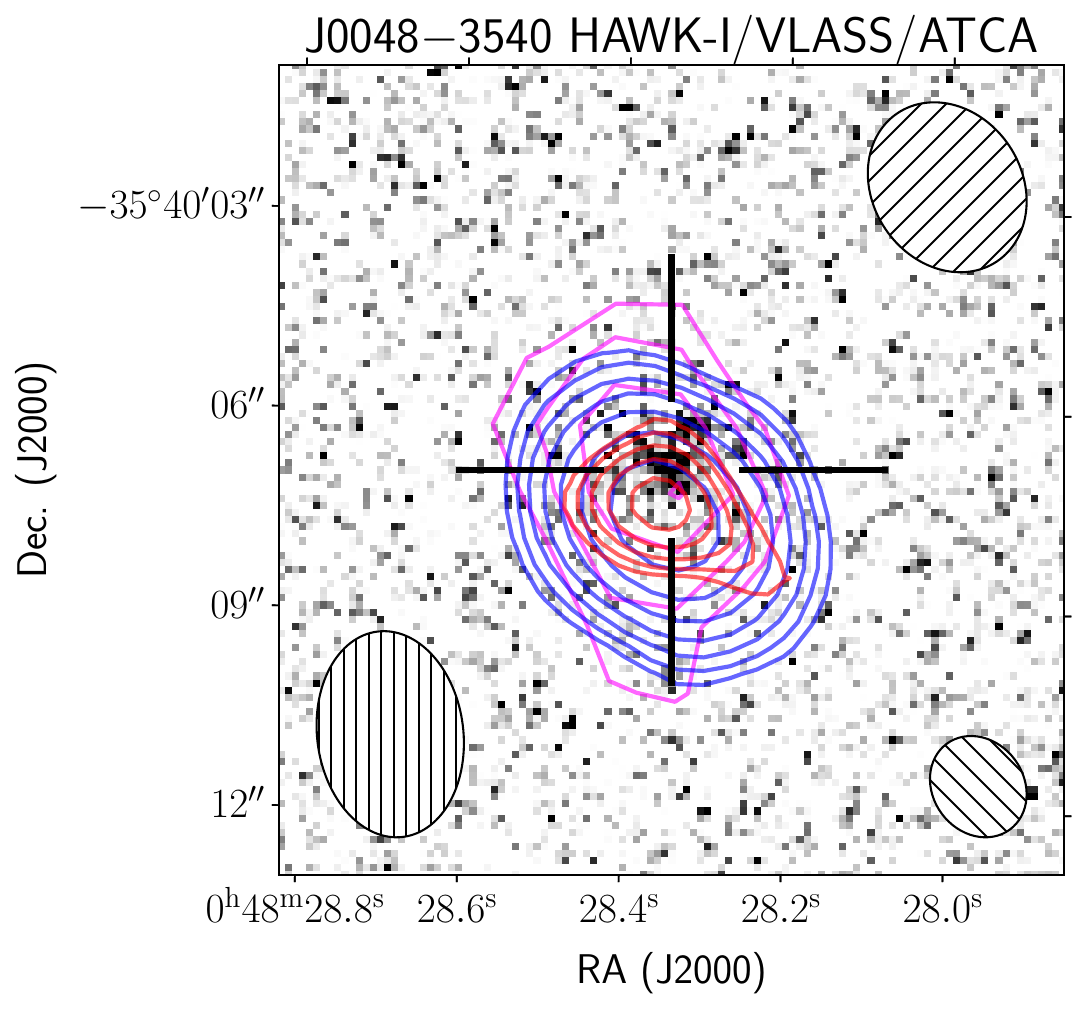}
\end{minipage}
\caption{Radio contours overlaid on HAWK-I $K_{\rm s}$-band images for the 30 sources not shown in Figure~\ref{fig:overlays}, again ordered by right ascension. The panels are formatted in the same way as those in Figure~\ref{fig:overlays}, except that the contours for J0309$-$3526 start at $3\sigma$ rather than $5\sigma$ (Table 5 in B22). Additionally, crosshairs mark the position of each host galaxy detection. Further relevant information is as follows. We used the new single-epoch images from VLASS for J1030$+$0135, J1032$+$0339 and J1037$-$0325. For J0007$-$3040, J0034$-$3112 and J1037$-$0325, we show 1367.5-MHz contours (in orange) from RACS-mid, while for J2311$-$3359 we show 887.5-MHz contours (in turquoise) from the ASKAP study by \citet[][]{gurkan22}. For the single-epoch images from VLASS, the lowest contour levels ($5\sigma$) are 0.75 (J1030$+$0135), 0.65 (J1032$+$0339) and 0.80 (J1037$-$0325) mJy\,beam$^{-1}$. For the RACS-mid data as well as the ASKAP data from \citet[][]{gurkan22}, the lowest contour levels ($5\sigma$) are 0.85 (J0007$-$3040), 0.70 (J0034$-$3112), 0.95 (J1037$-$0325) and 0.185 (J2311$-$3359) mJy\,beam$^{-1}$. The hatching style for the RACS-mid/ASKAP synthesised beams is horizontal. Note that the host galaxy of J0240$-$3206 is only detected above our $3\sigma$ threshold in a 1.5\arcsec\:diameter aperture.}  
\label{fig:overlays_appendix}
\end{figure*}

\setcounter{figure}{3} 
\begin{figure*}
\begin{minipage}[t]{0.48\textwidth}
\includegraphics[height=6.5cm]{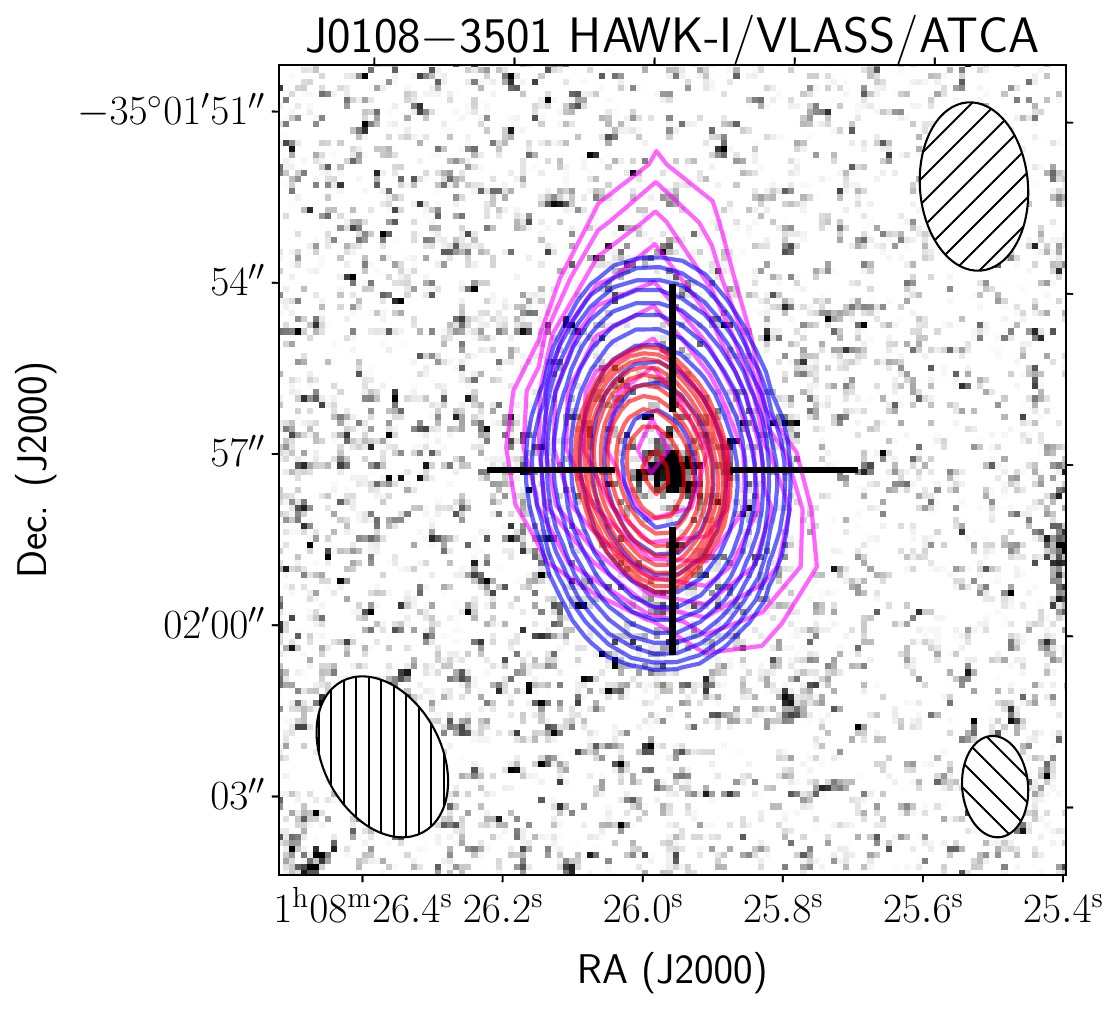}
\end{minipage}%
\hspace{0.02\linewidth}%
\begin{minipage}[t]{0.48\textwidth}
\includegraphics[height=6.5cm]{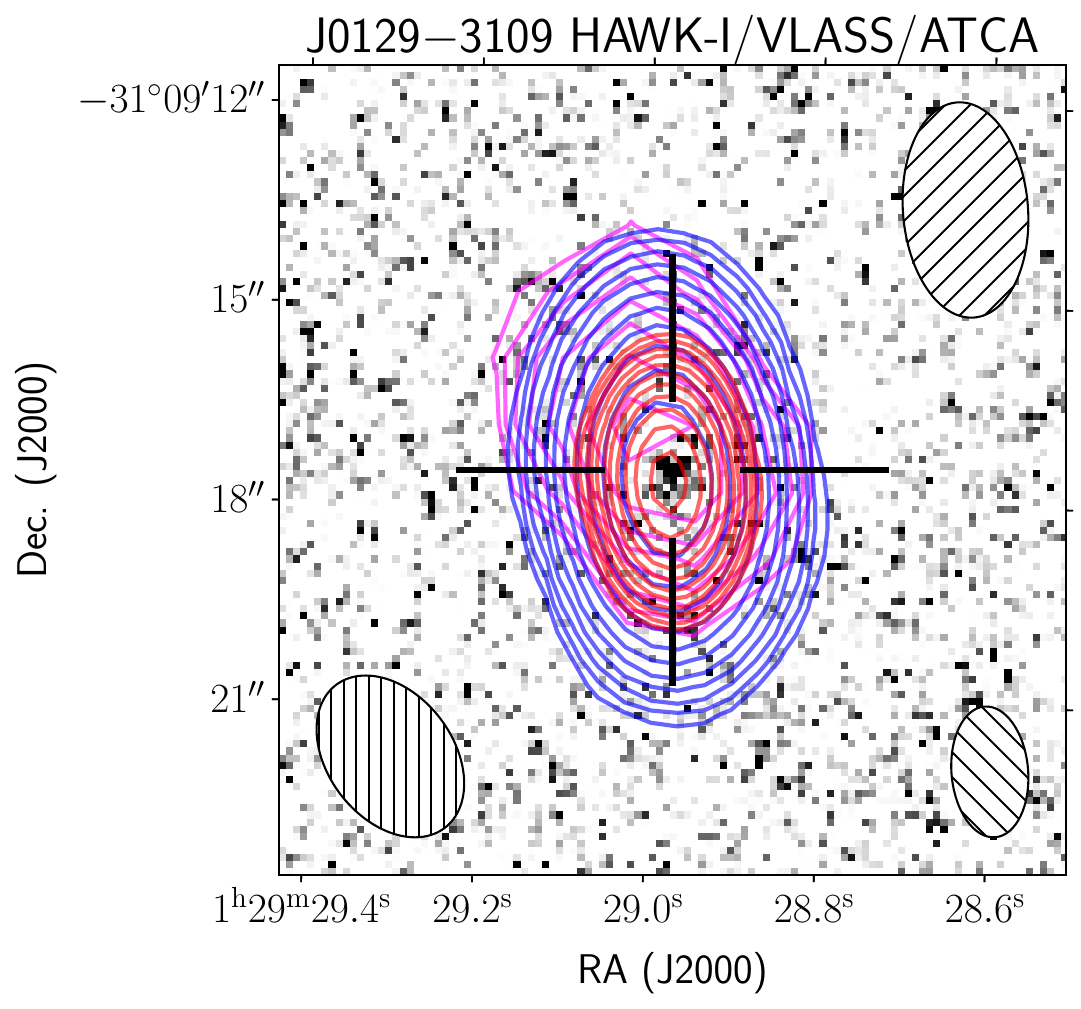}
\end{minipage}
\begin{minipage}[t]{0.48\textwidth}
\includegraphics[height=6.5cm]{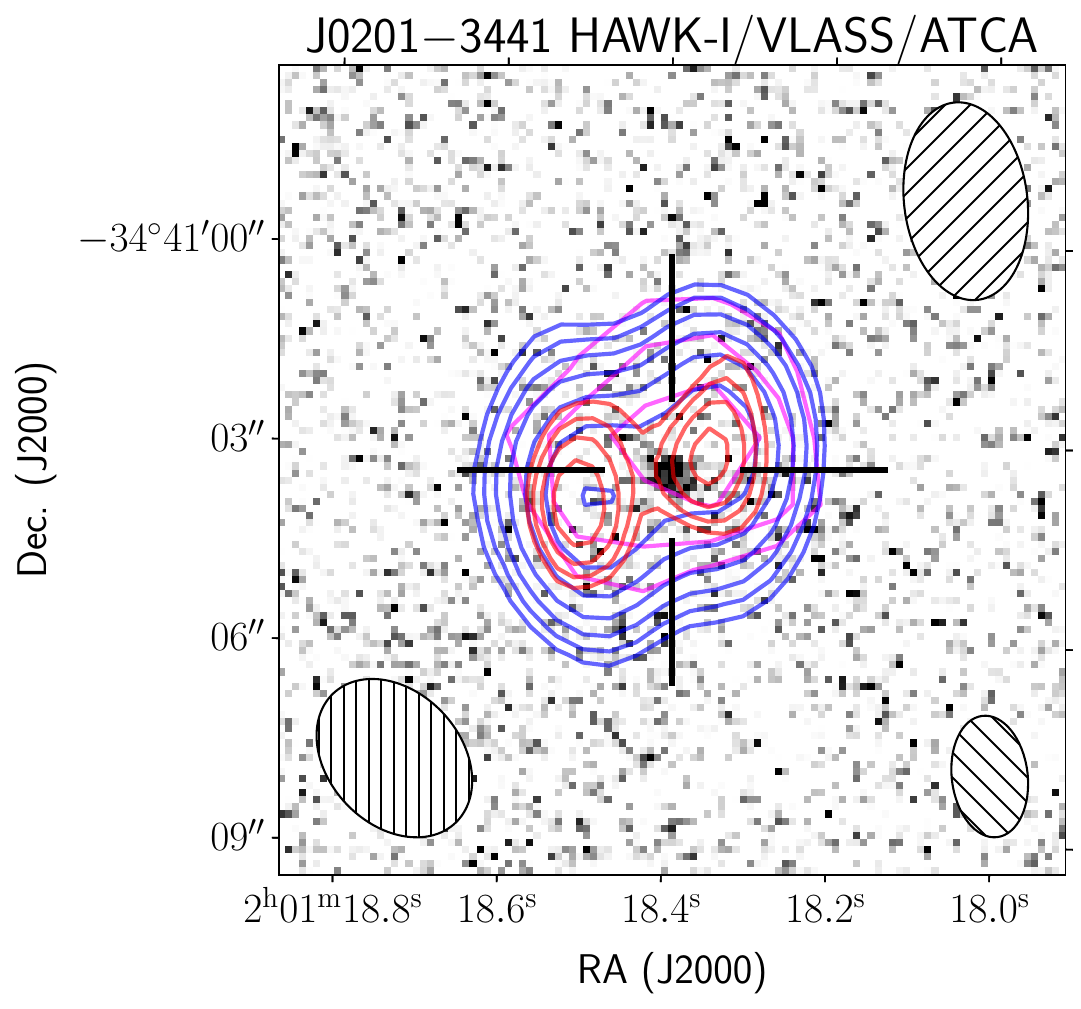}
\end{minipage}%
\hspace{0.02\linewidth}%
\begin{minipage}[t]{0.48\textwidth}
\includegraphics[height=6.5cm]{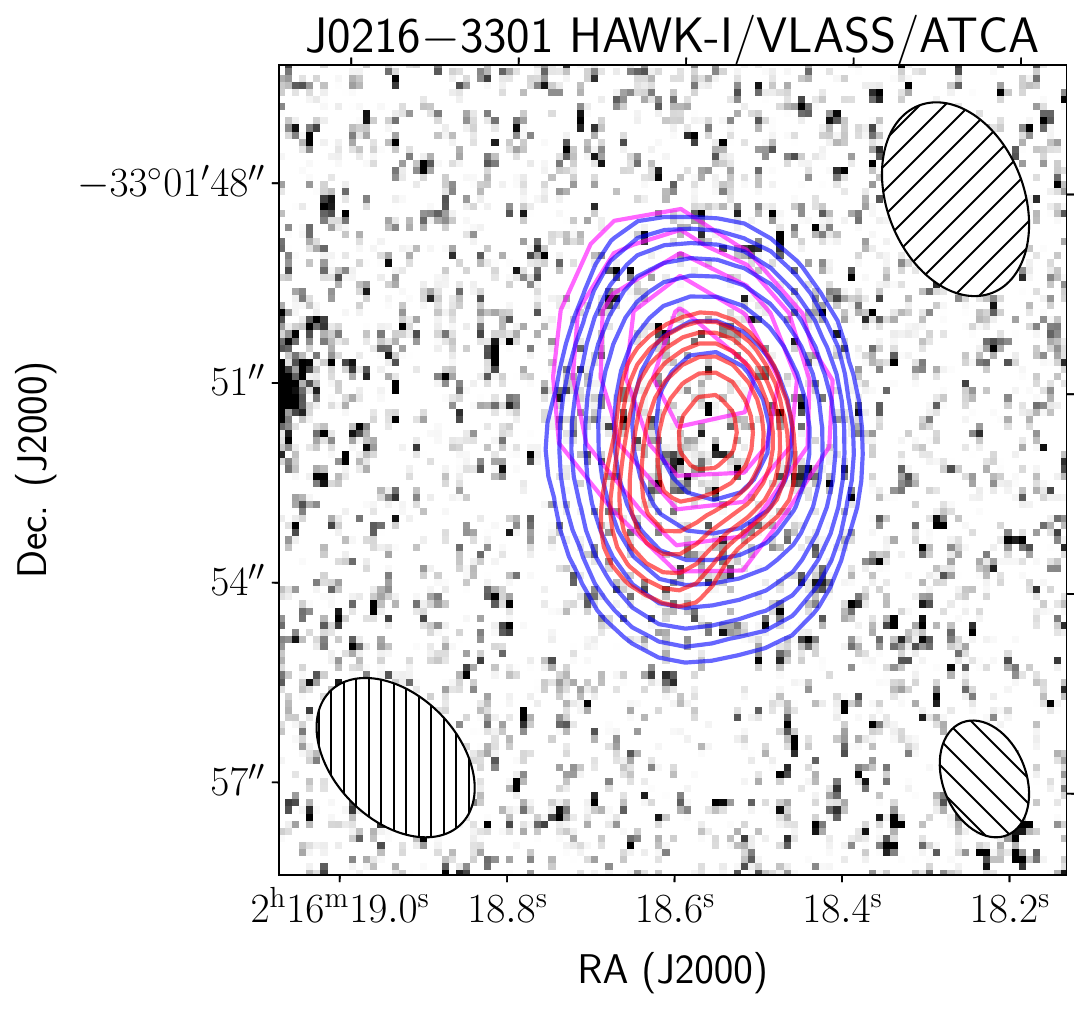}
\end{minipage}
\begin{minipage}[t]{0.48\textwidth}
\includegraphics[height=6.5cm]{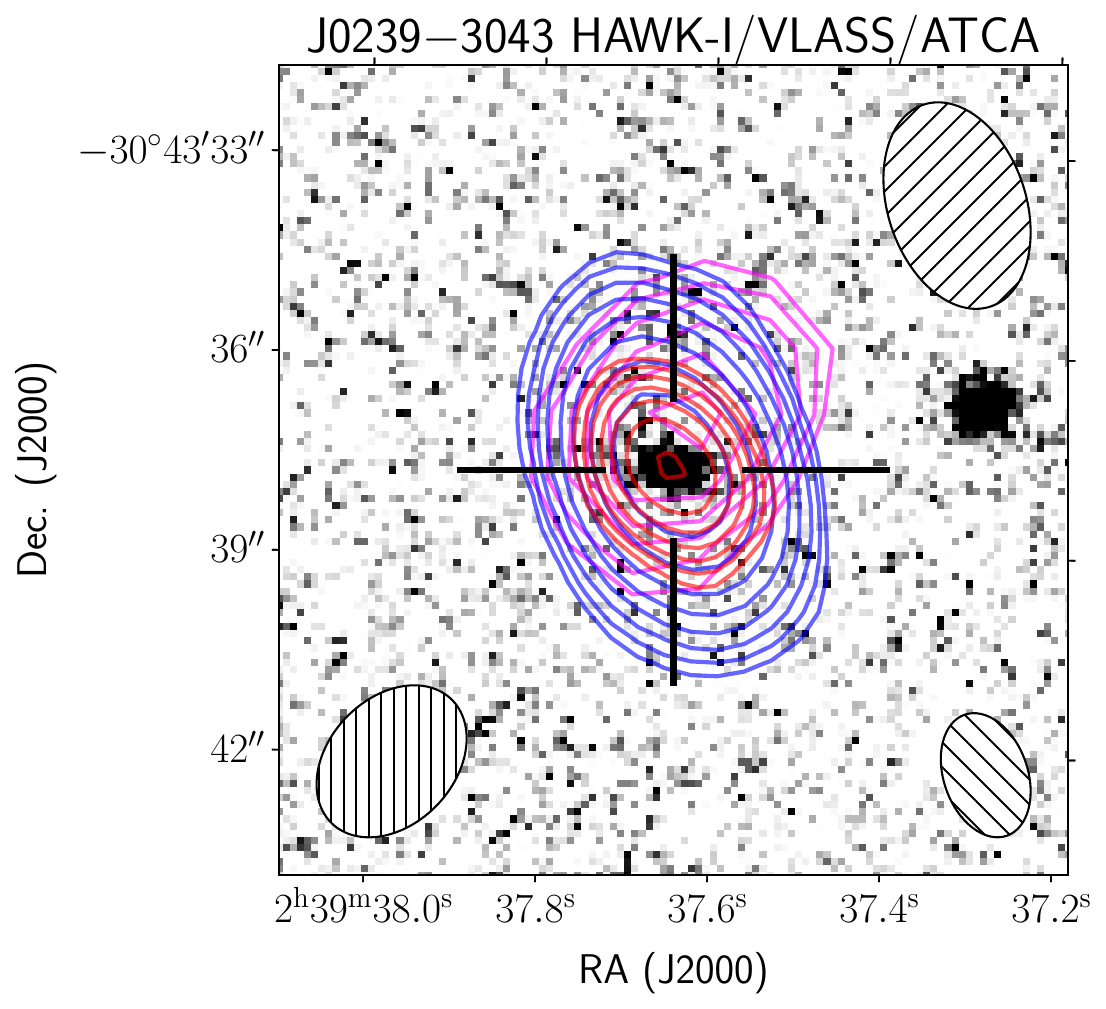}
\end{minipage}%
\hspace{0.02\linewidth}%
\begin{minipage}[t]{0.48\textwidth}
\includegraphics[height=6.5cm]{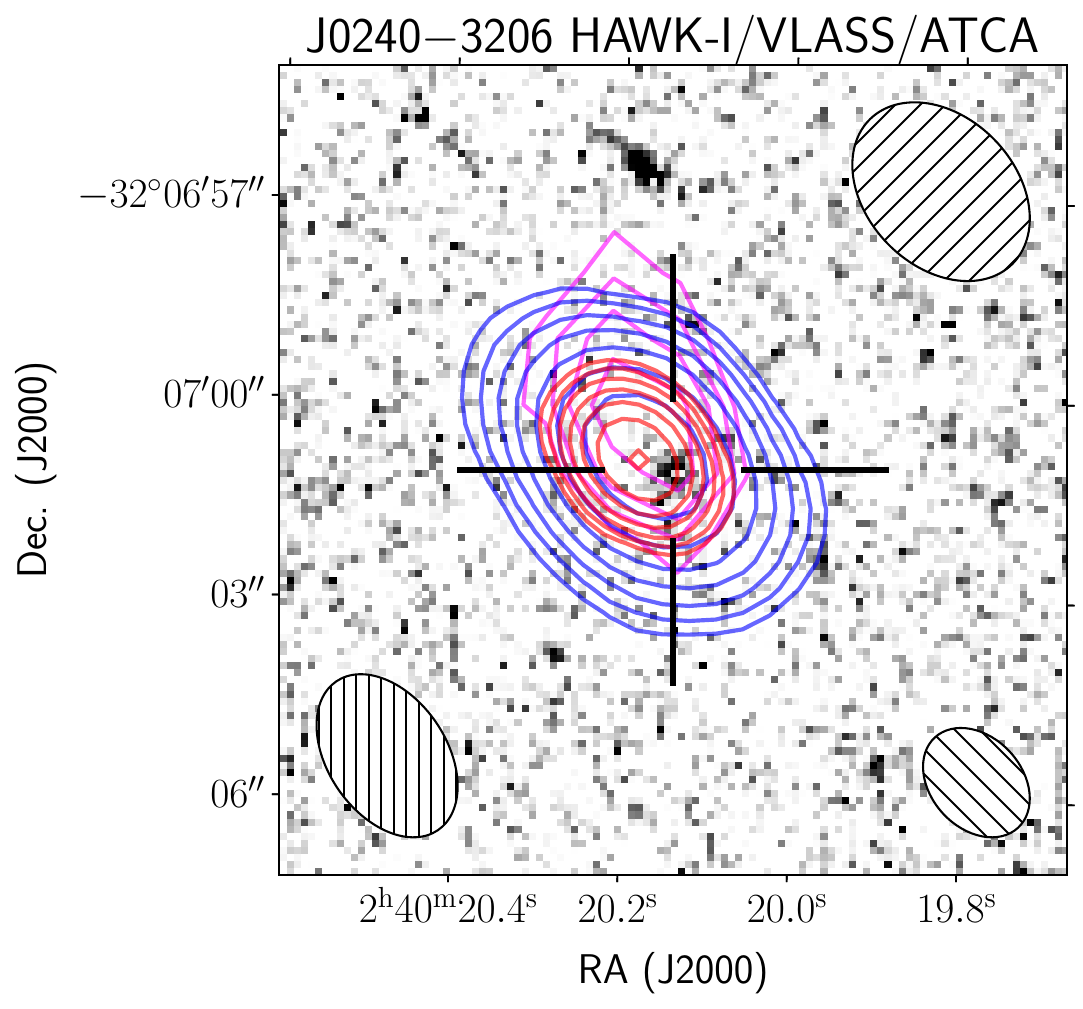}
\end{minipage}
\caption{{\em - continued.}} 
\end{figure*}

\setcounter{figure}{3} 
\begin{figure*}
\begin{minipage}[t]{0.48\textwidth}
\includegraphics[height=6.5cm]{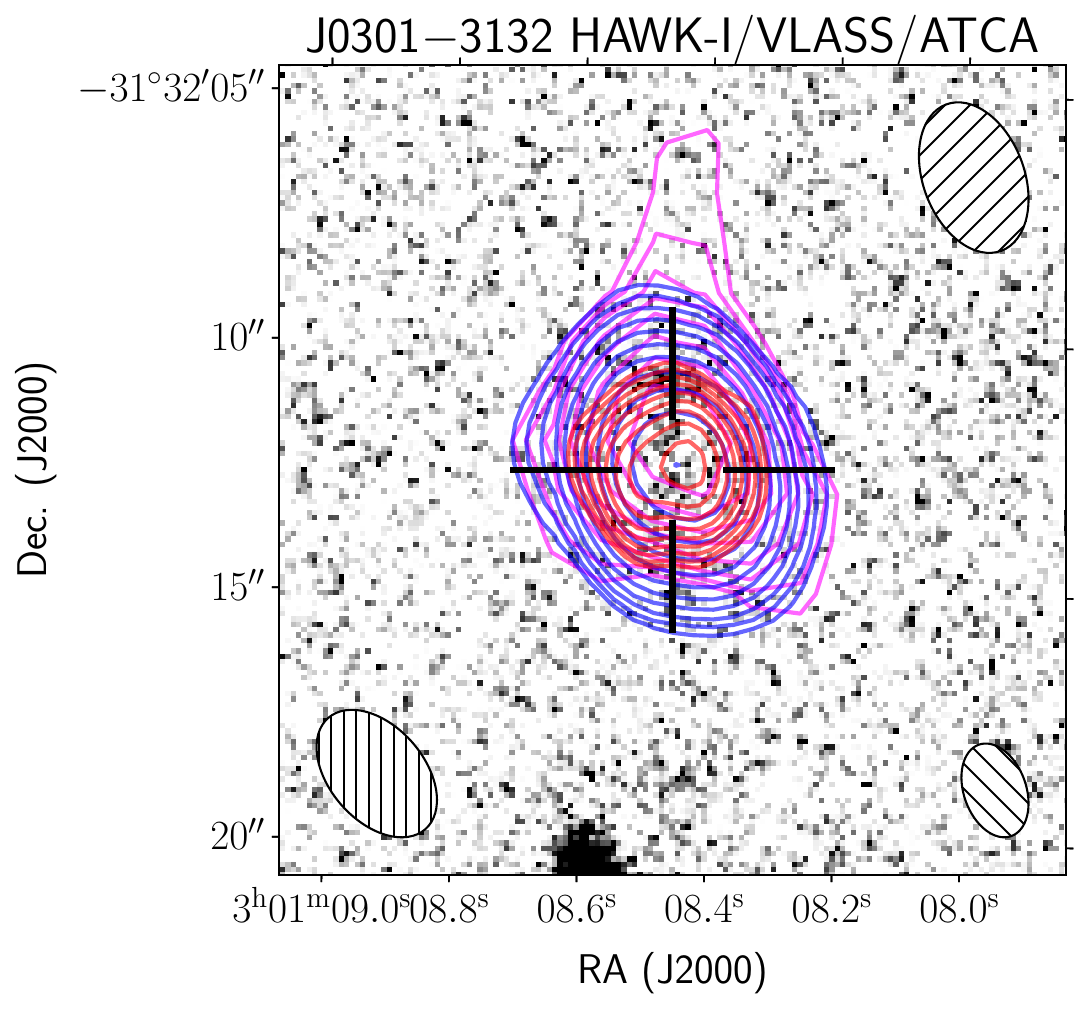}
\end{minipage}%
\hspace{0.02\linewidth}%
\begin{minipage}[t]{0.48\textwidth}
\includegraphics[height=6.5cm]{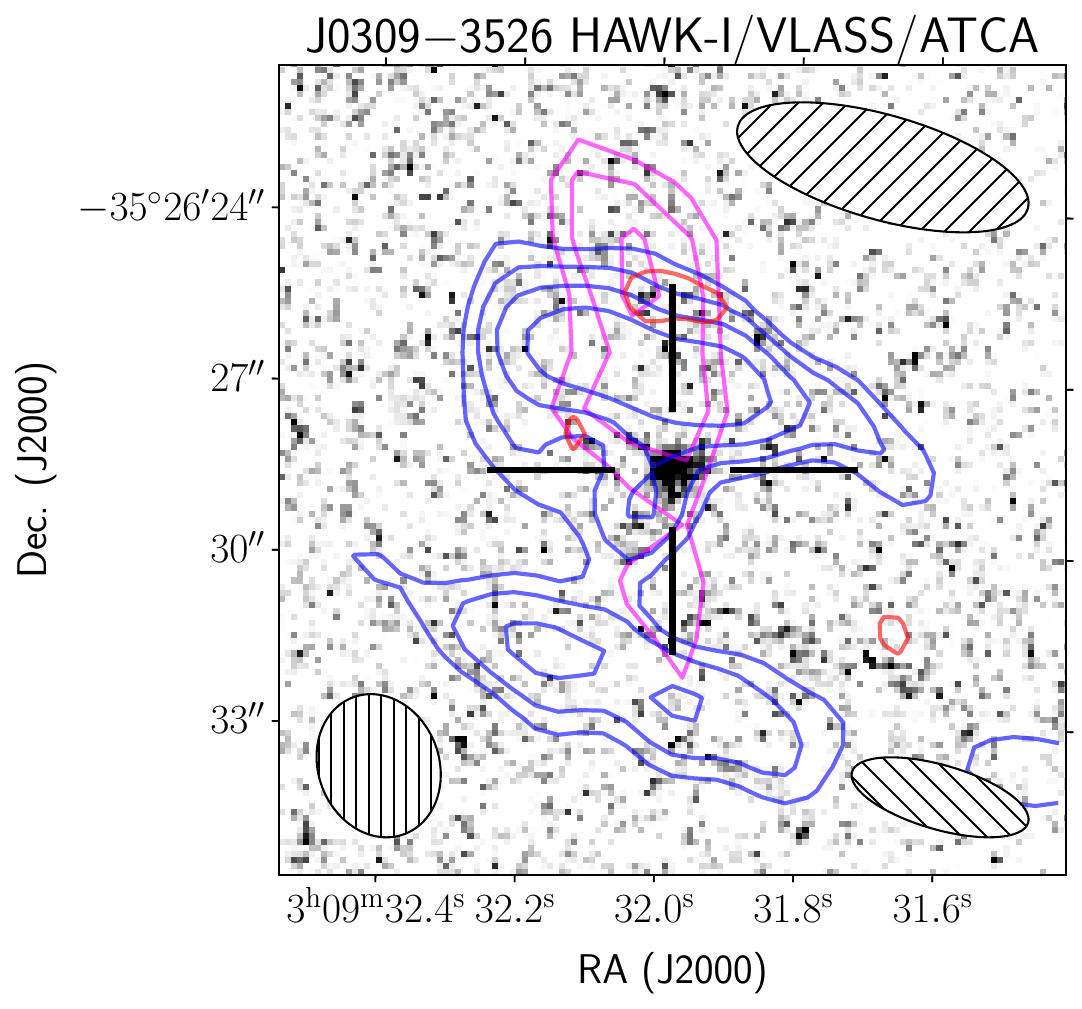}
\end{minipage}
\begin{minipage}[t]{0.48\textwidth}
\includegraphics[height=6.5cm]{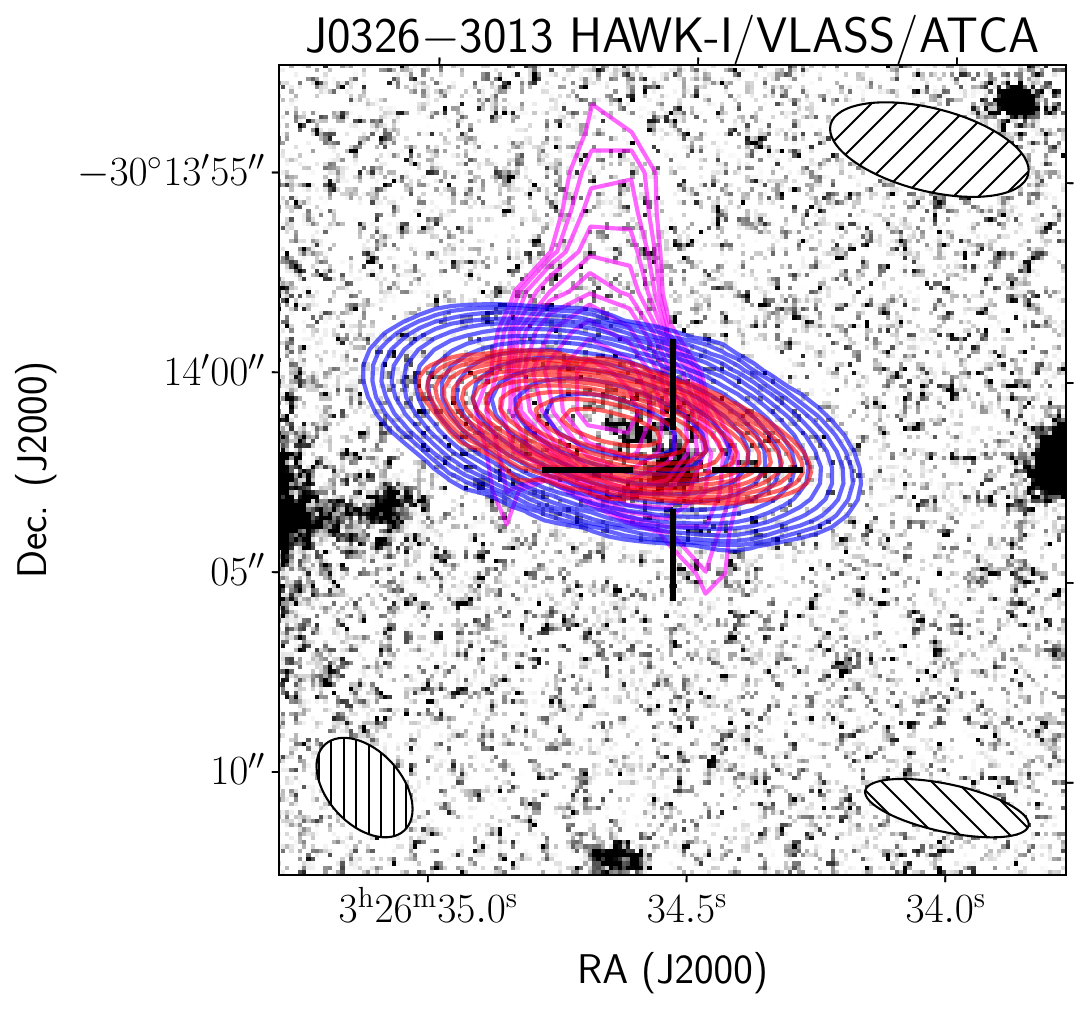}
\end{minipage}%
\hspace{0.02\linewidth}%
\begin{minipage}[t]{0.48\textwidth}
\includegraphics[height=6.5cm]{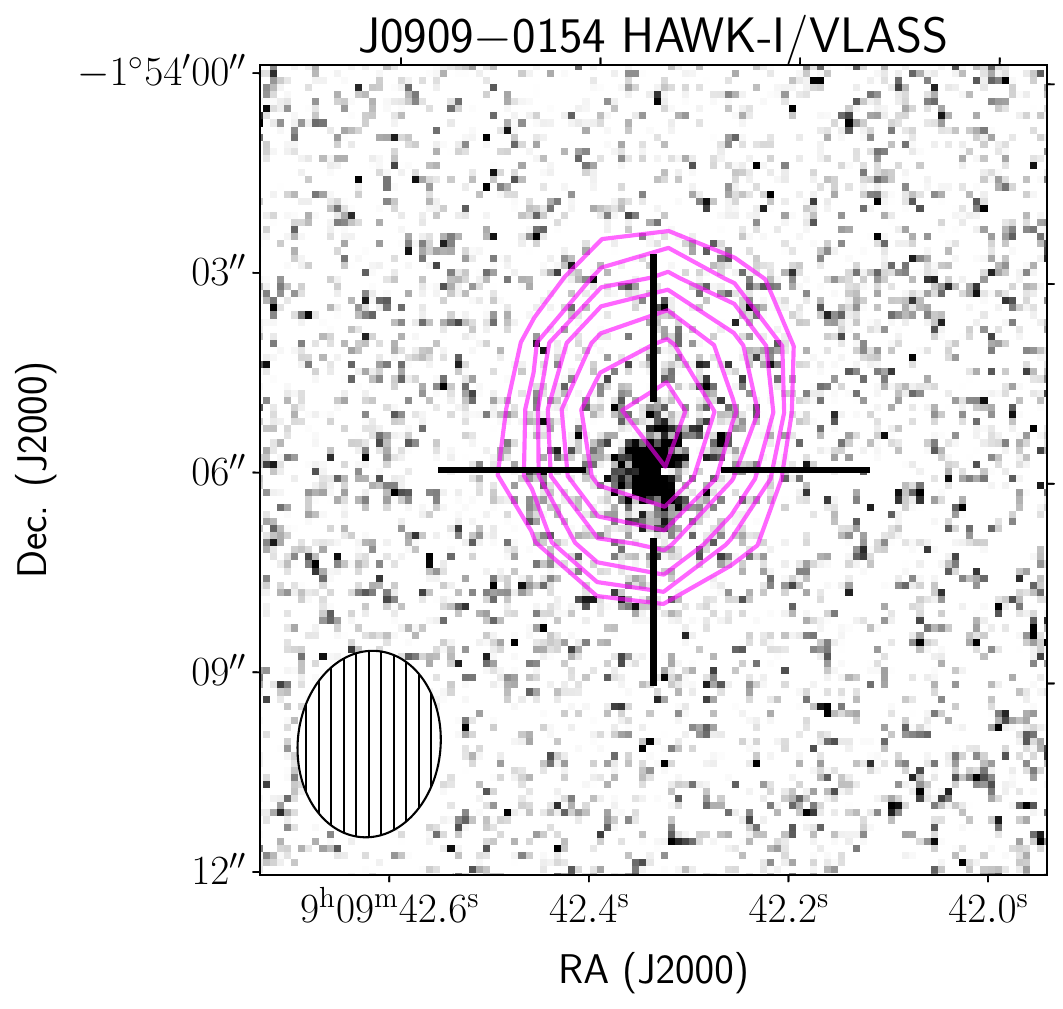}
\end{minipage}
\begin{minipage}[t]{0.48\textwidth}
\includegraphics[height=6.5cm]{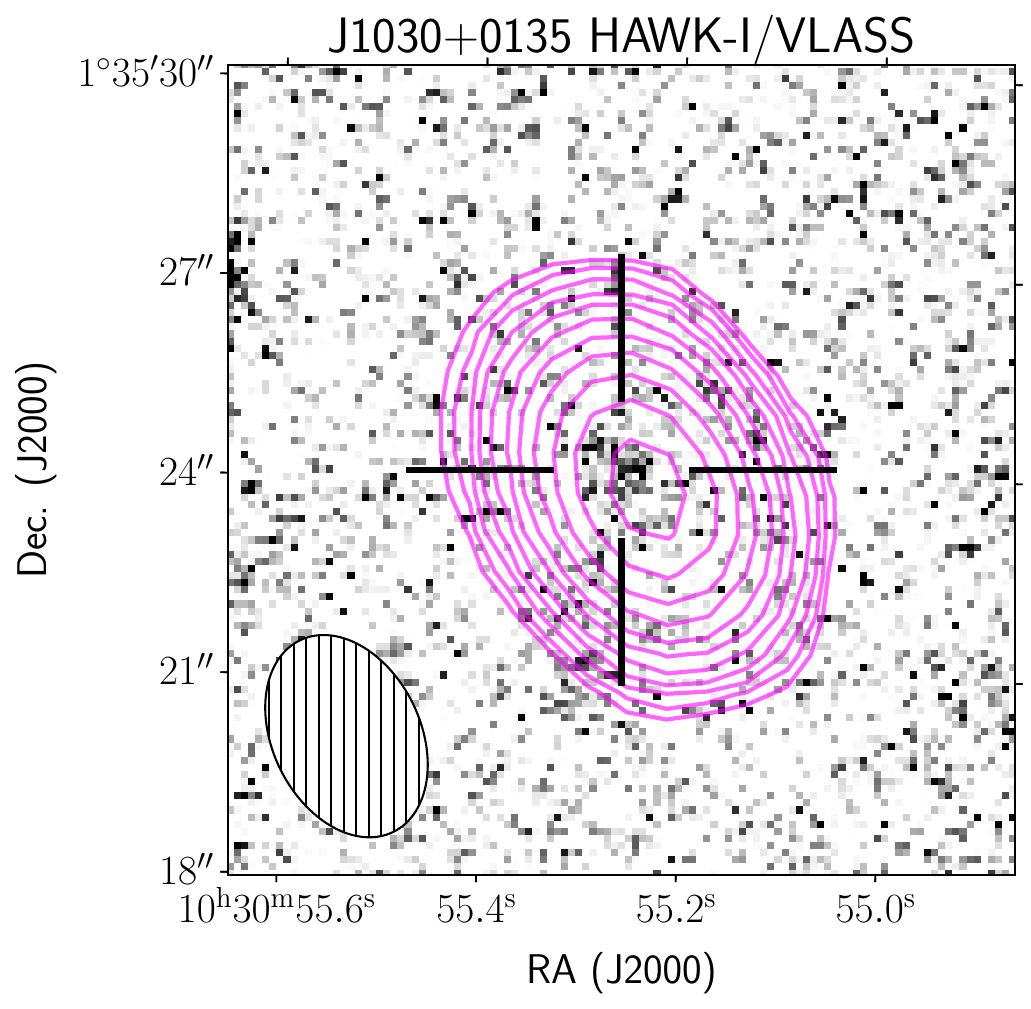}
\end{minipage}%
\hspace{0.02\linewidth}%
\begin{minipage}[t]{0.48\textwidth}
\includegraphics[height=6.5cm]{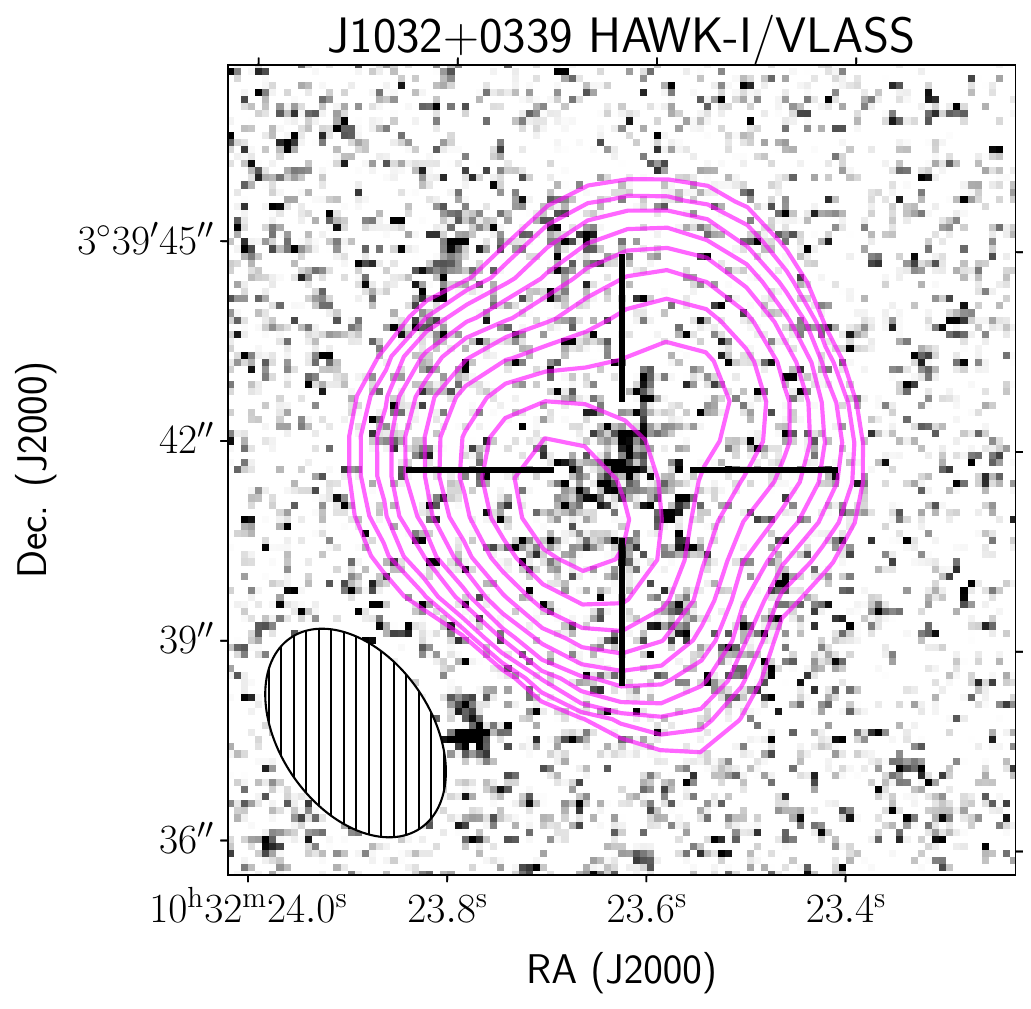}
\end{minipage}
\caption{{\em - continued.}} 
\end{figure*}

\setcounter{figure}{3} 
\begin{figure*}
\begin{minipage}[t]{0.48\textwidth}
\includegraphics[height=6.5cm]{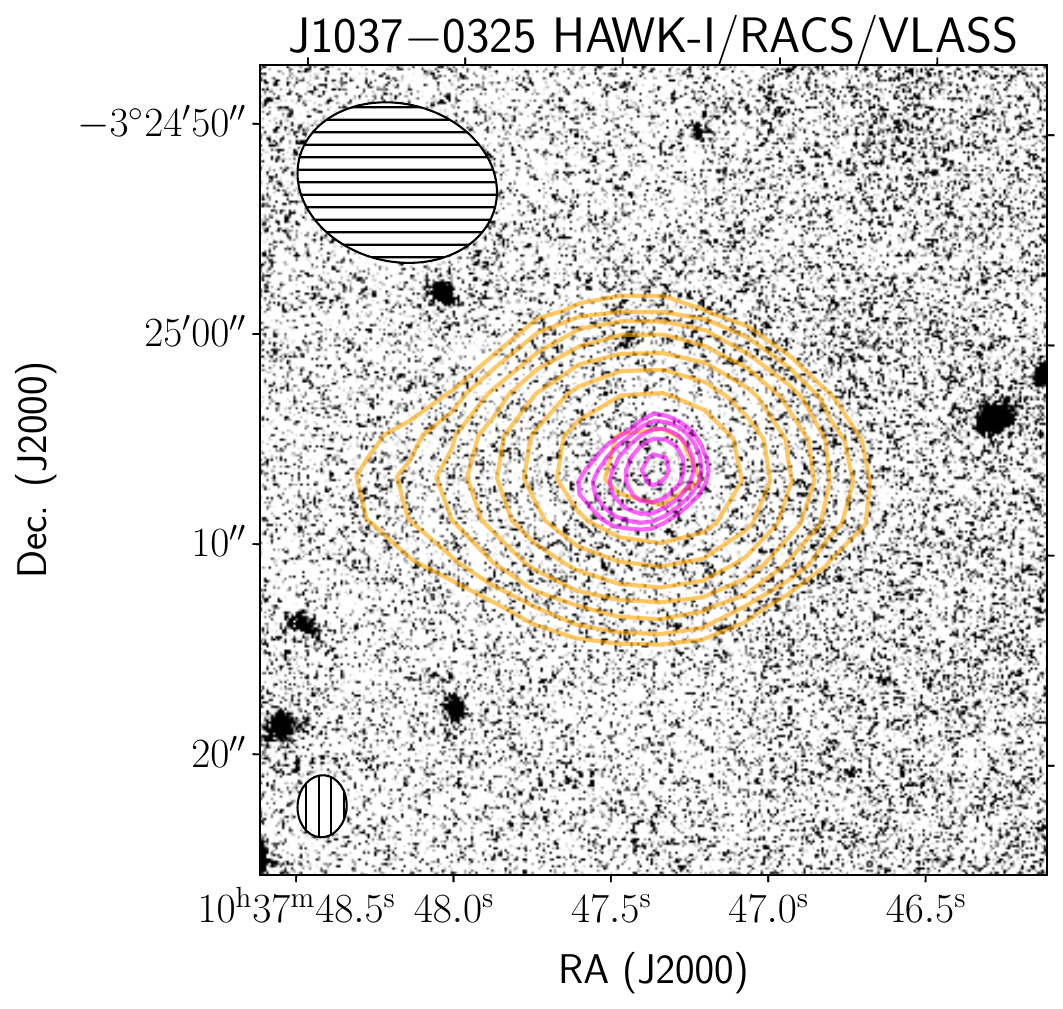}
\end{minipage}
\hspace{0.02\linewidth}%
\begin{minipage}[t]{0.48\textwidth}
\includegraphics[height=6.5cm]{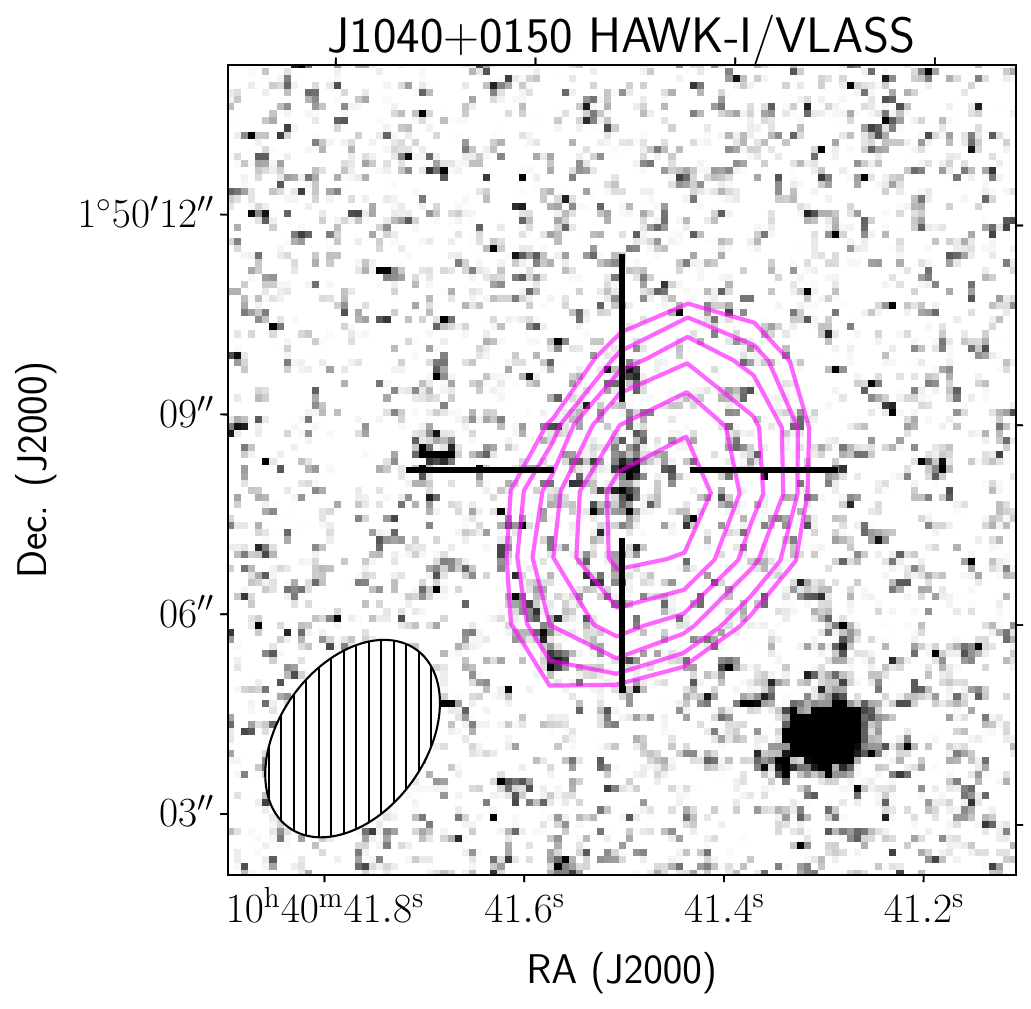}
\end{minipage}
\begin{minipage}[t]{0.48\textwidth}
\includegraphics[height=6.5cm]{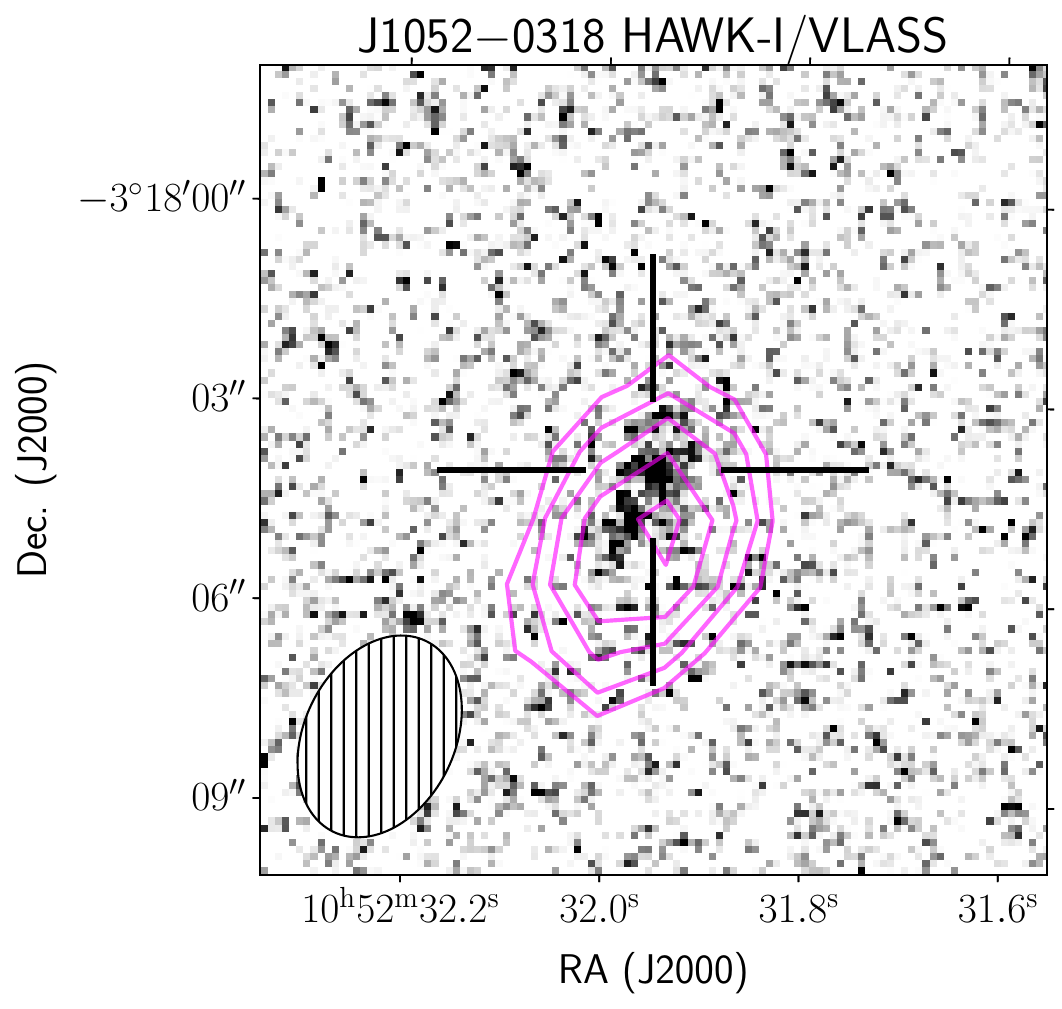}
\end{minipage}%
\hspace{0.02\linewidth}%
\begin{minipage}[t]{0.48\textwidth}
\includegraphics[height=6.5cm]{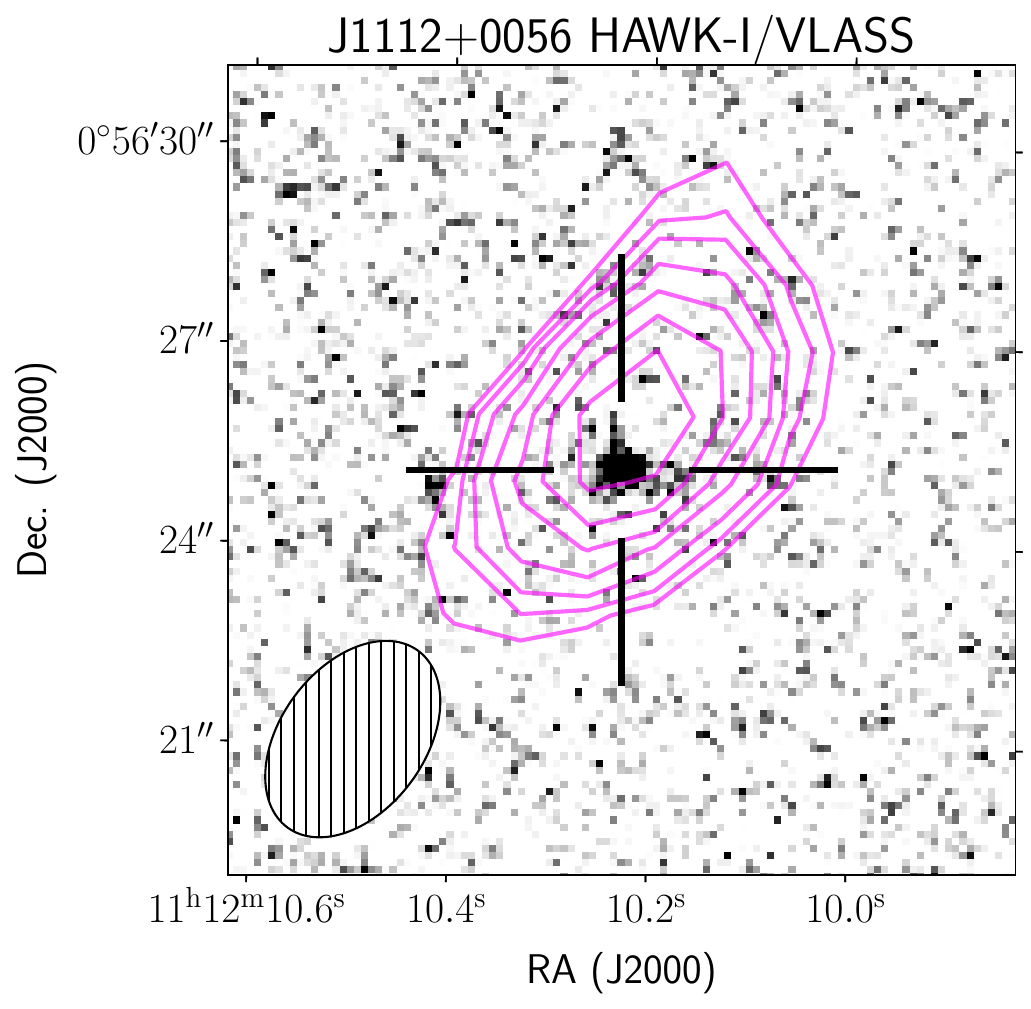}
\end{minipage}
\begin{minipage}[t]{0.48\textwidth}
\includegraphics[height=6.5cm]{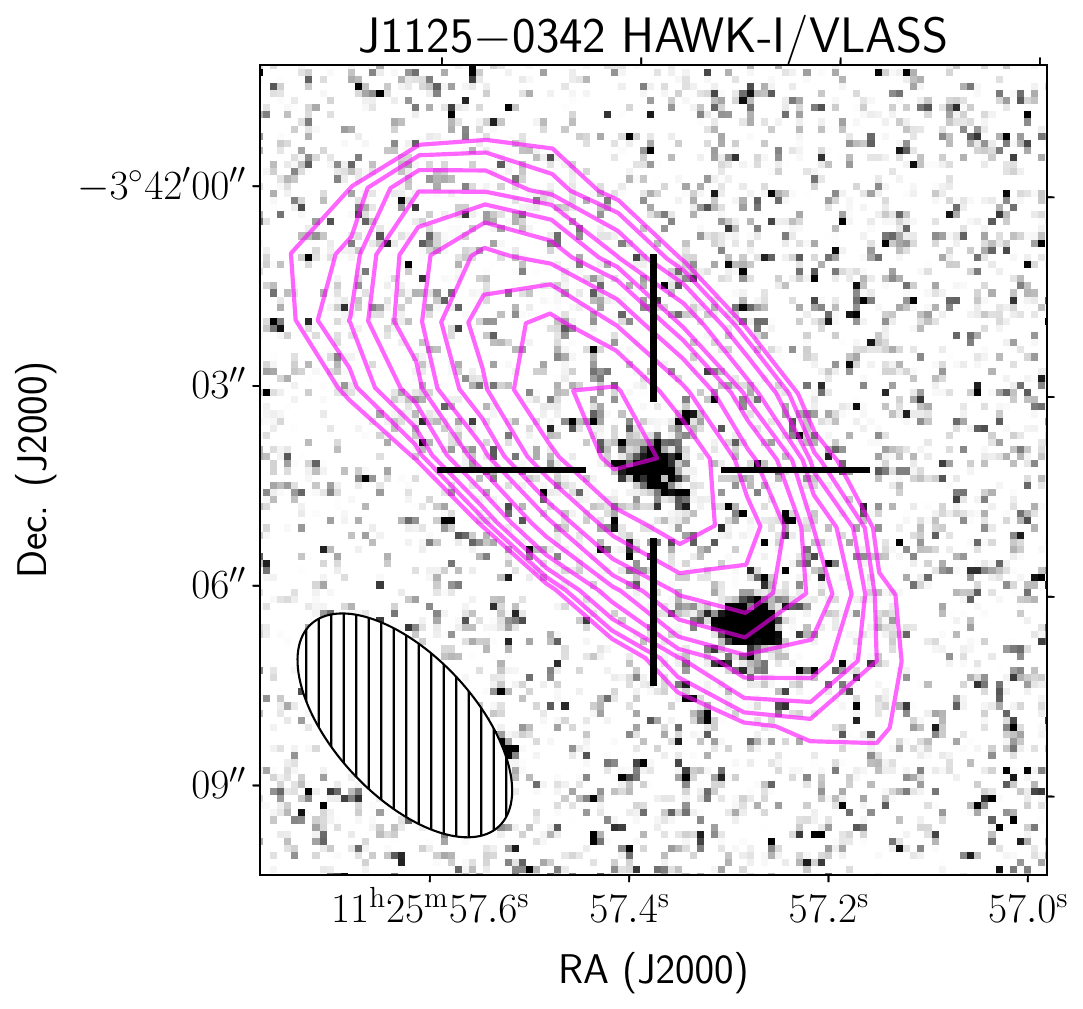}
\end{minipage}%
\hspace{0.02\linewidth}%
\begin{minipage}[t]{0.48\textwidth}
\includegraphics[height=6.5cm]{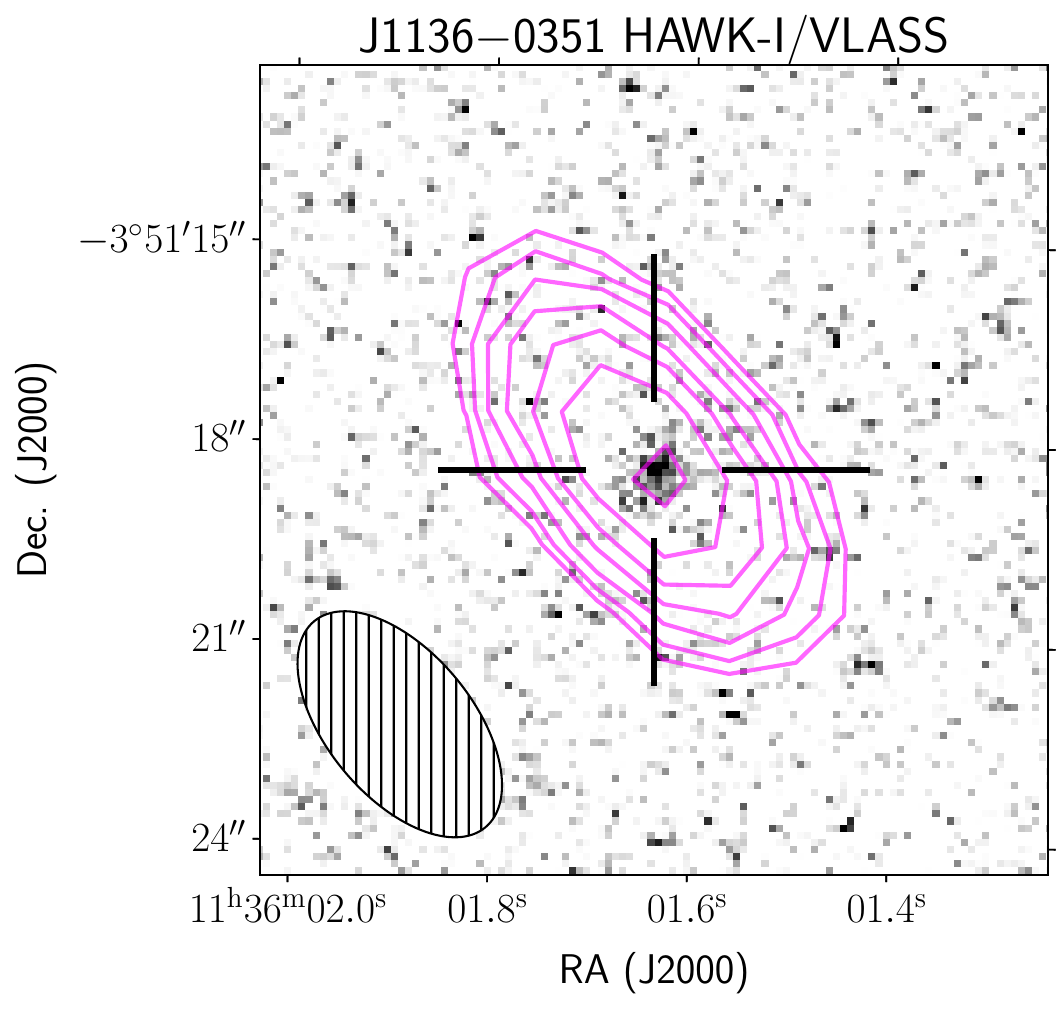}
\end{minipage}
\caption{{\em - continued.}}
\end{figure*}

\setcounter{figure}{3} 
\begin{figure*}
\begin{minipage}[t]{0.48\textwidth}
\includegraphics[height=6.5cm]{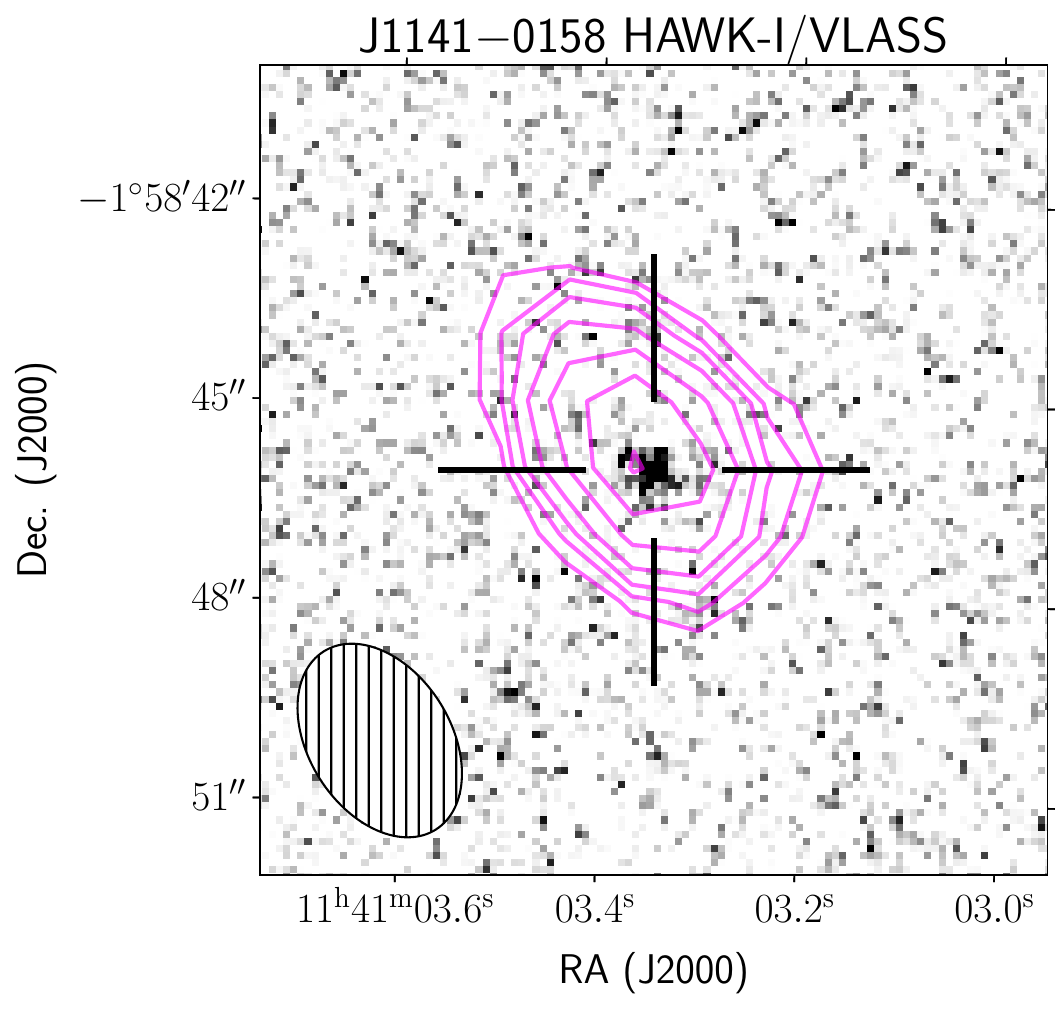}
\end{minipage}
\hspace{0.02\linewidth}%
\begin{minipage}[t]{0.48\textwidth}
\includegraphics[height=6.5cm]{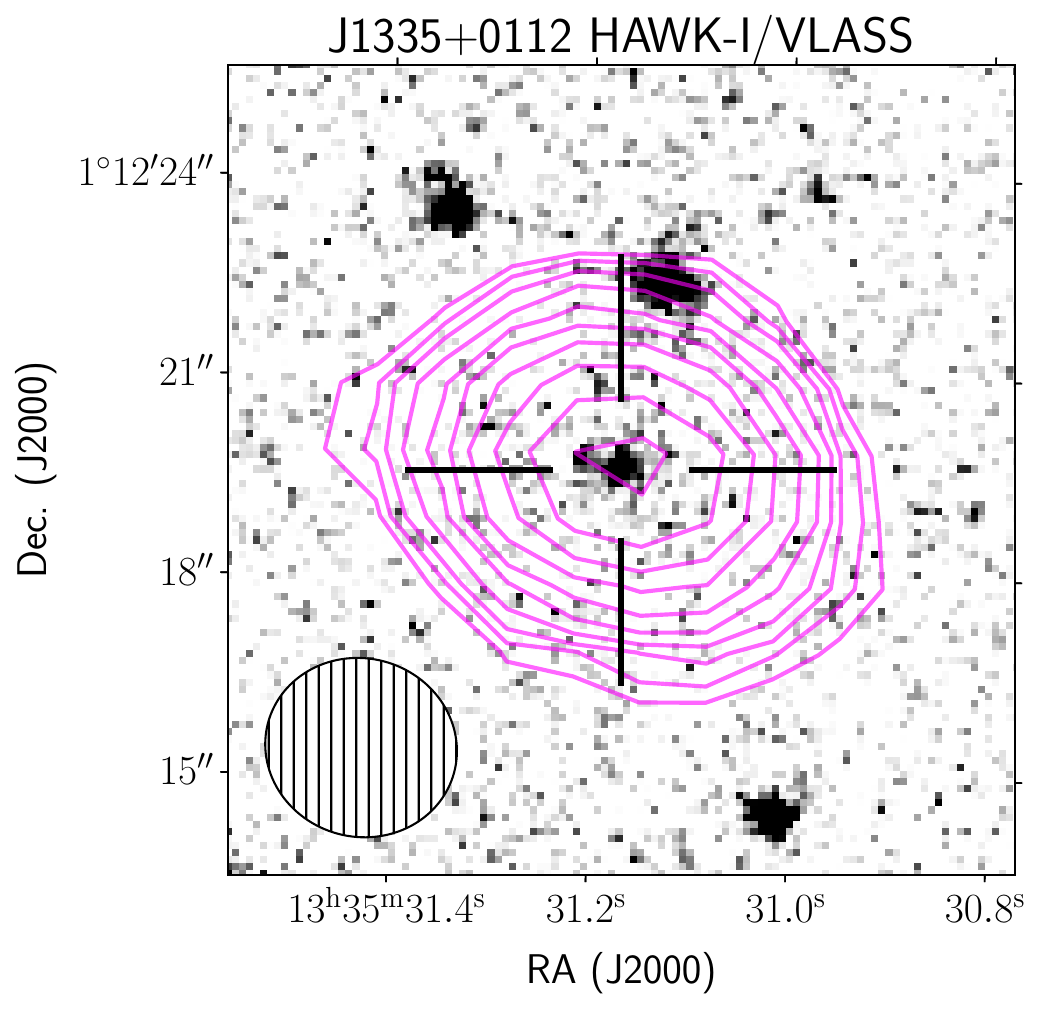}
\end{minipage}
\begin{minipage}[t]{0.48\textwidth}
\includegraphics[height=6.5cm]{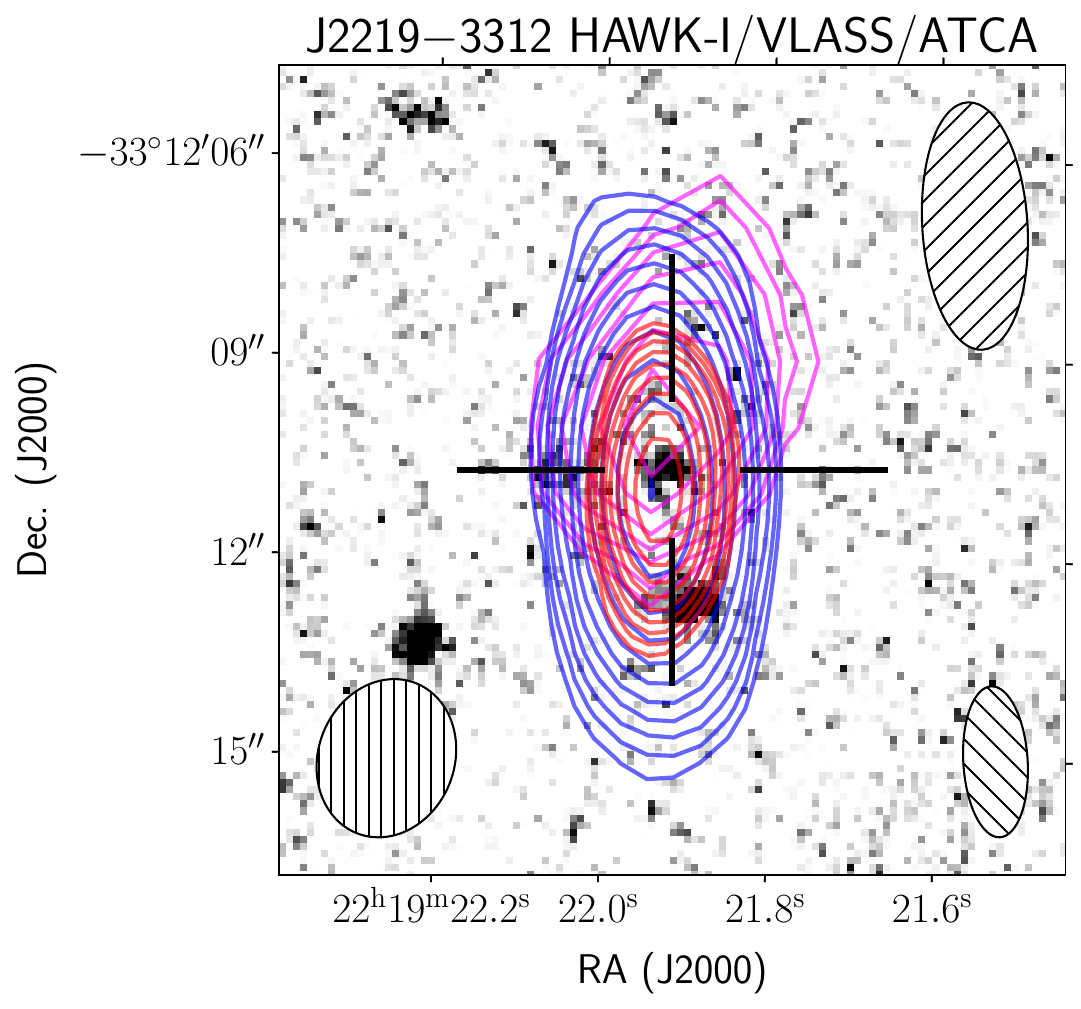}
\end{minipage}%
\hspace{0.02\linewidth}%
\begin{minipage}[t]{0.48\textwidth}
\includegraphics[height=6.5cm]{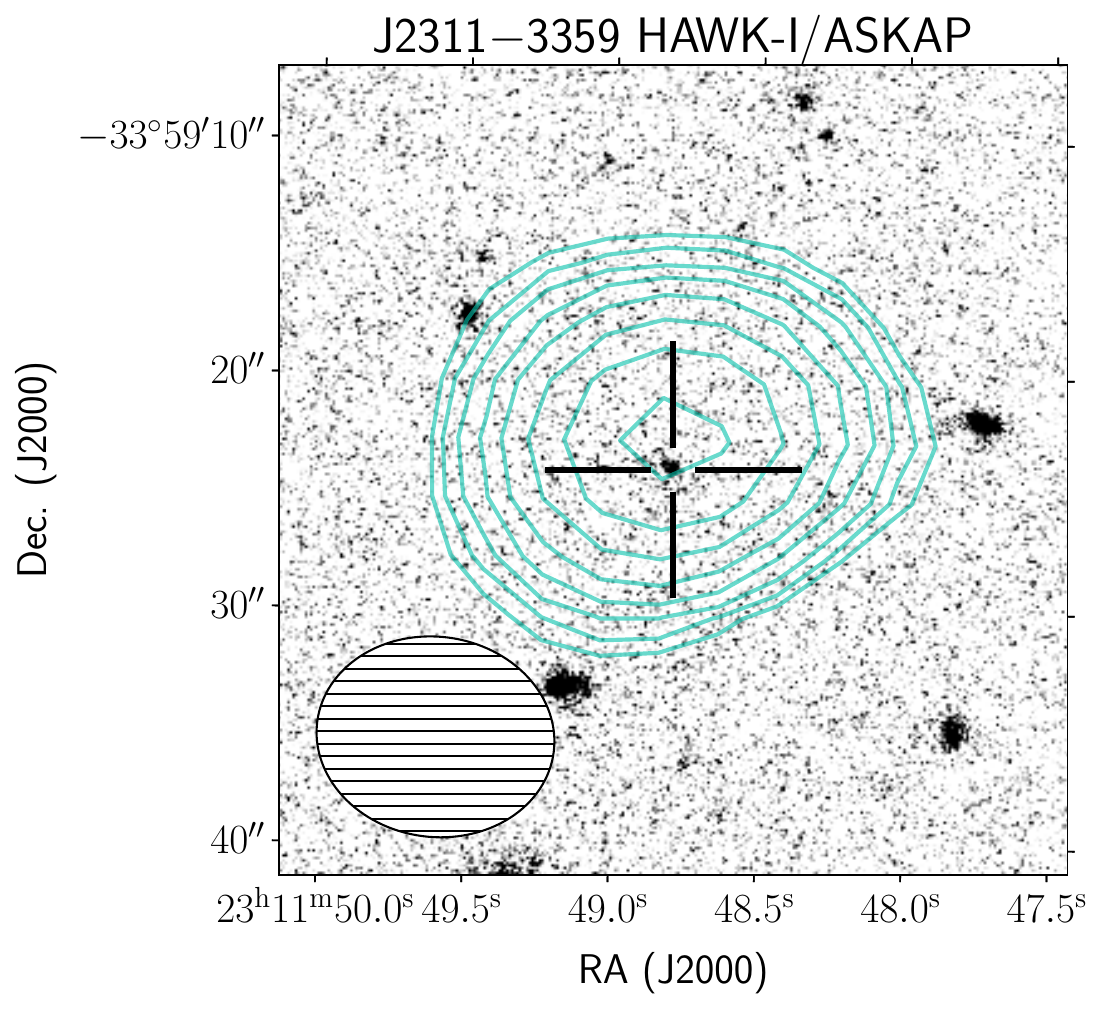}
\end{minipage}
\begin{minipage}[t]{0.48\textwidth}
\includegraphics[height=6.5cm]{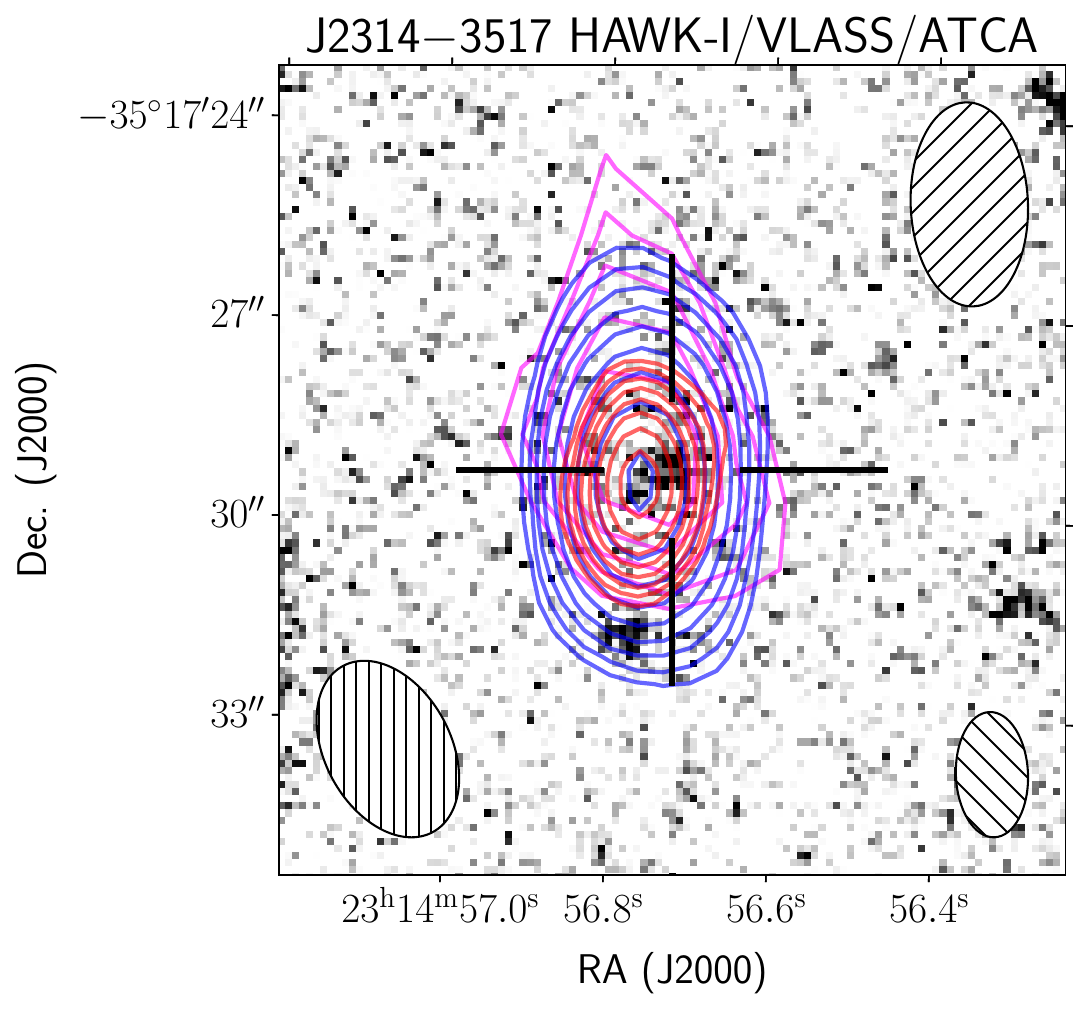}
\end{minipage}%
\hspace{0.02\linewidth}%
\begin{minipage}[t]{0.48\textwidth}
\includegraphics[height=6.5cm]{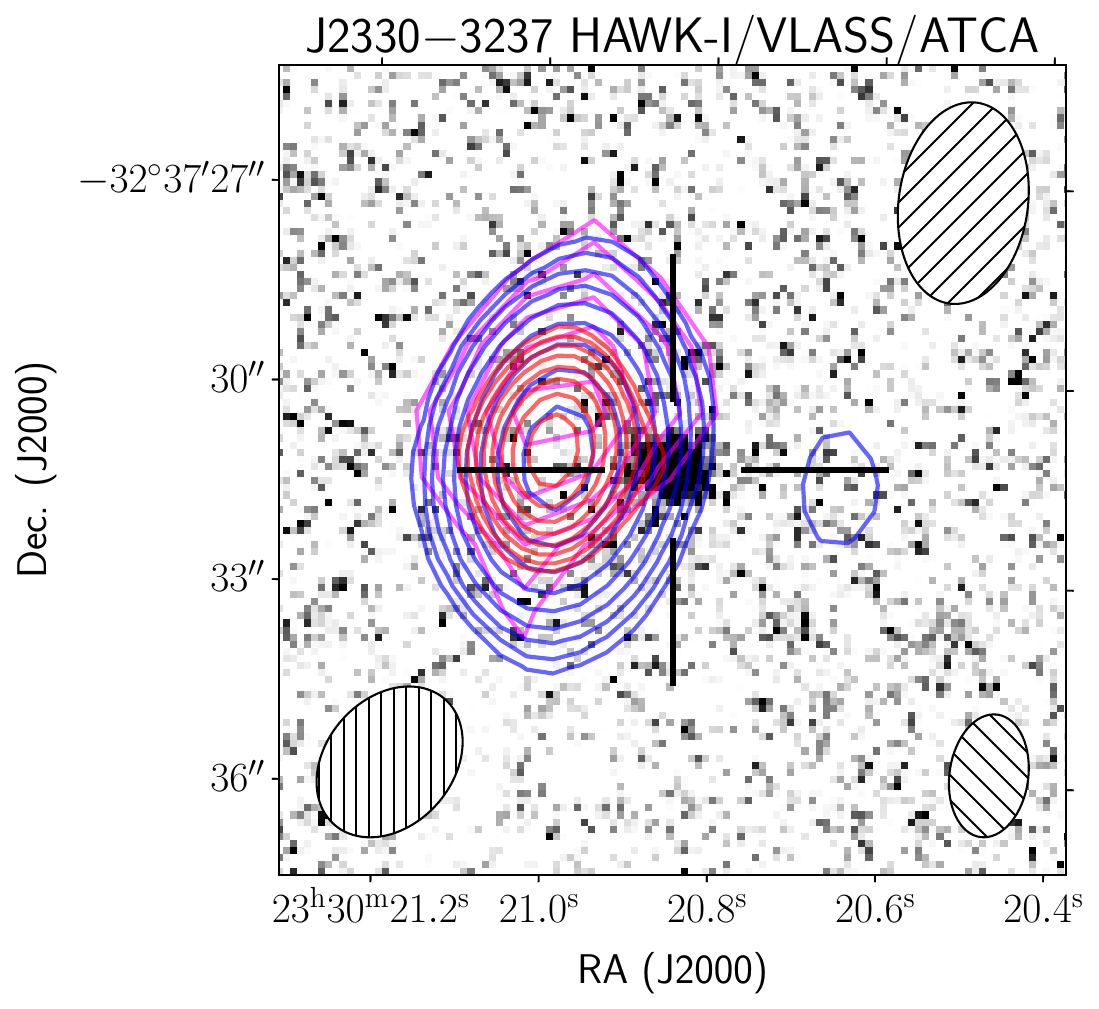}
\end{minipage}
\caption{{\em - continued.}}
\end{figure*}

\begin{figure*}
\begin{minipage}[t]{0.48\textwidth}
\includegraphics[height=6.5cm]{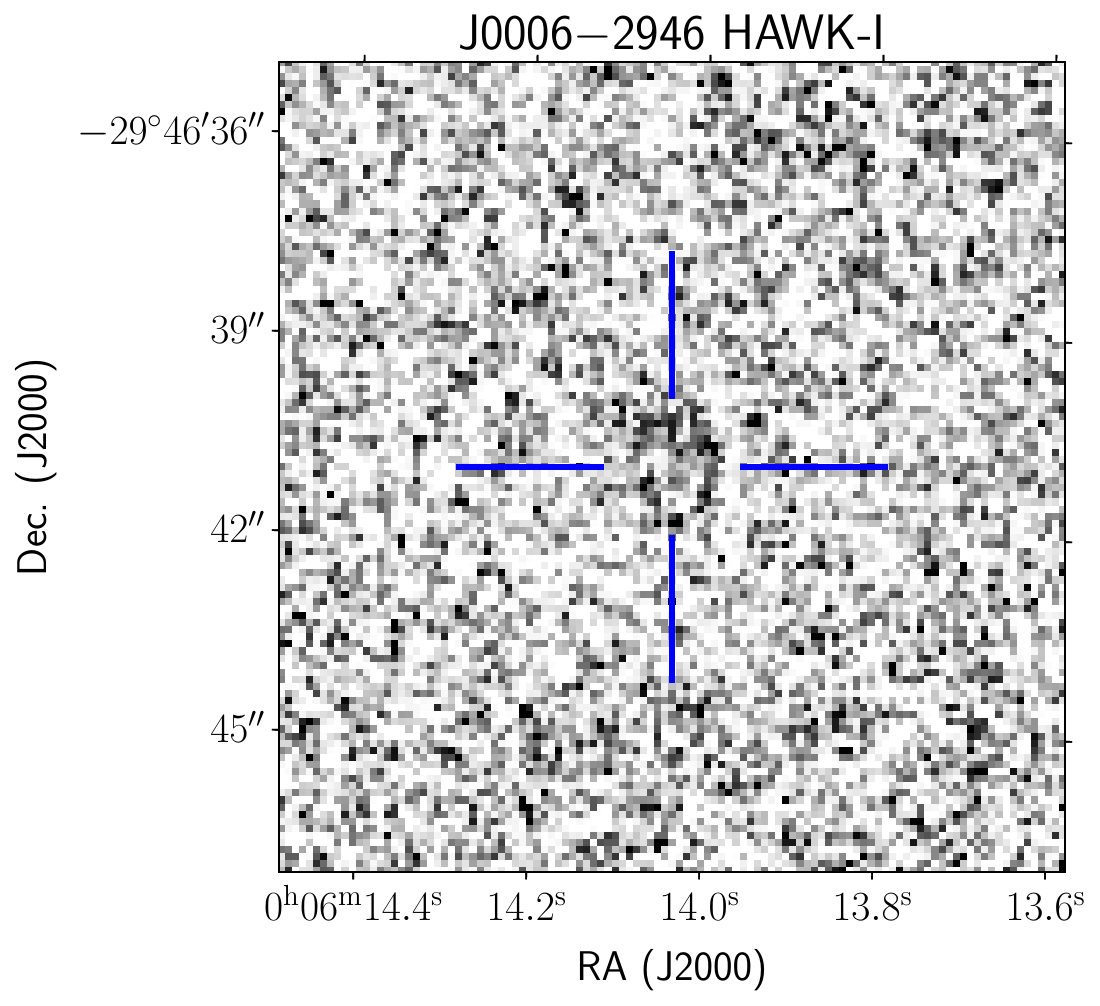}
\end{minipage}%
\hspace{0.02\linewidth}%
\begin{minipage}[t]{0.48\textwidth}
\includegraphics[height=6.5cm]{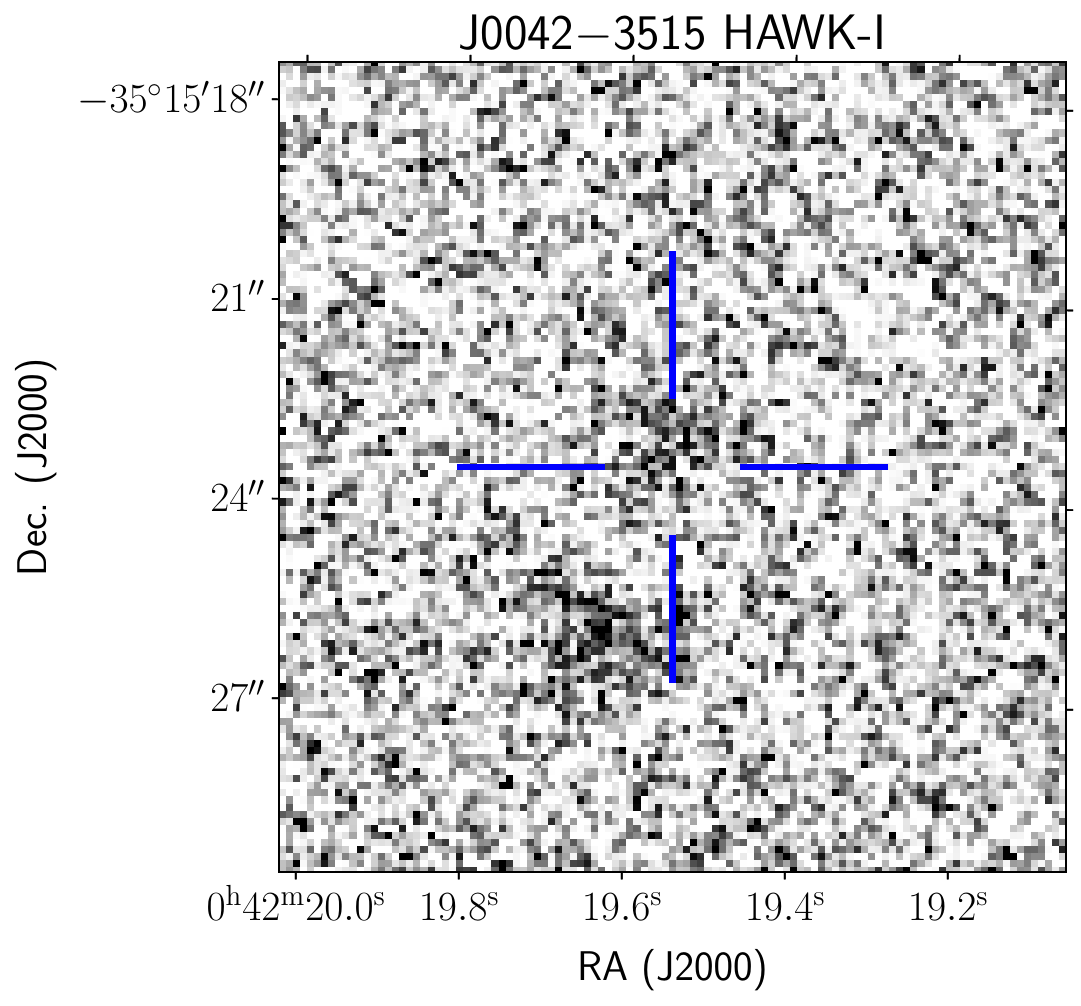}
\end{minipage}
\begin{minipage}[t]{0.48\textwidth}
\includegraphics[height=6.5cm]{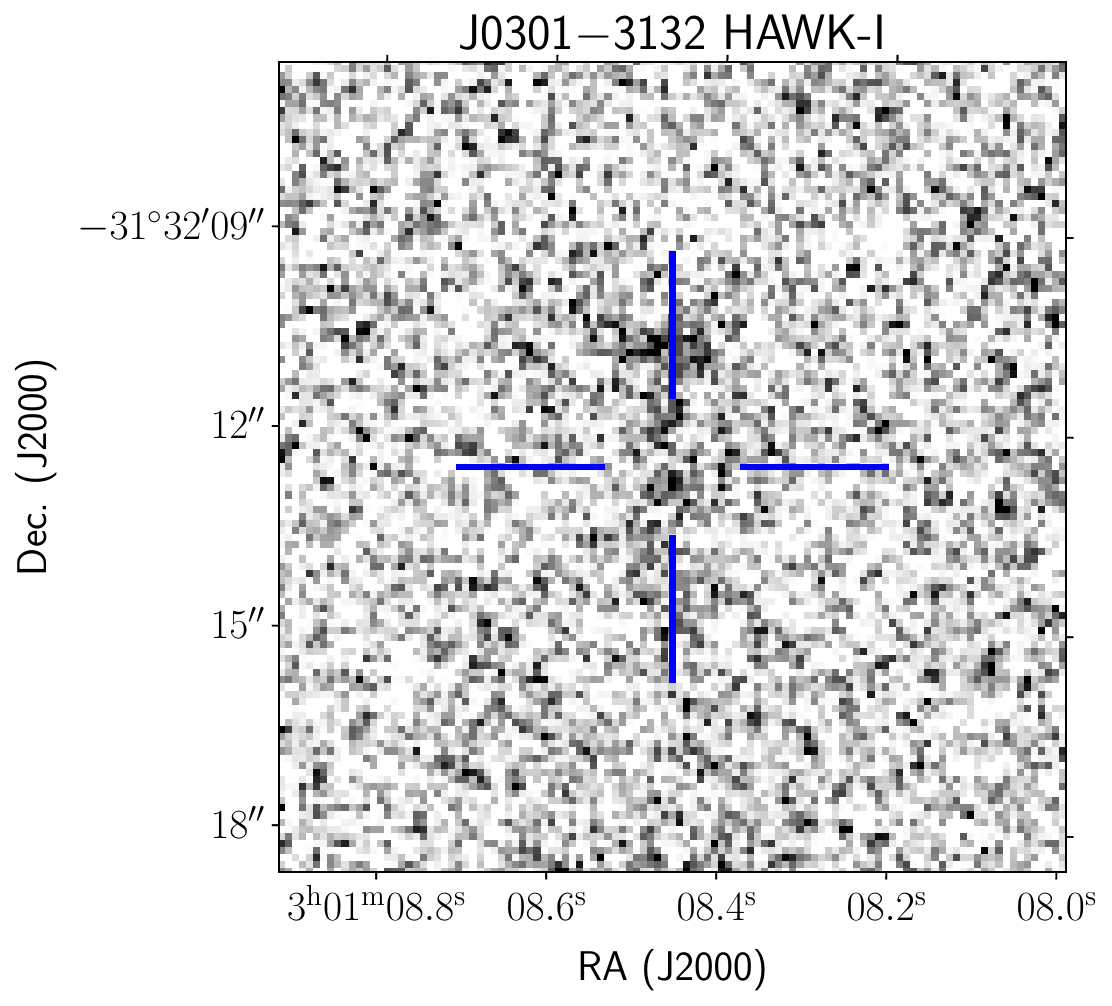}
\end{minipage}%
\hspace{0.02\linewidth}%
\begin{minipage}[t]{0.48\textwidth}
\includegraphics[height=6.5cm]{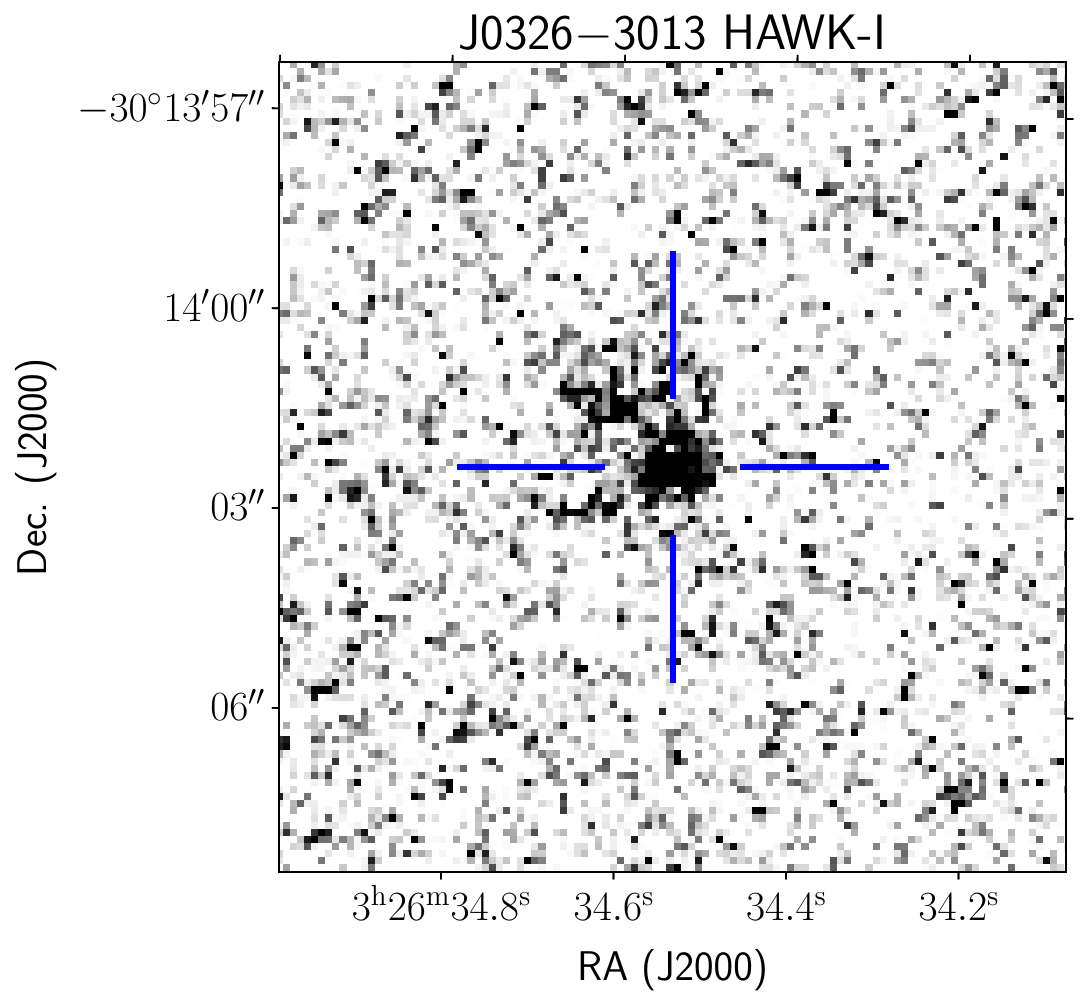}
\end{minipage}
\caption{A selection of the HAWK-I images shown in Figure~\ref{fig:overlays_appendix} without radio contours overlaid, so as to allow a clearer view of the host galaxy detections (marked with crosshairs, as in Figure~\ref{fig:overlays_appendix}). The contrast levels have been modified as well. See the notes on individual sources in this appendix for further details.}
\label{fig:overlays_appendix_2}
\end{figure*}

For all 35 targets for which HAWK-I data are presented in this paper, notes on individual sources are as follows. \\ 
{\bf J0002$-$3514:} The host galaxy, near the radio centroid, was detected by {\sc sextractor}. \\
{\bf J0006$-$2946:} There is evidence of faint, diffuse $K_{\rm s}$-band emission near the radio centroid, almost ring-like in appearance. While this emission is not bright enough to meet our connected pixel S/N criterion for a {\sc sextractor} detection, measuring the integrated flux density with {\sc photutils} in a 2\arcsec\:diameter aperture centred on the ATCA 9-GHz radio centroid yielded S/N $= 4.8$. The HAWK-I image without radio contours overlaid is shown in Figure~\ref{fig:overlays_appendix_2}; the crosshairs mark the position of the ATCA radio centroid.\\
{\bf J0007$-$3040:} This was a target of particular interest in the B22 study given its curved, ultra-steep broadband radio spectrum. The RACS-mid morphology of this source is extended (largest angular size LAS 4\farcs8; deconvolved position angle 36\degr\:measured north through east; \citealt[][]{duchesne23,duchesne24}) and suggestive of an incipient double (i.e. it would be resolved into two components at higher resolution). Therefore, the high-resolution VLASS and ATCA contours very likely map out the north-eastern lobe only. The revised LAS from RACS-mid is larger than the value reported in B22 given the hitherto undetected extension to the south-west. The north-eastern lobe is extended towards the host galaxy, which was detected by {\sc sextractor}. A 2\arcsec\:diameter aperture does not enclose all of the emission and the magnitude is therefore underestimated (in the sense that it is too faint and hence a larger value). The magnitude in a 3\arcsec\:diameter aperture, which encloses all of the flux density, is $K_{\rm s} = 22.34 \pm 0.23$\,mag.\\
{\bf J0034$-$3112:}  RACS-mid reveals that this source is an asymmetric double with LAS $ \approx 12\farcs5$ \citep[][]{duchesne23,duchesne24}; the high-resolution VLASS and ATCA contours map out the western lobe only. The eastern lobe is detected at about the 6$\sigma$ level. The revised LAS falls well outside of the B22 selection criterion and is sufficiently large such that this source is very unlikely to be a UHzRG. There is evidence of a faint host galaxy identification between the lobes and closer to the brighter western lobe. It is possible that the $K_{\rm s}$-band morphology is diffuse and with several components, but a deeper image is needed to better characterise this possible extension. While the $K_{\rm s}$-band emission is not bright enough to meet our connected pixel S/N criterion for a {\sc sextractor} detection, measuring the integrated flux density with {\sc photutils} in a 2\arcsec\:diameter aperture centred on the brightest component yielded S/N $= 4.6$. However, a 2\arcsec\:diameter aperture potentially does not enclose all of the emission and the magnitude is therefore possibly significantly underestimated. The magnitude in a 6\arcsec\:diameter aperture, which encloses all of the potential flux density from the host galaxy, is $K_{\rm s} = 21.81 \pm 0.42$\,mag. \\
{\bf J0042$-$3515:} There is evidence of faint, diffuse $K_{\rm s}$-band emission near the radio centroid. While this emission is not bright enough to meet our connected pixel S/N criterion for a {\sc sextractor} detection, measuring the integrated flux density with {\sc photutils} in a 2\arcsec\:diameter aperture centred on the ATCA 5.5-GHz radio centroid (the 9-GHz detection has relatively low S/N; B22) yielded S/N $= 5.1$. The HAWK-I image without radio contours overlaid is shown in Figure~\ref{fig:overlays_appendix_2}; the crosshairs mark the position of the ATCA 5.5-GHz radio centroid. As can be seen in this figure, there is another nearby diffuse source to the south-east (which is not enclosed by the aperture). \\
{\bf J0048$-$3540:} The host galaxy, near the radio centroid, was detected by {\sc sextractor}. The host is extended with two distinct components. A 2\arcsec\:diameter aperture does not enclose all of the emission and is therefore underestimated. The magnitude in a 3\arcsec\:diameter aperture, which encloses all of the flux density, is $K_{\rm s} = 21.56 \pm 0.14$\,mag. \\
{\bf J0053$-$3256:} The host galaxy was not detected. While it tentatively appears that there is faint $K_{\rm s}$-band emission near the radio centroid, an integrated flux density measurement with {\sc photutils} in a 2\arcsec\:diameter aperture centred on the ATCA 5.5-GHz radio centroid (the 9-GHz detection has relatively low S/N; B22) did not meet our detection threshold (S/N $= 1.5$). This is a high-priority target with $K_{\rm s} > 23.2$ ($3\sigma$); moreover, in a 1.5\arcsec\:diameter aperture, $K_{\rm s} > 23.6$ ($3\sigma$).  \\
{\bf J0108$-$3501:} The host galaxy, near the radio centroid, was detected by {\sc sextractor}. \\
{\bf J0129$-$3109:} The host galaxy, near the radio centroid, was detected by {\sc sextractor}. \\
{\bf J0201$-$3441:} The host galaxy, between the radio lobes, was detected by {\sc sextractor}. \\
{\bf J0216$-$3301:} While there is a hint of a potential host galaxy identification between the radio lobes of this incipient double source, an integrated flux density measurement with {\sc photutils} in a 2\arcsec\:diameter aperture centred on this $K_{\rm s}$-band emission was below our detection threshold (S/N $= 2.1$). This target has $K_{\rm s} > 22.8$ ($3\sigma$); moreover, in a 1.5\arcsec\:diameter aperture, $K_{\rm s} > 23.1$ ($3\sigma$). \\
{\bf J0239$-$3043:} The host galaxy, very close to the radio centroid, was detected by {\sc sextractor}. \\
{\bf J0240$-$3206:} While there is a hint of a potential host galaxy identification near the position of the radio centroid, an integrated flux density measurement with {\sc photutils} in a 2\arcsec\:diameter aperture centred on this $K_{\rm s}$-band emission was below our detection threshold (S/N $= 2.2$). The host galaxy magnitude limit is $K_{\rm s} > 23.3$ ($3\sigma$). However, the compact nature of this candidate host galaxy, coupled with the excellent seeing (0\farcs37), allowed a detection to be made above the $3\sigma$ threshold in a 1.5\arcsec\:diameter aperture. In this case, S/N $= 3.1$ and $K_{\rm s} = 23.65 \pm 0.35$\,mag.\\
{\bf J0301$-$3132:} There is evidence of faint, diffuse $K_{\rm s}$-band emission near the radio centroid. While this emission is not bright enough to meet our connected pixel S/N criterion for a {\sc sextractor} detection, measuring the integrated flux density with {\sc photutils} in a 2\arcsec\:diameter aperture centred on the ATCA 9-GHz radio centroid yielded S/N $= 4.5$. The HAWK-I image without radio contours overlaid is shown in Figure~\ref{fig:overlays_appendix_2}; the crosshairs mark the position of the ATCA radio centroid. The magnitude from the 2\arcsec\:diameter aperture might be significantly underestimated if the brighter diffuse emission to the north of the radio centroid is also associated with the host galaxy; for example, the magnitude in a 4\arcsec\:diameter aperture is $K_{\rm s} = 21.95 \pm 0.26$\,mag. A deeper $K_{\rm s}$-band image is needed to better characterise the diffuse emission near the radio centroid.\\
{\bf J0309$-$3526:} The host galaxy, near the possible core of this multi-component radio source (discussed in Section 5.5 in B22), was detected by {\sc sextractor}. \\
{\bf J0326$-$3013:} We interpret the $K_{\rm s}$-band emission near the radio centroid, detected by {\sc sextractor}, as a host galaxy that is extended and with two distinct components. A 2\arcsec\:diameter aperture does not enclose all of the flux density from these components and the magnitude is therefore significantly underestimated. The magnitude in a 5\arcsec\:diameter aperture, enclosing all of the emission, is $K_{\rm s} = 21.23 \pm 0.20$\,mag; this value is approximately 0.8\,mag brighter than the point-source $5\sigma$ limit from VIKING estimated in B22 (i.e. which did not take into account extended emission). Given the offset between the radio and $K_{\rm s}$-band emission (1\farcs9; Table~\ref{tab:magnitudes}), the former may not be as compact as calculated in B22. VLBI would be useful to determine the underlying radio morphology. Another possibility is that there are two galaxies close in projection, with the fainter emission to the north-east and closer to the radio centroid being the host galaxy. The HAWK-I image without radio contours overlaid is shown in Figure~\ref{fig:overlays_appendix_2}.\\ 
{\bf J0909$-$0154:} The host galaxy, near the radio centroid, was detected by {\sc sextractor}. A 2\arcsec\:diameter aperture does not enclose all of the extended emission and the magnitude is therefore underestimated. The magnitude in a 3\arcsec\:diameter aperture, which encloses all of the emission, is $K_{\rm s} = 21.48 \pm 0.11$\,mag; this value is brighter by approximately 0.2\,mag than the VIKING $5\sigma$ point-source limit estimated in B22 (i.e. which did not take into account extended emission).\\ 
{\bf J1030$+$0135:} There is evidence of faint, diffuse $K_{\rm s}$-band emission near the radio centroid. While this emission is not bright enough to meet our connected pixel S/N criterion for a {\sc sextractor} detection, measuring the integrated flux density with {\sc photutils} in a 2\arcsec\:diameter aperture centred on this emission yielded S/N $= 5.5$. \\
{\bf J1032$+$0339:} The host galaxy, near the radio centroid of this incipient double, was detected by {\sc sextractor}. A 2\arcsec\:diameter aperture does not enclose all of the extended emission and the magnitude is therefore underestimated. The magnitude in a 3\arcsec\:diameter aperture, which encloses all of the emission, is $K_{\rm s} = 21.82 \pm 0.15$\,mag.\\ 
{\bf J1033$+$0107:} While there is a hint of a potential host galaxy identification near the position of the radio centroid, an integrated flux density measurement with {\sc photutils} in a 2\arcsec\:diameter aperture centred on the VLASS radio centroid was below our detection threshold (S/N $= 1.7$). This is a high-priority target with $K_{\rm s} > 23.5$ ($3\sigma$); moreover, in a 1.5\arcsec\:diameter aperture, $K_{\rm s} > 23.9$ ($3\sigma$). \\ 
{\bf J1037$-$0325:} The host galaxy was not detected, but the RACS-mid data suggest an extended radio source with LAS $= 6\farcs6$ \citep[][]{duchesne23,duchesne24} and thus significantly larger than the value reported in B22 (indeed beyond the LAS selection criterion used in that study). These data confirm a discussion in Section 5.2.1 in B22 about the potential extension of this source in the radio that is not apparent in the high-resolution VLASS contours. While the revised LAS suggests that this source is unlikely to be a UHzRG, it is still of interest that the host galaxy was not detected. We determined that $K_{\rm s} > 23.3$ ($3\sigma$; 2\arcsec\:diameter aperture). We also investigated whether this source could be one of the lobes of a giant radio galaxy, but we could not find evidence of another candidate lobe in RACS-mid. \\ 
{\bf J1040$+$0150:} There is evidence of faint, diffuse $K_{\rm s}$-band emission near the radio centroid. While this emission is not bright enough to meet our connected pixel S/N criterion for a {\sc sextractor} detection, measuring the integrated flux density with {\sc photutils} in a 2\arcsec\:diameter aperture centred on this emission yielded S/N $= 5.5$. The magnitude from the 2\arcsec\:diameter aperture might be underestimated depending on the amount of diffuse emission that is associated with the host galaxy; for example, the magnitude in a 4\arcsec\:diameter aperture is $K_{\rm s} = 22.06 \pm 0.30$\,mag. A deeper $K_{\rm s}$-band image is needed to better characterise the diffuse emission near the radio centroid. \\ 
{\bf J1052$-$0318:} The host galaxy, near the radio centroid, was detected by {\sc sextractor}. The host is extended with possible multiple components. A 2\arcsec\:diameter aperture does not enclose all of the extended emission and the magnitude is therefore underestimated. The magnitude in a 3\arcsec\:diameter aperture, which encloses all of the emission, is $K_{\rm s} = 21.80 \pm 0.15$\,mag.\\
{\bf J1112$+$0056:} The host galaxy, near the radio centroid, was detected by {\sc sextractor}. \\  
{\bf J1125$-$0342:} The host galaxy, near the radio centroid, was detected by {\sc sextractor}. \\ 
{\bf J1127$-$0332:} The host galaxy was not detected; this is a high-priority target with $K_{\rm s} > 23.2$ ($3\sigma$; 2\arcsec\:diameter aperture); moreover, in a 1.5\arcsec\:diameter aperture, $K_{\rm s} > 23.6$ ($3\sigma$).  \\ 
{\bf J1136$-$0351:} The host galaxy, near the radio centroid, was detected by {\sc sextractor}. \\ 
{\bf J1141$-$0158:} The host galaxy, near the radio centroid, was detected by {\sc sextractor}. \\ 
{\bf J1335$+$0112:} The host galaxy, very close to the radio centroid, was detected by {\sc sextractor}. \\ 
{\bf J1443$+$0229:} The host galaxy was not detected; this is a high-priority target with $K_{\rm s} > 23.5$ ($3\sigma$; 2\arcsec\:diameter aperture); moreover, in a 1.5\arcsec\:diameter aperture, $K_{\rm s} > 23.9$ ($3\sigma$).  \\  
{\bf J2219$-$3312:} The host galaxy, near the radio centroid, was detected by {\sc sextractor}. Note that a 2\arcsec\:diameter aperture only encloses the source that we have identified as the host. \\ 
{\bf J2311$-$3359:} The host galaxy, near the radio centroid, was detected by {\sc sextractor}. Moreover, \citet[][]{gurkan22} reported a radio LAS $=5\farcs2$, confirming that this source does not meet the B22 LAS selection criterion (see Sections 2.3 and 5.5 in B22). \\ 
{\bf J2314$-$3517:} The host galaxy, near the radio centroid, was detected by {\sc sextractor}. \\ 
{\bf J2326$-$3028:} The host galaxy was not detected; this is a high-priority target with $K_{\rm s} > 23.4$ ($3\sigma$; 2\arcsec\:diameter aperture); moreover, in a 1.5\arcsec\:diameter aperture, $K_{\rm s} > 23.8$ ($3\sigma$). \\ 
{\bf J2330$-$3237:} The host galaxy, located between the two lobes (as indicated by the blue 5.5-GHz contours in Figure~\ref{fig:overlays_appendix}) of this asymmetric radio source, was detected by {\sc sextractor}. The host is closer to the brighter lobe to the east. The HAWK-I detection confirms the suggestion in B22 of a hint of a $K_{\rm s}$-band detection of the host, at the same position, in VIKING. \\ 

\end{document}